\documentclass[entropy,article,accept,moreauthors,pdftex]{mdpi}

\firstpage{1}
\pubvolume{21}
\issuenum{8}
\articlenumber{778}
\pubyear{2019}
\copyrightyear{2019}
\history{Received: 5 July 2019; Accepted: 5 August 2019; Published: 8 August 2019}
\updates{yes}

\newcommand{\plotpathdalphapxypxpy}{(0.000000cm,1.473134cm)--
(0.413600cm,1.535186cm)--
(0.805200cm,1.601703cm)--
(1.179200cm,1.673028cm)--
(1.540000cm,1.749678cm)--
(1.889800cm,1.831828cm)--
(2.235200cm,1.920849cm)--
(2.578400cm,2.017244cm)--
(2.926000cm,2.122889cm)--
(3.284600cm,2.239993cm)--
(3.665200cm,2.372538cm)--
(4.087600cm,2.528113cm)--
(4.613400cm,2.730880cm)--
(5.810200cm,3.203160cm)--
(6.314000cm,3.395062cm)--
(6.600000cm,3.500000cm)}
\newcommand{\plotpathjalpha}{(0.000000cm,0.000000cm)--
(0.143000cm,0.059254cm)--
(0.270600cm,0.119543cm)--
(0.387200cm,0.182056cm)--
(0.492800cm,0.246040cm)--
(0.589600cm,0.312063cm)--
(0.679800cm,0.381153cm)--
(0.761200cm,0.450939cm)--
(0.836000cm,0.522409cm)--
(0.906400cm,0.597226cm)--
(0.970200cm,0.672428cm)--
(1.029600cm,0.749814cm)--
(1.084600cm,0.828816cm)--
(1.135200cm,0.908706cm)--
(1.183600cm,0.992532cm)--
(1.249600cm,1.117013cm)--
(1.313400cm,1.237034cm)--
(1.361800cm,1.320821cm)--
(1.408000cm,1.392894cm)--
(1.454200cm,1.457097cm)--
(1.502600cm,1.516499cm)--
(1.555400cm,1.573181cm)--
(1.612600cm,1.626399cm)--
(1.674200cm,1.675758cm)--
(1.744600cm,1.723916cm)--
(1.823800cm,1.769664cm)--
(1.911800cm,1.812263cm)--
(2.013000cm,1.852972cm)--
(2.129600cm,1.891524cm)--
(2.266000cm,1.928163cm)--
(2.426600cm,1.962757cm)--
(2.618000cm,1.995358cm)--
(2.849000cm,2.025981cm)--
(3.130600cm,2.054532cm)--
(3.480400cm,2.081124cm)--
(3.922600cm,2.105745cm)--
(4.492400cm,2.128360cm)--
(5.244800cm,2.148981cm)--
(6.265600cm,2.167577cm)--
(6.600000cm,2.172200cm)}
\newcommand{\plotpathdeltaalphapxypxpy}{(0.000000cm,2.647648cm)--
(0.114400cm,2.688042cm)--
(0.224400cm,2.719041cm)--
(0.330000cm,2.741035cm)--
(0.433400cm,2.754788cm)--
(0.534600cm,2.760493cm)--
(0.635800cm,2.758330cm)--
(0.737000cm,2.748203cm)--
(0.838200cm,2.730126cm)--
(0.939400cm,2.704220cm)--
(1.042800cm,2.669908cm)--
(1.150600cm,2.626084cm)--
(1.262800cm,2.572308cm)--
(1.379400cm,2.508367cm)--
(1.504800cm,2.431487cm)--
(1.643400cm,2.338224cm)--
(1.804000cm,2.221558cm)--
(2.010800cm,2.062057cm)--
(2.532200cm,1.647488cm)--
(2.752200cm,1.479504cm)--
(2.945800cm,1.339171cm)--
(3.128400cm,1.214543cm)--
(3.304400cm,1.102181cm)--
(3.480400cm,0.997735cm)--
(3.656400cm,0.901254cm)--
(3.834600cm,0.811542cm)--
(4.017200cm,0.727644cm)--
(4.204200cm,0.649726cm)--
(4.397800cm,0.577050cm)--
(4.600200cm,0.509109cm)--
(4.813600cm,0.445606cm)--
(5.038000cm,0.386947cm)--
(5.277800cm,0.332449cm)--
(5.535200cm,0.282213cm)--
(5.812400cm,0.236381cm)--
(6.113800cm,0.194816cm)--
(6.446000cm,0.157325cm)--
(6.600000cm,0.142463cm)}
\newcommand{\plotpathkalpha}{(0.000000cm,2.647648cm)--
(0.198000cm,2.643761cm)--
(0.382800cm,2.632501cm)--
(0.558800cm,2.614071cm)--
(0.726000cm,2.588816cm)--
(0.884400cm,2.557215cm)--
(1.036200cm,2.519272cm)--
(1.183600cm,2.474641cm)--
(1.324400cm,2.424285cm)--
(1.460800cm,2.367798cm)--
(1.592800cm,2.305460cm)--
(1.722600cm,2.236401cm)--
(1.850200cm,2.160671cm)--
(1.975600cm,2.078413cm)--
(2.101000cm,1.988229cm)--
(2.226400cm,1.890096cm)--
(2.354000cm,1.782252cm)--
(2.486000cm,1.662681cm)--
(2.631200cm,1.522810cm)--
(2.807200cm,1.344256cm)--
(3.143800cm,0.994581cm)--
(3.280200cm,0.859477cm)--
(3.394600cm,0.753497cm)--
(3.498000cm,0.665362cm)--
(3.594800cm,0.590534cm)--
(3.689400cm,0.525105cm)--
(3.786200cm,0.466051cm)--
(3.887400cm,0.412346cm)--
(3.997400cm,0.362191cm)--
(4.118400cm,0.315310cm)--
(4.252600cm,0.271573cm)--
(4.404400cm,0.230488cm)--
(4.576000cm,0.192510cm)--
(4.771800cm,0.157699cm)--
(4.996200cm,0.126314cm)--
(5.258000cm,0.098245cm)--
(5.570400cm,0.073443cm)--
(5.951000cm,0.052081cm)--
(6.428400cm,0.034301cm)--
(6.600000cm,0.029611cm)}
\newcommand{\plotpointmi}{(2.200000cm,1.911411cm)}

\usepackage{amssymb}
\usepackage{bbm}
\usepackage{mleftright}

\newcommand{\Prv}[1]{\Pr[#1]}
\newcommand{\abs}[1]{\lvert#1\rvert}
\newcommand{\argmax}{\operatorname*{argmax}}
\newcommand{\bigexpectation}[1]{\operatorname{E}\bigl[#1\bigr]}
\newcommand{\binaryemptyleft}{\mkern-\medmuskip{}}
\newcommand{\card}[1]{\lvert#1\rvert}
\newcommand{\errorexponentpxy}{\mathsf{E}_\mathsf{P}}
\newcommand{\errorexponentqxqy}{\mathsf{E}_\mathsf{Q}}
\newcommand{\figref}[1]{Figure~\ref{#1}}

\newcommand{\kernedqxcommaqy}{Q_X,\vthinspace Q_Y}
\newcommand{\lmaref}[1]{Lemma~\ref{#1}}
\newcommand{\logcard}[1]{\log \vthinspace \card{#1}}
\newcommand{\markov}{\mathrel{\multimap}\joinrel\mathrel{-}\joinrel\mathrel{\mkern-6mu}\joinrel\mathrel{-}}
\newcommand{\mat}[1]{\mathsf{#1}}
\newcommand{\mdpicite}[2][\error]{\cite{#2} (#1)}
\newcommand{\minvphantomsup}{\operatorname*{min\vphantom{sup}}}
\newcommand{\mulspace}{\vthinspace}
\newcommand{\norm}[1]{\lVert#1\rVert}
\newcommand{\propref}[1]{Proposition~\ref{#1}}
\newcommand{\ratex}{\mathsf{R}_\mathsf{X}}
\newcommand{\ratey}{\mathsf{R}_\mathsf{Y}}
\newcommand{\reals}{\mathbb{R}}
\newcommand{\scexponentpxy}{\mathsf{S\hspace{-0.1em}C}_\mathsf{P}}
\newcommand{\secref}[1]{Section~\ref{#1}}
\newcommand{\sepsubandsup}{}
\newcommand{\set}[1]{\mathcal{#1}}
\newcommand{\supp}{\operatorname{supp}}
\newcommand{\thmref}[1]{Theorem~\ref{#1}}
\newcommand{\trans}[1]{#1^\mathsf{T}}
\newcommand{\vect}[1]{\mathrm{#1}}
\newcommand{\vthinnegspace}{\hspace{-0.083em}}
\newcommand{\vthinspace}{\hspace{0.083em}}

\Title{Two Measures of Dependence}

\Author{Amos Lapidoth\href{https://orcid.org/0000-0002-4742-5124}{\orcidicon} and Christoph Pfister *}

\AuthorNames{Amos Lapidoth and Christoph Pfister}

\address{Signal and Information Processing Laboratory, ETH Zurich, 8092 Zurich, Switzerland}

\corres{\hangafter=1 \hangindent=1.05em \hspace{-0.82em}Correspondence: pfister@isi.ee.ethz.ch}

\abstract{Two families of dependence measures between random variables are introduced.
They~are based on the R\'enyi divergence of order $\alpha$ and the relative $\alpha$-entropy, respectively, and both dependence measures reduce to Shannon's mutual information when their order $\alpha$ is one.
The~first measure shares many properties with the mutual information, including the data-processing inequality, and~can be related to the optimal error exponents in composite hypothesis testing.
The second measure does not satisfy the data-processing inequality, but appears naturally in the context of distributed task~encoding.}


\begin{document}

\section{Introduction}

The solutions to many information-theoretic problems can be expressed using Shannon's information measures such as entropy, relative entropy, and mutual information.
Other problems require R\'enyi's information measures, which generalize Shannon's.
In this paper, we analyze two R\'enyi measures of dependence, $J_\alpha(X;Y)$ and $K_\alpha(X;Y)$, between random variables $X$ and $Y$ taking values in the finite sets $\set{X}$ and $\set{Y}$, with $\alpha \in [0,\infty]$ being a parameter.
(Our notation is similar to the one used for the mutual information: technically, $J_\alpha(\cdot)$ and $K_\alpha(\cdot)$ are functions not of $X$ and $Y$, but of their joint probability mass function (PMF) $P_{XY}$.)
For $\alpha \in [0,\infty]$, we define $J_\alpha(X;Y)$ and $K_\alpha(X;Y)$ as
\begin{align}
J_\alpha(X;Y) &\triangleq \min_{(\kernedqxcommaqy) \in \set{P}(\set{X}) \times \set{P}(\set{Y})} D_\alpha(P_{XY}\|Q_X Q_Y),\label{eq:defjalpha}\\
K_\alpha(X;Y) &\triangleq \min_{(\kernedqxcommaqy) \in \set{P}(\set{X}) \times \set{P}(\set{Y})} \Delta_\alpha(P_{XY}\|Q_X Q_Y),\label{eq:defkalpha}
\end{align}
where $\set{P}(\set{X})$ and $\set{P}(\set{Y})$ denote the set of all PMFs over $\set{X}$ and $\set{Y}$, respectively;
$D_\alpha(P\|Q)$ denotes the R\'enyi divergence of order $\alpha$ (see \eqref{eq:defrenyidivergence} ahead); and
$\Delta_\alpha(P\|Q)$ denotes the relative $\alpha$-entropy (see \eqref{eq:defrelativealphaentropy} ahead).
As shown in \propref{prop:jalphavskalpha}, $J_\alpha(X;Y)$ and $K_\alpha(X;Y)$ are in fact closely related.

The measures $J_\alpha(X;Y)$ and $K_\alpha(X;Y)$ have the following operational meanings (see \secref{sec:operationalmeanings}):
$J_\alpha(X;Y)$ is related to the optimal error exponents in testing whether the observed independent and identically distributed (IID) samples were generated according to the joint PMF $P_{XY}$ or an unknown product PMF;
and $K_\alpha(X;Y)$ appears as a penalty term in the sum-rate constraint of distributed task~encoding.

The measures $J_\alpha(X;Y)$ and $K_\alpha(X;Y)$ share many properties with Shannon's mutual information~\cite{Shannon1948}, and both are equal to the mutual information when $\alpha$ is one.
Except for some special cases, we~have no closed-form expressions for $J_\alpha(X;Y)$ or $K_\alpha(X;Y)$.
As illustrated in \figref{fig:jandkvsmarginals}, unless $\alpha$ is one, the minimum in the definitions of $J_\alpha(X;Y)$ and $K_\alpha(X;Y)$ is typically not achieved by $Q_X = P_X$ and $Q_Y = P_Y$.
(When $\alpha$ is one, then the minimum is always achieved by $Q_X = P_X$ and $Q_Y = P_Y$; this follows from \propref{prop:minqxqydpxyqxqy} and the fact that $D_1(P_{XY}\|Q_X Q_Y) = \Delta_1(P_{XY}\|Q_X Q_Y) = D(P_{XY}\|Q_X Q_Y)$.)

\begin{figure}[H]
\centering
\begin{tikzpicture}[>=stealth]
\draw[densely dashed] (2.2cm,0cm)--\plotpointmi;
\begin{scope}
\clip (0cm,0cm) rectangle (6.6cm+1mm,3.5cm+1mm);
\draw[gray,thick] \plotpathdalphapxypxpy;
\draw[thick] \plotpathjalpha;
\end{scope}
\node[inner sep=0mm,gray,anchor=west] at (2.3cm,3cm) {\vphantom{$0$}\smash{$D_\alpha(P_{XY}\|P_X P_Y)$}};
\node[inner sep=0mm] at (5.5cm,1.9cm) {\vphantom{$0$}\smash{$J_\alpha(X;Y)$}};
\draw[->] (0cm,-1mm)--(0cm,3.7cm);
\draw[->] (0cm,0cm)--(7.2cm,0cm);
\draw (2.2cm,0cm)--(2.2cm,-1mm);
\draw (4.4cm,0cm)--(4.4cm,-1mm);
\draw (6.6cm,0cm)--(6.6cm,-1mm);
\node[inner sep=0mm,anchor=north] at (0cm,-2mm) {\vphantom{$0$}\smash{$0$}};
\node[inner sep=0mm,anchor=north] at (2.2cm,-2mm) {\vphantom{$0$}\smash{$1$}};
\node[inner sep=0mm,anchor=north] at (4.4cm,-2mm) {\vphantom{$0$}\smash{$2$}};
\node[inner sep=0mm,anchor=north] at (6.6cm,-2mm) {\vphantom{$0$}\smash{$3$}};
\node[inner sep=0mm,anchor=north] at (7.2cm,-2mm) {\vphantom{$0$}\smash{$\alpha$}};
\begin{scope}[xshift=8.1cm]
\draw[densely dashed] (2.2cm,0cm)--\plotpointmi;
\begin{scope}
\clip (0cm,0cm) rectangle (6.6cm+1mm,3.5cm+1mm);
\draw[gray,thick] \plotpathdeltaalphapxypxpy;
\draw[thick] \plotpathkalpha;
\end{scope}
\node[inner sep=0mm] at (1.1cm,2cm) {\vphantom{$0$}\smash{$K_\alpha(X;Y)$}};
\node[inner sep=0mm,gray] at (4.4cm,1.35cm) {\vphantom{$0$}\smash{$\Delta_\alpha(P_{XY}\|P_X P_Y)$}};
\draw[->] (0cm,-1mm)--(0cm,3.7cm);
\draw[->] (0cm,0cm)--(7.2cm,0cm);
\draw (2.2cm,0cm)--(2.2cm,-1mm);
\draw (4.4cm,0cm)--(4.4cm,-1mm);
\draw (6.6cm,0cm)--(6.6cm,-1mm);
\node[inner sep=0mm,anchor=north] at (0cm,-2mm) {\vphantom{$0$}\smash{$0$}};
\node[inner sep=0mm,anchor=north] at (2.2cm,-2mm) {\vphantom{$0$}\smash{$1$}};
\node[inner sep=0mm,anchor=north] at (4.4cm,-2mm) {\vphantom{$0$}\smash{$2$}};
\node[inner sep=0mm,anchor=north] at (6.6cm,-2mm) {\vphantom{$0$}\smash{$3$}};
\node[inner sep=0mm,anchor=north] at (7.2cm,-2mm) {\vphantom{$0$}\smash{$\alpha$}};
\end{scope}
\end{tikzpicture}
\caption{(\textbf{Left}) $J_\alpha(X;Y)$ and $D_\alpha(P_{XY}\|P_X P_Y)$ versus $\alpha$.
(\textbf{Right}) $K_\alpha(X;Y)$ and $\Delta_\alpha(P_{XY}\|P_X P_Y)$ versus $\alpha$.
In both plots, $X$ is Bernoulli with $\Pr(X = 1) = 0.2$, and $Y$ is equal to $X$.}
\label{fig:jandkvsmarginals}
\end{figure}

The rest of this paper is organized as follows.
In \secref{sec:relatedwork}, we review other generalizations of the mutual information.
In \secref{sec:operationalmeanings}, we discuss the operational meanings of $J_\alpha(X;Y)$ and $K_\alpha(X;Y)$.
In~\secref{sec:preliminaries}, we recall the required R\'enyi information measures and prove some preparatory results.
In \secref{sec:twomeasuresofdependence}, we state the properties of $J_\alpha(X;Y)$ and $K_\alpha(X;Y)$.
In \secref{sec:proofs}, we prove these properties.

\section{Related Work}
\label{sec:relatedwork}

The measure $J_\alpha(X;Y)$ was discovered independently from the authors of the present paper by Tomamichel and Hayashi \mdpicite[Equation~(58)]{TomamichelHayashi},
who, for the case when $\alpha > \frac{1}{2}$, derived some of its properties in \mdpicite[Appendix~A-C]{TomamichelHayashi}.

Other R\'enyi-based measures of dependence appeared in the past.
Notable are those by Sibson \cite{SibsonInformation}, Arimoto \cite{ArimotoEntropy}, and Csisz\'ar \cite{CsiszarInformation}, respectively denoted by $I_\alpha^\mathsf{s}(\cdot)$, $I_\alpha^\mathsf{a}(\cdot)$, and $I_\alpha^\mathsf{c}(\cdot)$:
\begin{align}
I_\alpha^\mathsf{s}(X;Y) &\triangleq \frac{\alpha}{\alpha - 1} \log \sum_y \mleft[\sum_x P(x) \mulspace P(y|x)^\alpha\mright]^\frac{1}{\alpha}\label{eq:sibsonrmd}\\
&= \min_{Q_Y} D_\alpha(P_{XY}\|P_X Q_Y),\label{eq:sibsonvsrenyidiv}\\
I_\alpha^\mathsf{a}(X;Y) &\triangleq H_\alpha(X) - H_\alpha(X|Y)\\[-0.75ex]
&= \frac{\alpha}{\alpha - 1} \log \sum_y \mleft[\sum_x \frac{P(x)^\alpha}{\sum_{x' \in \set{X}} P(x')^\alpha} \mulspace P(y|x)^\alpha\mright]^\frac{1}{\alpha},\label{eq:arimotodefvsezero}\\
I_\alpha^\mathsf{c}(X;Y) &\triangleq \min_{Q_Y} \sum_x P(x) \mulspace D_\alpha(P_{Y|X=x}\|Q_Y),\label{eq:csiszarrenyimi}
\end{align}
where, throughout the paper, $\log(\cdot)$ denotes the base-2 logarithm;
$D_\alpha(P\|Q)$ denotes the R\'enyi divergence of order $\alpha$ (see \eqref{eq:defrenyidivergence} ahead);
$H_\alpha(X)$ denotes the R\'enyi entropy of order $\alpha$ (see \eqref{eq:defhalpha} ahead);
and $H_\alpha(X|Y)$ denotes the Arimoto--R\'enyi conditional entropy \cite{ArimotoEntropy,FehrBerens,SasonVerduArimotoBayesian}, which is defined for positive $\alpha$ other than one as
\begin{align}
H_\alpha(X|Y) \triangleq \frac{\alpha}{1 - \alpha} \log \sum_y \mleft[\sum_x P(x,y)^\alpha\mright]^\frac{1}{\alpha}.\label{eq:defhalphaxgy}
\end{align}
(Equation \eqref{eq:sibsonvsrenyidiv} follows from \propref{prop:infqydalphapxyqxqy} ahead, and \eqref{eq:arimotodefvsezero} follows from \eqref{eq:defhalpha} and \eqref{eq:defhalphaxgy}.)
An overview of $I_\alpha^\mathsf{s}(\cdot)$, $I_\alpha^\mathsf{a}(\cdot)$, and $I_\alpha^\mathsf{c}(\cdot)$ is provided in \cite{VerduAlphaMutual}.
Another R\'enyi-based measure of dependence can be found in~\mdpicite[Equation~(19)]{TridenskiZamirIngber}:
\begin{align}
I_\alpha^\mathsf{t}(X;Y) \triangleq D_\alpha(P_{XY}\|P_X P_Y).
\end{align}

The relation between $I_\alpha^\mathsf{c}(X;Y)$, $J_\alpha(X;Y)$, and $I_\alpha^\mathsf{s}(X;Y)$ for $\alpha > 1$ was established recently:

\begin{Proposition}[{\mdpicite[Theorem~IV.1]{AishwaryaMadiman}}]\label{prop:jalphasandwich}
For every PMF $P_{XY}$ and every $\alpha > 1$,
\begin{align}
I_\alpha^\mathsf{c}(X;Y) &\le J_\alpha(X;Y)\label{eq:jalphasandwicha}\\
&\le I_\alpha^\mathsf{s}(X;Y).\label{eq:jalphasandwichb}
\end{align}
\end{Proposition}

\begin{proof}
This is proved in \cite{AishwaryaMadiman} for a measure-theoretic setting.
Here, we specialize the proof to finite alphabets.
We first prove \eqref{eq:jalphasandwicha}:
\begin{align}
J_\alpha(X;Y) &= \min_{Q_Y} \min_{Q_X} D_\alpha(P_{XY}\|Q_X Q_Y)\label{eq:jalphagecsiszara}\\
&= \min_{Q_Y} \frac{\alpha}{\alpha - 1} \log \sum_x \mleft[\sum_y P(x,y)^\alpha \mulspace Q_Y(y)^{1 - \alpha}\mright]^\frac{1}{\alpha}\label{eq:jalphagecsiszarb}\\
&= \min_{Q_Y} \frac{\alpha}{\alpha - 1} \log \sum_x P(x) \mleft[\sum_y P(y|x)^\alpha \mulspace Q_Y(y)^{1 - \alpha}\mright]^\frac{1}{\alpha}\\
&\ge \min_{Q_Y} \frac{\alpha}{\alpha - 1} \sum_x P(x) \log \mleft[\sum_y P(y|x)^\alpha \mulspace Q_Y(y)^{1 - \alpha}\mright]^\frac{1}{\alpha}\label{eq:jalphagecsiszard}\\
&= \min_{Q_Y} \sum_x P(x) \frac{1}{\alpha - 1} \log \sum_y P(y|x)^\alpha \mulspace Q_Y(y)^{1 - \alpha}\\
&= I_\alpha^\mathsf{c}(X;Y),\label{eq:jalphagecsiszarf}
\end{align}
where \eqref{eq:jalphagecsiszara} follows from the definition of $J_\alpha(X;Y)$ in \eqref{eq:defjalpha};
\eqref{eq:jalphagecsiszarb} follows from \propref{prop:infqydalphapxyqxqy} ahead with the roles of $Q_X$ and $Q_Y$ swapped;
\eqref{eq:jalphagecsiszard} follows from Jensen's inequality because $\log(\cdot)$ is concave and because $\frac{\alpha}{\alpha - 1} > 0$;
and \eqref{eq:jalphagecsiszarf} follows from the definition of $I_\alpha^\mathsf{c}(X;Y)$ in \eqref{eq:csiszarrenyimi}.

We next prove \eqref{eq:jalphasandwichb}:
\begin{align}
J_\alpha(X;Y) &= \min_{Q_X,Q_Y} D_\alpha(P_{XY}\|Q_X Q_Y)\label{eq:jalphalesibsona}\\
&\le \min_{Q_Y} D_\alpha(P_{XY}\|P_X Q_Y)\\
&= I_\alpha^\mathsf{s}(X;Y),\label{eq:jalphalesibsonc}
\end{align}
where \eqref{eq:jalphalesibsona} follows from the definition of $J_\alpha(X;Y)$ in \eqref{eq:defjalpha}, and \eqref{eq:jalphalesibsonc} follows from \eqref{eq:sibsonvsrenyidiv}.
\end{proof}

Many of the above R\'enyi information measures coincide when they are maximized over $P_X$ with $P_{Y|X}$ held fixed: for every conditional PMF $P_{Y|X}$ and every positive $\alpha$ other than one,
\begin{align}
\max_{P_X} I_\alpha^\mathsf{a}(P_X P_{Y|X}) &= \max_{P_X} I_\alpha^\mathsf{s}(P_X P_{Y|X})\label{eq:maxpxarimoto}\\
&= \max_{P_X} I_\alpha^\mathsf{c}(P_X P_{Y|X}),\label{eq:maxpxcsiszar}
\end{align}
where $P_X P_{Y|X}$ denotes the joint PMF of $X$ and $Y$;
\eqref{eq:maxpxarimoto} follows from \mdpicite[Lemma~1]{ArimotoEntropy};
and \eqref{eq:maxpxcsiszar} follows from \mdpicite[Proposition~1]{CsiszarInformation}.
It was recently established that, for $\alpha > 1$, this is also true for $J_\alpha(X;Y)$:

\begin{Proposition}[{\mdpicite[Theorem~V.1]{AishwaryaMadiman}}]
For every conditional PMF $P_{Y|X}$ and every $\alpha > 1$,
\begin{align}
\max_{P_X} J_\alpha(P_X P_{Y|X}) = \max_{P_X} I_\alpha^\mathsf{s}(P_X P_{Y|X}).\label{eq:equalalphacapacities}
\end{align}
\end{Proposition}

\begin{proof}
By \propref{prop:jalphasandwich}, we have for all $\alpha > 1$
\begin{align}
\max_{P_X} I_\alpha^\mathsf{c}(P_X P_{Y|X}) &\le \max_{P_X} J_\alpha(P_X P_{Y|X})\label{eq:maxpxjalphasandwicha}\\
&\le \max_{P_X} I_\alpha^\mathsf{s}(P_X P_{Y|X}).\label{eq:maxpxjalphasandwichb}
\end{align}
By \eqref{eq:maxpxcsiszar}, the left-hand side (LHS) of \eqref{eq:maxpxjalphasandwicha} is equal to the right-hand side (RHS) of \eqref{eq:maxpxjalphasandwichb}, so \eqref{eq:maxpxjalphasandwicha} and~\eqref{eq:maxpxjalphasandwichb}~both hold with equality.
\end{proof}

Dependence measures can also be based on the $f$-divergence $D_f(P\|Q)$ \cite{CsiszarShields,LieseVajda,fDivergenceInequalities}.
Every convex function $f\colon (0,\infty) \to \reals$ satisfying $f(1) = 0$ induces a dependence measure, namely
\begin{align}
I_f(X;Y) &\triangleq D_f(P_{XY}\|P_X P_Y)\\
&= \sum_{x,y} P(x) \mulspace P(y) \mulspace f\mleft(\frac{P(x,y)}{P(x) \mulspace P(y)}\mright),\label{eq:fdivdepmeasure}
\end{align}
where \eqref{eq:fdivdepmeasure} follows from the definition of the $f$-divergence.
(For $f(t) = t \log t$, $I_f(X;Y)$ is the mutual information.)
Such dependence measures are used for example in \cite{JiaoHanWeissman}, and a construction equivalent to \eqref{eq:fdivdepmeasure} is studied in \cite{ZivZakai}.

\section{Operational Meanings}
\label{sec:operationalmeanings}

In this section, we discuss the operational meaning of $J_\alpha(X;Y)$ in hypothesis testing (\secref{sec:opmeanjalpha}) and of $K_\alpha(X;Y)$ in distributed task encoding (\secref{sec:opmeankalpha}).

\subsection{Testing Against Independence and $J_\alpha(X;Y)$}
\label{sec:opmeanjalpha}

Consider the hypothesis testing problem of guessing whether an observed sequence of pairs was drawn IID from some given joint PMF $P_{XY}$ or IID from some unknown product distribution.
Thus, based on a sequence of pairs of random variables $\{(X_i,Y_i)\}_{i=1}^n$, two hypotheses have to be distinguished:
\begin{enumerate}[midpenalty=10000]
\item[0)]
Under the null hypothesis, $(X_1,Y_1),\ldots,(X_n,Y_n)$ are IID according to $P_{XY}$.
\item[1)]
Under the alternative hypothesis, $(X_1,Y_1),\ldots,(X_n,Y_n)$ are IID according to some unknown PMF of the form $Q_{XY} = Q_X Q_Y$, where $Q_X$ and $Q_Y$ are arbitrary PMFs over $\set{X}$ and $\set{Y}$, respectively.
\end{enumerate}

Associated with every deterministic test $T_n\colon \set{X}^n \times \set{Y}^n \to \{0,1\}$ and pair $(Q_X,Q_Y)$ are the type-I error probability $P_{XY}^{\times n}[T_n(X^n,Y^n) = 1]$ and the type-II error probability $(Q_X Q_Y)^{\times n}[T_n(X^n,Y^n) = 0]$,
where $R_{XY}^{\times n}[\set{A}]$ denotes the probability of an event $\set{A}$ when $\{(X_i,Y_i)\}_{i=1}^n$ are IID according to $R_{XY}$.
We seek sequences of tests whose worst-case type-II error probability decays exponentially faster than $2^{-n \errorexponentqxqy}$.
To be more specific, for a fixed $\errorexponentqxqy \in \reals$, denote by $\set{T}(\errorexponentqxqy)$ the set of all sequences of deterministic tests $\{T_n\}_{n=1}^\infty$ for which
\begin{align}
\liminf_{n \to \infty} \min_{\kernedqxcommaqy} -\frac{1}{n} \log \bigl((Q_X Q_Y)^{\times n}[T_n(X^n,Y^n) = 0]\bigr) > \errorexponentqxqy,\label{eq:hypotestrateqxqyconstraint}
\end{align}
where $\log(\cdot)$ denotes the base-2 logarithm.
Note that \eqref{eq:hypotestrateqxqyconstraint} implies---but is not equivalent to---that for $n$~sufficiently large, $(Q_X Q_Y)^{\times n}[T_n(X^n,Y^n) = 0] \le 2^{-n \mulspace \errorexponentqxqy}$ for all $(Q_X,Q_Y) \in \set{P}(\set{X}) \times \set{P}(\set{Y})$.
For a fixed $\errorexponentqxqy \in \reals$, the optimal type-I error exponent that can be asymptotically achieved under the constraint \eqref{eq:hypotestrateqxqyconstraint} is given by
\begin{align}
\errorexponentpxy(\errorexponentqxqy) \triangleq \sup_{\{T_n\}_{n=1}^\infty \in \set{T}(\errorexponentqxqy)} \liminf_{n \to \infty} -\frac{1}{n} \log \bigl(P_{XY}^{\times n}[T_n(X^n,Y^n) = 1]\bigr).
\end{align}

The measure $J_\alpha(X;Y)$ appears as follows:
In \mdpicite[first part of (57)]{TomamichelHayashi}, it is shown that for $\errorexponentqxqy$ sufficiently close to $I(X;Y)$,
\begin{align}
\errorexponentpxy(\errorexponentqxqy) = \sup_{\alpha \in (\frac{1}{2},1]} \frac{1 - \alpha}{\alpha} \mulspace (J_\alpha(X;Y) - \errorexponentqxqy),\label{eq:hypotesterrorexponentclosemi}
\end{align}
and in \mdpicite[Theorem~3]{TestingAgainstIndependence}, it is shown that for all $\errorexponentqxqy \in \reals$,
\begin{align}
\errorexponentpxy^{**}(\errorexponentqxqy) = \sup_{\alpha \in (0,1]} \frac{1 - \alpha}{\alpha} \mulspace (J_\alpha(X;Y) - \errorexponentqxqy),\label{eq:hypotesterrorexponentbiconjugate}
\end{align}
where $\errorexponentpxy^{**}(\cdot)$ denotes the Fenchel biconjugate of $\errorexponentpxy(\cdot)$.
In general, the Fenchel biconjugation cannot be omitted because sometimes \mdpicite[Equation~(11) and Example~14]{TestingAgainstIndependence}
\begin{align}
\errorexponentpxy(\errorexponentqxqy) \ne \errorexponentpxy^{**}(\errorexponentqxqy).\label{eq:hypotestnebiconjugate}
\end{align}

For large values of $\errorexponentqxqy$, the optimal type-I error tends to one as $n$ tends to infinity.
In this case, the type-I strong-converse exponent \cite{HanKobayashi,Nakagawa}, which is defined for a sequence of tests $\{T_n\}_{n=1}^\infty$ as
\begin{align}
\scexponentpxy \triangleq \limsup_{n \to \infty} -\frac{1}{n} \log \bigl(1 - P_{XY}^{\times n}[T_n(X^n,Y^n) = 1]\bigr),\label{eq:hypotestscexpfixedtns}
\end{align}
measures how fast the type-I error tends to one as $n$ tends to infinity (smaller values correspond to lower error probabilities).
For a fixed $\errorexponentqxqy \in \reals$, the optimal type-I strong-converse exponent that can be asymptotically achieved under the constraint \eqref{eq:hypotestrateqxqyconstraint} is given by
\begin{align}
\scexponentpxy(\errorexponentqxqy) \triangleq \inf_{\{T_n\}_{n=1}^\infty \in \set{T}(\errorexponentqxqy)} \limsup_{n \to \infty} -\frac{1}{n} \log \bigl(1 - P_{XY}^{\times n}[T_n(X^n,Y^n) = 1]\bigr).\label{eq:hypotestscexp}
\end{align}
In \mdpicite[second part of (57)]{TomamichelHayashi}, it is shown that for $\errorexponentqxqy$ sufficiently close to $I(X;Y)$,
\begin{align}
\scexponentpxy(\errorexponentqxqy) = \sup_{\alpha > 1} \frac{1 - \alpha}{\alpha} \mulspace (J_\alpha(X;Y) - \errorexponentqxqy).\label{eq:hypoteststrongconverseexponent}
\end{align}
Here, the same $\frac{1 - \alpha}{\alpha} \mulspace (J_\alpha(X;Y) - \errorexponentqxqy)$ expression appears as in \eqref{eq:hypotesterrorexponentclosemi} and \eqref{eq:hypotesterrorexponentbiconjugate}, but with a different set of $\alpha$'s to optimize over.

\subsection{Distributed Task Encoding and $K_\alpha(X;Y)$}
\label{sec:opmeankalpha}

The task-encoding problem studied in \cite{TaskEncoding} can be extended to a distributed setting as follows~\cite{DistributedTaskEncoding}:
A source $\{(X_i,Y_i)\}_{i=1}^\infty$ emits pairs of random variables $(X_i,Y_i)$ taking values in a finite alphabet $\set{X} \times \set{Y}$.
For a fixed rate pair $(\ratex,\ratey) \in \reals_{\ge 0}^2$ and a positive integer $n$, the sequences $\{X_i\}_{i=1}^n$ and $\{Y_i\}_{i=1}^n$ are described separately using $\lfloor 2^{n \ratex} \rfloor$ and $\lfloor 2^{n \ratey} \rfloor$ labels, respectively.
The decoder produces a list comprising all the pairs $(x^n,y^n)$ whose description matches the given labels, and the goal is to minimize the $\rho$-th moment of the list size as $n$~tends to infinity (for some $\rho > 0$).

For a fixed $\rho > 0$, a rate pair $(\ratex,\ratey) \in \reals_{\ge 0}^2$ is called achievable if there exists a sequence of encoders $\{(f_n,g_n)\}_{n=1}^\infty$,
\begin{align}
f_n\colon \set{X}^n &\to \{1,\ldots,\lfloor 2^{n \ratex} \rfloor\},\\*
g_n\colon \set{Y}^n &\to \{1,\ldots,\lfloor 2^{n \ratey} \rfloor\},
\end{align}
such that the $\rho$-th moment of the list size tends to one as $n$ tends to infinity, i.e.,
\begin{align}
\lim_{n \to \infty} \bigexpectation{\card{\set{L}(X^n,Y^n)}^\rho} = 1,
\end{align}
where
\begin{align}
\set{L}(x^n,y^n) \triangleq \{(\tilde{x}^n,\tilde{y}^n) \in \set{X}^n \times \set{Y}^n : f_n(\tilde{x}^n) = f_n(x^n) \,\land\, g_n(\tilde{y}^n) = g_n(y^n)\}.
\end{align}

For a memoryless source and a fixed $\rho > 0$, rate pairs in the interior of the region $\set{R}(\rho)$ defined next are achievable, while those outside $\set{R}(\rho)$ are not achievable \mdpicite[Theorem~1]{DistributedTaskEncoding}.
The region $\set{R}(\rho)$ is defined as the set of all rate pairs $(\ratex,\ratey)$ satisfying the following inequalities simultaneously:
\begin{align}
\ratex &\ge H_\frac{1}{1+\rho}(X),\\
\ratey &\ge H_\frac{1}{1+\rho}(Y),\\
\ratex + \ratey &\ge H_\frac{1}{1+\rho}(X,Y) + K_\frac{1}{1+\rho}(X;Y),\label{eq:taskencodingratexplusybound}
\end{align}
where $H_\alpha(X)$ denotes the R\'enyi entropy of order $\alpha$ (see \eqref{eq:defhalpha} ahead).

To better understand the role of $K_\alpha(X;Y)$, suppose that the sequences $\{X_i\}_{i=1}^n$ and $\{Y_i\}_{i=1}^n$ were allowed to be described jointly using $\lfloor 2^{n \ratex} \rfloor \cdot \lfloor 2^{n \ratey} \rfloor \approx 2^{n (\ratex + \ratey)}$ labels.
Then, by \mdpicite[Theorem~I.2]{TaskEncoding}, all rate pairs $(\ratex,\ratey) \in \reals_{\ge 0}^2$ satisfying the following inequality with strict inequality would be achievable, while those not satisfying the inequality would not:
\begin{align}
\ratex + \ratey \ge H_\frac{1}{1+\rho}(X,Y).\label{eq:taskencodingratexplusyjoint}
\end{align}
Comparing \eqref{eq:taskencodingratexplusybound} and \eqref{eq:taskencodingratexplusyjoint}, we see that the measure $K_\alpha(X;Y)$ appears as a penalty term on the sum-rate constraint incurred by requiring that the sequences be described separately as opposed to jointly.

\section{Preliminaries}
\label{sec:preliminaries}

Throughout the paper, $\log(\cdot)$ denotes the base-2 logarithm, $\set{X}$ and $\set{Y}$ are finite sets, $P_{XY}$ denotes a joint PMF over $\set{X} \times \set{Y}$, $Q_X$ denotes a PMF over $\set{X}$, and $Q_Y$ denotes a PMF over $\set{Y}$.
We use $P$ and $Q$ as generic PMFs over a finite set $\set{X}$.
We denote by $\supp(P) \triangleq \{x \in \set{X} : P(x) > 0\}$ the support of $P$, and by $\set{P}(\set{X})$ the set of all PMFs over $\set{X}$.
When clear from the context, we often omit sets and subscripts: for example, we write $\min_{\kernedqxcommaqy}$ for $\min_{(\kernedqxcommaqy) \in \set{P}(\set{X}) \times \set{P}(\set{Y})}$, $\sum_x$ for $\sum_{x \in \set{X}}$, $P(x)$ for $P_X(x)$, and $P(y|x)$ for $P_{Y|X}(y|x)$.
Whenever a conditional probability $P(y|x)$ is undefined because $P(x) = 0$, we define \smash{$P(y|x) \triangleq 1 / \card{\set{Y}}$}.
We denote by $\mathbbm{1}\{\mathsf{condition}\}$ the indicator function that is one if the condition is satisfied and zero otherwise.
In the definitions below, we use the following conventions:
\begin{align}
\frac{0}{0} = 0, \qquad
\frac{p}{0} = \infty \quad \forall \vthinspace p > 0, \qquad
0 \log 0 = 0, \qquad
\beta \log 0 = -\infty \quad \forall \vthinspace \beta > 0.\label{eq:zeroinftyconventions}
\end{align}

The R\'enyi entropy of order $\alpha$ \cite{RenyiEntropyDivergence} is defined for positive $\alpha$ other than one as
\begin{align}
H_\alpha(X) \triangleq \frac{1}{1 - \alpha} \log \sum_x P(x)^\alpha.\label{eq:defhalpha}
\end{align}
For $\alpha$ being zero, one, or infinity, we define by continuous extension of \eqref{eq:defhalpha}
\begin{align}
H_0(X) &\triangleq \logcard{\supp(P)},\\
H_1(X) &\triangleq H(X),\\
H_\infty(X) &\triangleq -\log \max_x P(x),\label{eq:defhinfinity}
\end{align}
where $H(X)$ is the Shannon entropy.
With this extension to $\alpha \in \{0,1,\infty\}$, the R\'enyi entropy satisfies the following basic properties:

\begin{Proposition}[\cite{CsiszarInformation}]\label{prop:renyientropyproperties}
Let $P$ be a PMF.
Then,
\begin{enumerate}[beginpenalty=10000,label=(\roman*)]
\item
For all $\alpha \in [0,\infty]$, $H_\alpha(X) \le \logcard{\set{X}}$. If $\alpha \in (0,\infty]$, then $H_\alpha(X) = \logcard{\set{X}}$ if and only if $X$ is distributed uniformly over $\set{X}$.
\item
The mapping $\alpha \mapsto H_\alpha(X)$ is nonincreasing on $[0,\infty]$.
\item
The mapping $\alpha \mapsto H_\alpha(X)$ is continuous on $[0,\infty]$.
\end{enumerate}
\end{Proposition}

The relative entropy (or Kullback--Leibler divergence) is defined as
\begin{align}
D(P\|Q) \triangleq \sum_x P(x) \log \frac{P(x)}{Q(x)}.
\end{align}

The R\'enyi divergence of order $\alpha$ \cite{RenyiEntropyDivergence,VanErvenHarremoes} is defined for positive $\alpha$ other than one as
\begin{align}
D_\alpha(P\|Q) \triangleq \frac{1}{\alpha - 1} \log \sum_x P(x)^\alpha \mulspace Q(x)^{1 - \alpha},\label{eq:defrenyidivergence}
\end{align}
where we read $P(x)^\alpha \mulspace Q(x)^{1 - \alpha}$ as $P(x)^\alpha \vthinnegspace / Q(x)^{\alpha - 1}$ if $\alpha > 1$.
For $\alpha$ being zero, one, or infinity, we define by continuous extension of \eqref{eq:defrenyidivergence}
\begin{align}
D_0(P\|Q) &\triangleq -\log \sum_{x \in \supp(P)} Q(x),\\
D_1(P\|Q) &\triangleq D(P\|Q),\\
D_\infty(P\|Q) &\triangleq \log \max_x \frac{P(x)}{Q(x)}.
\end{align}
With this extension to $\alpha \in \{0,1,\infty\}$, the R\'enyi divergence satisfies the following basic properties:

\begin{Proposition}\label{prop:renyidivproperties}
Let $P$ and $Q$ be PMFs.
Then,
\begin{enumerate}[beginpenalty=10000,label=(\roman*)]
\item
For all $\alpha \in [0,1)$, $D_\alpha(P\|Q)$ is finite if and only if $\card{\supp(P) \cap \supp(Q)} > 0$.
For all $\alpha \in [1,\infty]$, $D_\alpha(P\|Q)$ is finite if and only if $\supp(P) \subseteq \supp(Q)$.
\item
For all $\alpha \in [0,\infty]$, $D_\alpha(P\|Q) \ge 0$.
If $\alpha \in (0,\infty]$, then $D_\alpha(P\|Q) = 0$ if and only if $P = Q$.
\item
For every $\alpha \in [0,\infty]$, the mapping $Q \mapsto D_\alpha(P\|Q)$ is continuous.
\item
The mapping $\alpha \mapsto D_\alpha(P\|Q)$ is nondecreasing on $[0,\infty]$.
\item
The mapping $\alpha \mapsto D_\alpha(P\|Q)$ is continuous on $[0,\infty]$.
\end{enumerate}
\end{Proposition}

\begin{proof}
Part~(i) follows from the definition of $D_\alpha(P\|Q)$ and the conventions \eqref{eq:zeroinftyconventions}, and Parts~(ii)--(v) are shown in \cite{VanErvenHarremoes}.
\end{proof}

The R\'enyi divergence for negative $\alpha$ is defined as
\begin{align}
D_\alpha(P\|Q) \triangleq \frac{1}{\alpha - 1} \log \sum_x \frac{Q(x)^{1 - \alpha}}{P(x)^{-\alpha}}.\label{eq:defrenyidivnegative}
\end{align}
(We use negative $\alpha$ only in \lmaref{lma:jalpharxezero}.
More about negative orders can be found in \mdpicite[Section~V]{VanErvenHarremoes}.
For~other applications of negative orders, see \mdpicite[Proof of Theorem~1 and Example~1]{GuessingImprovedBounds}.)

The relative $\alpha$-entropy \cite{KumarSundaresanA,KumarSundaresanB} is defined for positive $\alpha$ other than one as
\begin{align}
\Delta_\alpha(P\|Q) \triangleq \frac{\alpha}{1 - \alpha} \log \sum_x P(x) \mulspace Q(x)^{\alpha - 1} + \log \sum_x Q(x)^\alpha - \frac{1}{1 - \alpha} \log \sum_x P(x)^\alpha,\label{eq:defrelativealphaentropy}
\end{align}
where we read $P(x) \mulspace Q(x)^{\alpha - 1}$ as $P(x) / Q(x)^{1 - \alpha}$ if $\alpha < 1$.
The relative $\alpha$-entropy appears in mismatched guessing \cite{SundaresanGuessing}, mismatched source coding \mdpicite[Theorem~8]{SundaresanGuessing}, and mismatched task encoding \mdpicite[Section~IV]{TaskEncoding}.
It also arises in robust parameter estimation and constrained compression settings \mdpicite[Section~II]{KumarSundaresanB}.
For $\alpha$ being zero, one, or infinity, we define by continuous extension of \eqref{eq:defrelativealphaentropy}
\begin{align}
\Delta_0(P\|Q) &\triangleq \begin{cases}\log \frac{\card{\supp(Q)}}{\card{\supp(P)}} & \text{if $\supp(P) \subseteq \supp(Q)$,}\\
\infty & \text{otherwise,}\end{cases}\label{eq:relalphaentropyzero}\\
\Delta_1(P\|Q) &\triangleq D(P\|Q),\\
\Delta_\infty(P\|Q) &\triangleq \log \frac{\max_x P(x)}{\card{\argmax(Q)}^{-1} \sum_{x \in \argmax(Q)} P(x)},\label{eq:relalphaentropyinfty}
\end{align}
where $\argmax(Q) \triangleq \{x \in \set{X} : Q(x) = \max_{x' \in \set{X}} Q(x')\}$ and $\card{\argmax(Q)}$ is the cardinality of this set.
With this extension to $\alpha \in \{0,1,\infty\}$, the relative $\alpha$-entropy satisfies the following basic properties:

\begin{Proposition}\label{prop:relalphaentropyproperties}
Let $P$ and $Q$ be PMFs.
Then,
\begin{enumerate}[beginpenalty=10000,label=(\roman*)]
\item
For all $\alpha \in [0,1]$, $\Delta_\alpha(P\|Q)$ is finite if and only if $\supp(P) \subseteq \supp(Q)$.
For all $\alpha \in (1,\infty)$, $\Delta_\alpha(P\|Q)$ is finite if and only if $\card{\supp(P) \cap \supp(Q)} > 0$.
\item
For all $\alpha \in [0,\infty]$, $\Delta_\alpha(P\|Q) \ge 0$.
If $\alpha \in (0,\infty)$, then $\Delta_\alpha(P\|Q) = 0$ if and only if $P = Q$.
\item
For every $\alpha \in (0,\infty)$, the mapping $Q \mapsto \Delta_\alpha(P\|Q)$ is continuous.
\item
The mapping $\alpha \mapsto \Delta_\alpha(P\|Q)$ is continuous on $[0,\infty]$.
\end{enumerate}
\end{Proposition}

(Part~(i) differs from \mdpicite[Proposition~IV.1]{TaskEncoding}, where the conventions for $\alpha > 1$ differ from ours.
Our conventions are compatible with \cite{KumarSundaresanA,KumarSundaresanB}, and, as stated in Part~(iii), they result in the continuity of the mapping $Q \mapsto \Delta_\alpha(P\|Q)$.)

\begin{proof}[Proof~of~\propref{prop:relalphaentropyproperties}]
Part~(i) follows from the definition of $\Delta_\alpha(P\|Q)$ in \eqref{eq:defrelativealphaentropy} and the conventions \eqref{eq:zeroinftyconventions}.
For $\alpha \in (0,1) \cup (1,\infty)$, Part~(ii) follows from \mdpicite[Proposition~IV.1]{TaskEncoding};
for $\alpha = 1$, Part~(ii) holds because $\Delta_1(P\|Q) = D(P\|Q)$; and
for $\alpha \in \{0,\infty\}$, Part~(ii) follows from the definition of $\Delta_\alpha(P\|Q)$.
Part~(iii) follows from the definition of $\Delta_\alpha(P\|Q)$, and Part~(iv) follows from \mdpicite[Proposition~IV.1]{TaskEncoding}.
\end{proof}

In the rest of this section, we prove some auxiliary results that we need later (Propositions~\ref{prop:deltaalphadonedivalpha}--\ref{prop:infqydalphapxyqxqy}).
We first establish the relation between $D_\alpha(P\|Q)$ and $\Delta_\alpha(P\|Q)$.

\begin{Proposition}[{\mdpicite[Section~V, Property~4]{SundaresanGuessing}}]\label{prop:deltaalphadonedivalpha}
Let $P$ and $Q$ be PMFs, and let $\alpha > 0$.
Then,
\begin{align}
\Delta_\alpha(P\|Q) = D_\frac{1}{\alpha}(\widetilde{P}\|\widetilde{Q}),\label{eq:deltaalphadalpha}
\end{align}
where the PMFs $\widetilde{P}$ and $\widetilde{Q}$ are given by
\begin{align}
\widetilde{P}(x) &\triangleq \frac{P(x)^\alpha}{\sum_{x' \in \set{X}} P(x')^\alpha},\label{eq:ptiltedalpha}\\[0.5ex]
\widetilde{Q}(x) &\triangleq \frac{Q(x)^\alpha}{\sum_{x' \in \set{X}} Q(x')^\alpha}.\label{eq:qtiltedalpha}
\end{align}
\end{Proposition}

\begin{proof}
If $\alpha = 1$, then \eqref{eq:deltaalphadalpha} holds because $\widetilde{P} = P$, $\widetilde{Q} = Q$, and $\Delta_1(P\|Q) = D_1(P\|Q) = D(P\|Q)$.
Now let $\alpha \ne 1$.
Because $\widetilde{P}(x)$ and $\widetilde{Q}(x)$ are zero if and only if $P(x)$ and $Q(x)$ are zero, respectively, the LHS of \eqref{eq:deltaalphadalpha} is finite if and only if its RHS is finite.
If $D_{1/\alpha}(\widetilde{P}\|\widetilde{Q})$ is finite, then \eqref{eq:deltaalphadalpha} follows from a simple computation.
\end{proof}

\newpage
In light of \propref{prop:deltaalphadonedivalpha}, $J_\alpha(X;Y)$ and $K_\alpha(X;Y)$ are related as follows:

\begin{Proposition}\label{prop:jalphavskalpha}
Let $P_{XY}$ be a joint PMF, and let $\alpha > 0$.
Then,
\begin{align}
K_\alpha(X;Y) = J_\frac{1}{\alpha}(\widetilde{X};\widetilde{Y}),
\end{align}
where the joint PMF of $\widetilde{X}$ and $\widetilde{Y}$ is given by
\begin{align}
\widetilde{P}_{XY}(x,y) \triangleq \frac{P_{XY}(x,y)^\alpha}{\sum_{(x',y') \in \set{X} \times \set{Y}} P_{XY}(x',y')^\alpha}.\label{eq:pxytiltedalpha}
\end{align}
\end{Proposition}

\begin{proof}
Let $\alpha > 0$.
For fixed PMFs $Q_X$ and $Q_Y$, define the transformed PMFs $\widetilde{Q_X Q_Y}$, $\widetilde{Q}_X$, and $\widetilde{Q}_Y$ as
\begin{align}
\widetilde{Q_X Q_Y}(x,y) &\triangleq \frac{[Q_X(x) \mulspace Q_Y(y)]^\alpha}{\sum_{(x',y') \in \set{X} \times \set{Y}} [Q_X(x') \mulspace Q_Y(y')]^\alpha},\\[0.5ex]
\widetilde{Q}_X(x) &\triangleq \frac{Q_X(x)^\alpha}{\sum_{x' \in \set{X}} Q_X(x')^\alpha},\label{eq:qxtiltedalpha}\\[0.5ex]
\widetilde{Q}_Y(y) &\triangleq \frac{Q_Y(y)^\alpha}{\sum_{y' \in \set{Y}} Q_Y(y')^\alpha}.\label{eq:qytiltedalpha}
\end{align}
Then,
\begin{align}
K_\alpha(X;Y) &= \min_{\kernedqxcommaqy} \Delta_\alpha(P_{XY}\|Q_X Q_Y)\label{eq:relationjalphakalphaa}\\
&= \min_{\kernedqxcommaqy} D_\frac{1}{\alpha}(\widetilde{P}_{XY}\|\widetilde{Q_X Q_Y})\label{eq:relationjalphakalphab}\\
&= \min_{\kernedqxcommaqy} D_\frac{1}{\alpha}(\widetilde{P}_{XY}\|\widetilde{Q}_X \widetilde{Q}_Y)\label{eq:relationjalphakalphac}\\
&= \min_{\kernedqxcommaqy} D_\frac{1}{\alpha}(\widetilde{P}_{XY}\|Q_X Q_Y)\label{eq:relationjalphakalphad}\\
&= J_\frac{1}{\alpha}(\widetilde{X};\widetilde{Y}),\label{eq:relationjalphakalphae}
\end{align}
where \eqref{eq:relationjalphakalphaa} holds by the definition of $K_\alpha(X;Y)$;
\eqref{eq:relationjalphakalphab} follows from \propref{prop:deltaalphadonedivalpha};
\eqref{eq:relationjalphakalphac} holds because $\widetilde{Q_X Q_Y} = \widetilde{Q}_X \widetilde{Q}_Y$;
\eqref{eq:relationjalphakalphad} holds because the transformations \eqref{eq:qxtiltedalpha} and \eqref{eq:qytiltedalpha} are bijective on the set of PMFs over $\set{X}$ and $\set{Y}$, respectively; and
\eqref{eq:relationjalphakalphae} holds by the definition of $J_\alpha(X;Y)$.
\end{proof}

The next proposition provides a characterization of the mutual information that parallels the definitions of $J_\alpha(X;Y)$ and $K_\alpha(X;Y)$.
Because $D_1(P\|Q) = \Delta_1(P\|Q) = D(P\|Q)$, this also shows that $J_\alpha(X;Y)$ and $K_\alpha(X;Y)$ reduce to the mutual information when $\alpha$ is one.

\begin{Proposition}[{\mdpicite[Theorem~3.4]{PolyanskiyWu}}]\label{prop:minqxqydpxyqxqy}
Let $P_{XY}$ be a joint PMF.
Then, for all PMFs $Q_X$ and $Q_Y$,
\begin{align}
D(P_{XY}\|Q_X Q_Y) \ge D(P_{XY}\|P_X P_Y),\label{eq:dpxyqxqygemi}
\end{align}
with equality if and only if $Q_X = P_X$ and $Q_Y = P_Y$.
Thus,
\begin{align}
I(X;Y) = \min_{\kernedqxcommaqy} D(P_{XY}\|Q_X Q_Y).\label{eq:miasminimization}
\end{align}
\end{Proposition}

\begin{proof}
A simple computation reveals that
\begin{align}
D(P_{XY}\|Q_X Q_Y) = D(P_{XY}\|P_X P_Y) + D(P_X\|Q_X) + D(P_Y\|Q_Y),\label{eq:dpxyqxqydecomposition}
\end{align}
which implies \eqref{eq:dpxyqxqygemi} because $D(P\|Q) \ge 0$ with equality if and only if $P = Q$.
Thus, \eqref{eq:miasminimization} holds because $I(X;Y) = D(P_{XY}\|P_X P_Y)$.
\end{proof}

The last proposition of this section is about a precursor to $J_\alpha(X;Y)$, namely, the minimization of $D_\alpha(P_{XY}\|Q_X Q_Y)$ with respect to $Q_Y$ only, which can be carried out explicitly.
(This proposition extends~\mdpicite[Equation~(13)]{CsiszarInformation} and \mdpicite[Lemma~29]{TomamichelHayashi}.)

\begin{Proposition}\label{prop:infqydalphapxyqxqy}
Let $P_{XY}$ be a joint PMF and $Q_X$ a PMF.
Then, for every $\alpha \in (0,1) \cup (1,\infty)$,
\begin{align}
\min_{Q_Y} D_\alpha(P_{XY}\|Q_X Q_Y) = \frac{\alpha}{\alpha - 1} \log \sum_y \mleft[\sum_x P(x,y)^\alpha \mulspace Q_X(x)^{1 - \alpha}\mright]^\frac{1}{\alpha},\label{eq:infqydalphapxyqxqyvalue}
\end{align}
with the conventions of \eqref{eq:zeroinftyconventions}.
If the RHS of \eqref{eq:infqydalphapxyqxqyvalue} is finite, then the minimum is achieved uniquely by
\begin{align}
Q_Y^*(y) = \frac{\bigl[\sum_x P(x,y)^\alpha \mulspace Q_X(x)^{1 - \alpha}\bigr]^\frac{1}{\alpha}}{\sum_{y' \in \set{Y}}\bigl[\sum_x P(x,y')^\alpha \mulspace Q_X(x)^{1 - \alpha}\bigr]^\frac{1}{\alpha}}.\label{eq:infqydalphapxyqxqyminimizer}
\end{align}

For $\alpha = \infty$,
\begin{align}
\min_{Q_Y} D_\infty(P_{XY}\|Q_X Q_Y) = \log \sum_y \max_x \frac{P(x,y)}{Q_X(x)},\label{eq:infqydinftypxyqxqyvalue}
\end{align}
with the conventions of \eqref{eq:zeroinftyconventions}.
If the RHS of \eqref{eq:infqydinftypxyqxqyvalue} is finite, then the minimum is achieved uniquely by
\begin{align}
Q_Y^*(y) = \frac{\max_x [P(x,y) / Q_X(x)]}{\sum_{y' \in \set{Y}} \max_x [P(x,y') / Q_X(x)]}.\label{eq:infqydinftypxyqxqyminimizer}
\end{align}
\end{Proposition}

\begin{proof}
We first treat the case $\alpha \in (0,1) \cup (1,\infty)$.
If the RHS of \eqref{eq:infqydalphapxyqxqyvalue} is infinite, then the conventions imply that $D_\alpha(P_{XY}\|Q_X Q_Y)$ is infinite for every $Q_Y \in \set{P}(\set{Y})$, so \eqref{eq:infqydalphapxyqxqyvalue} holds.
Otherwise, if the RHS of~\eqref{eq:infqydalphapxyqxqyvalue} is finite, then the PMF $Q_Y^*$ given by \eqref{eq:infqydalphapxyqxqyminimizer} is well-defined, and a simple computation shows that for every $Q_Y \in \set{P}(\set{Y})$,
\begin{align}
D_\alpha(P_{XY}\|Q_X Q_Y) = \frac{\alpha}{\alpha - 1} \log \sum_y \mleft[\sum_x P(x,y)^\alpha \mulspace Q_X(x)^{1 - \alpha}\mright]^\frac{1}{\alpha} + D_\alpha(Q_Y^*\|Q_Y).\label{eq:infqydalphapxyqxqyc}
\end{align}
The only term on the RHS of \eqref{eq:infqydalphapxyqxqyc} that depends on $Q_Y$ is $D_\alpha(Q_Y^*\|Q_Y)$.
Because $D_\alpha(Q_Y^*\|Q_Y) \ge 0$ with equality if and only if $Q_Y = Q_Y^*$ (\propref{prop:renyidivproperties}), \eqref{eq:infqydalphapxyqxqyc} implies \eqref{eq:infqydalphapxyqxqyvalue} and \eqref{eq:infqydalphapxyqxqyminimizer}.

The case $\alpha = \infty$ is analogous: if the RHS of \eqref{eq:infqydinftypxyqxqyvalue} is infinite, then the LHS of \eqref{eq:infqydinftypxyqxqyvalue} is infinite, too;
and if the RHS of \eqref{eq:infqydinftypxyqxqyvalue} is finite, then the PMF $Q_Y^*$ given by \eqref{eq:infqydinftypxyqxqyminimizer} is well-defined, and a simple computation shows that for every $Q_Y \in \set{P}(\set{Y})$,
\begin{align}
D_\infty(P_{XY}\|Q_X Q_Y) = \log \sum_y \max_x \frac{P(x,y)}{Q_X(x)} + D_\infty(Q_Y^*\|Q_Y).\label{eq:infqydinftypxyqxqyc}
\end{align}
The only term on the RHS of \eqref{eq:infqydinftypxyqxqyc} that depends on $Q_Y$ is $D_\infty(Q_Y^*\|Q_Y)$.
Because $D_\infty(Q_Y^*\|Q_Y) \ge 0$ with equality if and only if $Q_Y = Q_Y^*$ (\propref{prop:renyidivproperties}), \eqref{eq:infqydinftypxyqxqyc} implies \eqref{eq:infqydinftypxyqxqyvalue} and \eqref{eq:infqydinftypxyqxqyminimizer}.
\end{proof}

\section{Two Measures of Dependence}
\label{sec:twomeasuresofdependence}

We state the properties of $J_\alpha(X;Y)$ in \thmref{thm:jalphaprop} and those of $K_\alpha(X;Y)$ in \thmref{thm:kalphaprop}.
The~enumeration labels in the theorems refer to the lemmas in \secref{sec:proofs} where the properties are proved.
(The enumeration labels are not consecutive because, in order to avoid forward references in the proofs, the order of the results in \secref{sec:proofs} is not the same as here.)

\begin{Theorem}\label{thm:jalphaprop}
Let $X$, $X_1$, $X_2$, $Y$, $Y_1$, $Y_2$, and $Z$ be random variables taking values in finite sets.
Then:
\begin{enumerate}[leftmargin=19mm,labelsep=2mm]
\item[(Lemma~\ref{lma:jalphawelldefinedfinite})]
For every $\alpha \in [0,\infty]$, the minimum in the definition of $J_\alpha(X;Y)$ exists and is finite.
\end{enumerate}
The following properties of the mutual information $I(X;Y)$ \mdpicite[Chapter~2]{CoverThomas} are also satisfied by $J_\alpha(X;Y)$:
\begin{enumerate}[leftmargin=19mm,labelsep=2mm]
\item[(Lemma~\ref{lma:jalphanonnegative})]
For all $\alpha \in [0,\infty]$, $J_\alpha(X;Y) \ge 0$.
If $\alpha \in (0,\infty]$, then $J_\alpha(X;Y) = 0$ if and only if $X$ and $Y$ are independent (nonnegativity).
\item[(Lemma~\ref{lma:jalphasymmetric})]
For all $\alpha \in [0,\infty]$, $J_\alpha(X;Y) = J_\alpha(Y;X)$ (symmetry).
\item[(Lemma~\ref{lma:jalphadpi})]
If $X \markov Y \markov Z$ form a Markov chain, then $J_\alpha(X;Z) \le J_\alpha(X;Y)$ for all $\alpha \in [0,\infty]$ (data-processing inequality).
\item[(Lemma~\ref{lma:jalphaadditive})]
If the pairs $(X_1,Y_1)$ and $(X_2,Y_2)$ are independent, then $J_\alpha(X_1,X_2;Y_1,Y_2) = J_\alpha(X_1;Y_1) + J_\alpha(X_2;Y_2)$ for all $\alpha \in [0,\infty]$ (additivity).
\item[(Lemma~\ref{lma:jalphacardinalityupperbound})]
For all $\alpha \in [0,\infty]$, $J_\alpha(X;Y) \le \logcard{\set{X}}$ with equality if and only if $\bigl(\alpha \in [\frac{1}{2},\infty]$, $X$ is distributed uniformly over $\set{X}$, and $H(X|Y) = 0\bigr)$.
\item[(Lemma~\ref{lma:jalphaconcavepx})]
For every $\alpha \in [1,\infty]$, $J_\alpha(X;Y)$ is concave in $P_X$ for fixed $P_{Y|X}$.
\end{enumerate}
Moreover:
\begin{enumerate}[leftmargin=19mm,labelsep=2mm]
\item[(Lemma~\ref{lma:jzeroiszero})]
$J_0(X;Y) = 0$.
\item[(Lemma~\ref{lma:jonehalfsingularvalue})]
Let $f\colon \{1,\ldots,\card{\set{X}}\} \to \set{X}$ and $g\colon \{1,\ldots,\card{\set{Y}}\} \to \set{Y}$ be bijective functions, and let $\mat{A}$ be the $\card{\set{X}} \times \card{\set{Y}}$ matrix whose Row-$i$ Column-$j$ entry $\mat{A}_{i,j}$ equals $\sqrt{P_{XY}(f(i),g(j))}$.
Then,
\begin{align}
J_\frac{1}{2}(X;Y) = -2 \log \sigma_1(\mat{A}),
\end{align}
where $\sigma_1(\mat{A})$ denotes the largest singular value of $\mat{A}$.
(Because the singular values of a matrix are invariant under row and column permutations, the result does not depend on $f$ or $g$.)
\item[(Lemma~\ref{lma:jonemutualinformation})]
$J_1(X;Y) = I(X;Y)$.
\item[(Lemma~\ref{lma:oneminusalphajalpharxy})]
For all $\alpha > 0$,
\begin{align}
(1 - \alpha) \mulspace J_\alpha(X;Y) = \min_{R_{XY} \in \set{P}(\set{X} \times \set{Y})} \bigl[(1 - \alpha) \mulspace D(R_{XY}\|R_X R_Y) + \alpha \mulspace D(R_{XY}\|P_{XY})\bigr].\label{eq:prvw-oneminusalphajalpharxy}
\end{align}
Thus, being the minimum of concave functions in $\alpha$, the mapping $\alpha \mapsto (1 - \alpha) \mulspace J_\alpha(X;Y)$ is concave on $(0,\infty)$.
\item[(Lemma~\ref{lma:jalphanondecreasing})]
The mapping $\alpha \mapsto J_\alpha(X;Y)$ is nondecreasing on $[0,\infty]$.
\item[(Lemma~\ref{lma:jalphacontinuous})]
The mapping $\alpha \mapsto J_\alpha(X;Y)$ is continuous on $[0,\infty]$.
\item[(Lemma~\ref{lma:jalphaselfinformation})]
If $X = Y$ with probability one, then
\begin{align}
J_\alpha(X;Y) = \begin{cases}\frac{\alpha}{1 - \alpha} \mulspace H_\infty(X) & \text{if $\alpha \in [0,\frac{1}{2}]$,}\\
H_\frac{\alpha}{2 \alpha - 1}(X) & \text{if $\alpha > \frac{1}{2}$,}\\
H_\frac{1}{2}(X) & \text{if $\alpha = \infty$.}\end{cases}\label{eq:prvw-jalphaselfinformation}
\end{align}
\end{enumerate}
The minimization problem in the definition of $J_\alpha(X;Y)$ has the following characteristics:
\begin{enumerate}[leftmargin=19mm,labelsep=2mm]
\item[(Lemma~\ref{lma:dalphapxyqxqyconvexgeonehalf})]
For every $\alpha \in [\frac{1}{2},\infty]$, the mapping $(Q_X,Q_Y) \mapsto D_\alpha(P_{XY}\|Q_X Q_Y)$ is convex, i.e., for all $\lambda,\lambda' \in [0,1]$ with $\lambda + \lambda' = 1$, all $Q_X,Q_X' \in \set{P}(\set{X})$, and all $Q_Y,Q_Y' \in \set{P}(\set{Y})$,
\begin{align}
D_\alpha\bigl(P_{XY}\|(\lambda Q_X + \lambda' Q_X') (\lambda Q_Y + \lambda' Q_Y')\bigr) \le \lambda \mulspace D_\alpha(P_{XY}\|Q_X Q_Y) + \lambda' \mulspace D_\alpha(P_{XY}\|Q_X' Q_Y').\label{eq:prvw-dalphapxyqxqyconvex}
\end{align}
For $\alpha \in [0,\frac{1}{2})$, the mapping need not be convex.
\item[(Lemma~\ref{lma:dalphapxyqxqyqxqyconsistency})]
Let $\alpha \in (0,1) \cup (1,\infty)$.
If $(Q_X^*,Q_Y^*)$ achieves the minimum in the definition of $J_\alpha(X;Y)$, then there exist positive normalization constants $c$ and $d$ such that
\begin{align}
Q_X^*(x) &= c \mleft[\sum_y P(x,y)^\alpha \mulspace Q_Y^*(y)^{1 - \alpha}\mright]^\frac{1}{\alpha} \quad \forall \vthinspace x \in \set{X},\label{eq:prvw-dalphapxyqxstarforqystar}\\*
Q_Y^*(y) &= d \mleft[\sum_x P(x,y)^\alpha \mulspace Q_X^*(x)^{1 - \alpha}\mright]^\frac{1}{\alpha} \quad \forall \vthinspace y \in \set{Y},\label{eq:prvw-dalphapxyqystarforqxstar}
\end{align}
with the conventions of \eqref{eq:zeroinftyconventions}.
The case $\alpha = \infty$ is similar: if $(Q_X^*,Q_Y^*)$ achieves the minimum in the definition of $J_\infty(X;Y)$, then there exist positive normalization constants $c$ and $d$ such that
\begin{align}
Q_X^*(x) &= c \max_y \frac{P(x,y)}{Q_Y^*(y)} \quad \forall \vthinspace x \in \set{X},\label{eq:prvw-dinftypxyqxstarforqystar}\\*
Q_Y^*(y) &= d \max_x \frac{P(x,y)}{Q_X^*(x)} \quad \forall \vthinspace y \in \set{Y},\label{eq:prvw-dinftypxyqystarforqxstar}
\end{align}
with the conventions of \eqref{eq:zeroinftyconventions}.
(If $\alpha = 1$, then $Q_X^* = P_X$ and $Q_Y^* = P_Y$ by \propref{prop:minqxqydpxyqxqy}.)
Thus, for all $\alpha \in (0,\infty]$, both inclusions $\supp(Q_X^*) \subseteq \supp(P_X)$ and $\supp(Q_Y^*) \subseteq \supp(P_Y)$ hold.\vphantom{\raisebox{-2pt}{$\frac{1}{2}$}}
\item[(Lemma~\ref{lma:jalphauniqueminimizer})]
For every $\alpha \in (\frac{1}{2},\infty]$, the mapping $(Q_X,Q_Y) \mapsto D_\alpha(P_{XY}\|Q_X Q_Y)$ has a unique minimizer.
This need not be the case when $\alpha \in [0,\frac{1}{2}]$.
\end{enumerate}
The measure $J_\alpha(X;Y)$ can also be expressed as follows:
\begin{enumerate}[leftmargin=19mm,labelsep=2mm]
\item[(Lemma~\ref{lma:jalphaminimizationqx})]
For all $\alpha \in (0,\infty]$,
\begin{align}
J_\alpha(X;Y) = \min_{Q_X} \phi_\alpha(Q_X),\label{eq:prvw-jalphaminimizationqx}
\end{align}
where $\phi_\alpha(Q_X)$ is defined as
\begin{align}
\phi_\alpha(Q_X) \triangleq \min_{Q_Y} D_\alpha(P_{XY}\|Q_X Q_Y)\label{eq:prvw-jalphaminimizationqxdefphi}
\end{align}
and is given explicitly as follows: for $\alpha \in (0,1) \cup (1,\infty)$,
\begin{align}
\phi_\alpha(Q_X) = \frac{\alpha}{\alpha - 1} \log \sum_y \mleft[\sum_x P(x,y)^\alpha \mulspace Q_X(x)^{1 - \alpha}\mright]^\frac{1}{\alpha},\label{eq:prvw-jalphaminimizationqxphi}
\end{align}
with the conventions of \eqref{eq:zeroinftyconventions}; and for $\alpha \in \{1,\infty\}$,
\begin{align}
\phi_1(Q_X) &= D(P_{XY}\|Q_X P_Y),\label{eq:prvw-joneminimizationqxphi}\\
\phi_\infty(Q_X) &= \log \sum_y \max_x \frac{P(x,y)}{Q_X(x)},\label{eq:prvw-jinftyminimizationqxphi}
\end{align}
with the conventions of \eqref{eq:zeroinftyconventions}.
For every $\alpha \in [\frac{1}{2},\infty]$, the mapping $Q_X \mapsto \phi_\alpha(Q_X)$ is convex.
For $\alpha \in (0,\frac{1}{2})$, the mapping need not be convex.
\item[(Lemma~\ref{lma:jalpharxypsi})]
For all $\alpha \in (0,1) \cup (1,\infty]$,
\begin{align}
J_\alpha(X;Y) = \begin{cases}\displaystyle \min_{R_{XY} \in \set{P}(\set{X} \times \set{Y})} \psi_\alpha(R_{XY}) & \text{if $\alpha \in (0,1)$,}\\
\displaystyle \max_{R_{XY} \in \set{P}(\set{X} \times \set{Y})} \psi_\alpha(R_{XY}) & \text{if $\alpha \in (1,\infty]$,}\end{cases}\label{eq:prvw-jalpharxypsi}
\end{align}
where
\begin{align}
\psi_\alpha(R_{XY}) \triangleq \begin{cases}\displaystyle D(R_{XY}\|R_X R_Y) + \frac{\alpha}{1 - \alpha} D(R_{XY}\|P_{XY}) & \text{if $\alpha \in (0,1) \cup (1,\infty)$,}\\[1ex]
\displaystyle D(R_{XY}\|R_X R_Y) - D(R_{XY}\|P_{XY}) & \text{if $\alpha = \infty$.}\end{cases}\label{eq:prvw-jalpharxypsidefpsi}
\end{align}
For every $\alpha \in (1,\infty]$, the mapping $R_{XY} \mapsto \psi_\alpha(R_{XY})$ is concave.
For all $\alpha \in (1,\infty]$ and all $R_{XY} \in \set{P}(\set{X} \times \set{Y})$, the statement $J_\alpha(X;Y) = \psi_\alpha(R_{XY})$ is equivalent to $\psi_\alpha(R_{XY}) = D_\alpha(P_{XY}\|R_X R_Y)$.
\item[(Lemma~\ref{lma:jalpharxezero})]
For all $\alpha \in (0,1) \cup (1,\infty)$,
\begin{align}
J_\alpha(X;Y) = \min_{R_X \ll P_X} \frac{1}{\alpha - 1} \Bigl[D_\frac{\alpha}{\alpha - 1}(P_X\|R_X) - \alpha \mulspace E_0\bigl(\tfrac{1 - \alpha}{\alpha},R_X\bigr)\Bigr],\label{eq:prvw-jalpharxezero}
\end{align}
where the minimization is over all PMFs $R_X$ satisfying $R_X \ll P_X$ $\bigl(\text{i.e., $\supp(R_X) \subseteq \supp(P_X)$}\bigr)$;
$D_\alpha(P\|Q)$ for negative $\alpha$ is given by \eqref{eq:defrenyidivnegative};
and Gallager's $E_0$ function \cite{Gallager} is defined as
\begin{align}
E_0(\rho,R_X) \triangleq -\log \sum_y \mleft[\sum_x R_X(x) \mulspace P(y|x)^\frac{1}{1 + \rho}\mright]^{1 + \rho}.
\end{align}
\end{enumerate}
\end{Theorem}

We now move on to the properties of $K_\alpha(X;Y)$.
Some of these properties are derived from their counterparts of $J_\alpha(X;Y)$ using the relation $K_\alpha(X;Y) = J_{1 / \alpha}(\widetilde{X};\widetilde{Y})$ described in \propref{prop:jalphavskalpha}.

\begin{Theorem}\label{thm:kalphaprop}
Let $X$, $X_1$, $X_2$, $Y$, $Y_1$, $Y_2$, and $Z$ be random variables taking values in finite sets.
Then:
\begin{enumerate}[leftmargin=19mm,labelsep=2mm]
\item[(Lemma~\ref{lma:kalphawelldefinedfinite})]
For every $\alpha \in [0,\infty]$, the minimum in the definition of $K_\alpha(X;Y)$ in \eqref{eq:defkalpha} exists and is finite.
\end{enumerate}
The following properties of the mutual information $I(X;Y)$ are also satisfied by $K_\alpha(X;Y)$:
\begin{enumerate}[leftmargin=19mm,labelsep=2mm]
\item[(Lemma~\ref{lma:kalphanonnegative})]
For all $\alpha \in [0,\infty]$, $K_\alpha(X;Y) \ge 0$.
If $\alpha \in (0,\infty)$, then $K_\alpha(X;Y) = 0$ if and only if $X$ and $Y$ are independent (nonnegativity).
\item[(Lemma~\ref{lma:kalphasymmetric})]
For all $\alpha \in [0,\infty]$, $K_\alpha(X;Y) = K_\alpha(Y;X)$ (symmetry).
\item[(Lemma~\ref{lma:kalphaadditive})]
If the pairs $(X_1,Y_1)$ and $(X_2,Y_2)$ are independent, then $K_\alpha(X_1,X_2;Y_1,Y_2) = K_\alpha(X_1;Y_1) + K_\alpha(X_2;Y_2)$ for all $\alpha \in [0,\infty]$ (additivity).
\item[(Lemma~\ref{lma:kalphacardinalityupperbound})]
For all $\alpha \in [0,\infty]$, $K_\alpha(X;Y) \le \logcard{\set{X}}$.
\end{enumerate}
Unlike the mutual information, $K_\alpha(X;Y)$ does not satisfy the data-processing inequality:
\begin{enumerate}[leftmargin=19mm,labelsep=2mm]
\item[(Lemma~\ref{lma:kalphanodpi})]
There exists a Markov chain $X \markov Y \markov Z$ for which $K_2(X;Z) > K_2(X;Y)$.
\end{enumerate}
Moreover:
\begin{enumerate}[leftmargin=19mm,labelsep=2mm]
\item[(Lemma~\ref{lma:kalphapowermean})]
For all $\alpha \in (0,\infty)$,
\begin{align}
K_\alpha(X;Y) + H_\alpha(X,Y) = \min_{\kernedqxcommaqy} -\log M_\frac{\alpha - 1}{\alpha}(Q_X,Q_Y),\label{eq:prvw-kalphaminpowermean}
\end{align}
where $M_\beta(Q_X,Q_Y)$ is the following weighted power mean \mdpicite[Chapter~III]{HandbookMeansInequalities}:
For $\beta \in \reals \setminus \{0\}$,
\begin{align}
M_\beta(Q_X,Q_Y) \triangleq \mleft[\sum_{x,y} P(x,y) [Q_X(x) \mulspace Q_Y(y)]^\beta\mright]^\frac{1}{\beta},\label{eq:prvw-defmbeta}
\end{align}
where for $\beta < 0$, we read $P(x,y) [Q_X(x) \mulspace Q_Y(y)]^\beta$ as $P(x,y) / [Q_X(x) \mulspace Q_Y(y)]^{-\beta}$ and use the conventions \eqref{eq:zeroinftyconventions};
and for $\beta = 0$, using the convention $0^0 = 1$,
\begin{align}
M_0(Q_X,Q_Y) \triangleq \prod_{x,y} [Q_X(x) \mulspace Q_Y(y)]^{P(x,y)}.\label{eq:prvw-defmzero}
\end{align}
\item[(Lemma~\ref{lma:kzerovalueandlimit})]
For $\alpha = 0$,
\begin{align}
K_0(X;Y) &= \log \frac{\card{\supp(P_X P_Y)}}{\card{\supp(P_{XY})}}\label{eq:prvw-kzerovalueandlimita}\\*[-0.5ex]
&\ge \min_{\kernedqxcommaqy} \log \max_{(x,y) \in \supp(P_{XY})} \frac{1}{Q_X(x) \mulspace Q_Y(y)} - \logcard{\supp(P_{XY})}\label{eq:prvw-kzerovalueandlimitb}\\
&= \lim_{\alpha \downarrow 0} K_\alpha(X;Y),\label{eq:prvw-kzerovalueandlimitc}
\end{align}
where in the RHS of \eqref{eq:prvw-kzerovalueandlimitb}, we use the conventions \eqref{eq:zeroinftyconventions}.
The inequality can be strict, so $\alpha \mapsto K_\alpha(X;Y)$ need not be continuous at $\alpha = 0$.
\item[(Lemma~\ref{lma:konemutualinformation})]
$K_1(X;Y) = I(X;Y)$.
\item[(Lemma~\ref{lma:ktwosingularvalue})]
Let $f\colon \{1,\ldots,\card{\set{X}}\} \to \set{X}$ and $g\colon \{1,\ldots,\card{\set{Y}}\} \to \set{Y}$ be bijective functions, and let $\mat{B}$ be the $\card{\set{X}} \times \card{\set{Y}}$ matrix whose Row-$i$ Column-$j$ entry $\mat{B}_{i,j}$ equals $P_{XY}(f(i),g(j))$.
Then,
\begin{align}
K_2(X;Y) = -2 \log \sigma_1(\mat{B}) - H_2(X,Y),
\end{align}
where $\sigma_1(\mat{B})$ denotes the largest singular value of $\mat{B}$.
(Because the singular values of a matrix are invariant under row and column permutations, the result does not depend on $f$ or $g$.)
\item[(Lemma~\ref{lma:kinftyiszero})]
$K_\infty(X;Y) = 0$.
\item[(Lemma~\ref{lma:kalphanotmonotonic})]
The mapping $\alpha \mapsto K_\alpha(X;Y)$ need not be monotonic on $[0,\infty]$.
\item[(Lemma~\ref{lma:kalphaplushalphanonincreasing})]
The mapping $\alpha \mapsto K_\alpha(X;Y) + H_\alpha(X,Y)$ is nonincreasing on $[0,\infty]$.
\item[(Lemma~\ref{lma:kalphacontinuousalpha})]
The mapping $\alpha \mapsto K_\alpha(X;Y)$ is continuous on $(0,\infty]$.
(See \lmaref{lma:kzerovalueandlimit} for the behavior at $\alpha = 0$.)
\item[(Lemma~\ref{lma:kalphaselfinformation})]
If $X = Y$ with probability one, then
\begin{align}
K_\alpha(X;Y) = \begin{cases}2 H_\frac{\alpha}{2 - \alpha}(X) - H_\alpha(X) & \text{if $\alpha \in [0,2)$,}\\
\frac{\alpha}{\alpha - 1} \mulspace H_\infty(X) - H_\alpha(X) & \text{if $\alpha \ge 2$,}\\
0 & \text{if $\alpha = \infty$.}\end{cases}\label{eq:prvw-kalphaselfinformation}
\end{align}
\item[(Lemma~\ref{lma:kalphauniqueminimizer})]
For every $\alpha \in (0,2)$, the mapping $(Q_X,Q_Y) \mapsto \Delta_\alpha(P_{XY}\|Q_X Q_Y)$ in the definition of $K_\alpha(X;Y)$ in \eqref{eq:defkalpha} has a unique minimizer.
This need not be the case when $\alpha \in \{0\} \cup [2,\infty]$.
\end{enumerate}
\end{Theorem}

\section{Proofs}
\label{sec:proofs}

In this section, we prove the properties of $J_\alpha(X;Y)$ and $K_\alpha(X;Y)$ stated in \secref{sec:twomeasuresofdependence}.

\begin{Lemma}\label{lma:jalphawelldefinedfinite}
For every $\alpha \in [0,\infty]$, the minimum in the definition of $J_\alpha(X;Y)$ exists and is finite.
\end{Lemma}

\begin{proof}
Let $\alpha \in [0,\infty]$.
Then $\inf_{\kernedqxcommaqy} D_\alpha(P_{XY}\|Q_X Q_Y)$ is finite because $D_\alpha(P_{XY}\|P_X P_Y)$ is finite and because the R\'enyi divergence is nonnegative.
The minimum exists because the set $\set{P}(\set{X}) \times \set{P}(\set{Y})$ is compact and the mapping $(Q_X,Q_Y) \mapsto D_\alpha(P_{XY}\|Q_X Q_Y)$ is continuous.
\end{proof}

\begin{Lemma}\label{lma:jalphanonnegative}
For all $\alpha \in [0,\infty]$, $J_\alpha(X;Y) \ge 0$.
If $\alpha \in (0,\infty]$, then $J_\alpha(X;Y) = 0$ if and only if $X$ and $Y$ are independent (nonnegativity).
\end{Lemma}

\begin{proof}
The nonnegativity follows from the definition of $J_\alpha(X;Y)$ because the R\'enyi divergence is nonnegative for $\alpha \in [0,\infty]$.
If $X$ and $Y$ are independent, then $P_{XY} = P_X P_Y$, and the choice $Q_X = P_X$ and $Q_Y = P_Y$ in the definition of $J_\alpha(X;Y)$ achieves $J_\alpha(X;Y) = 0$.
Conversely, if $J_\alpha(X;Y) = 0$, then there exist PMFs $Q_X^*$ and $Q_Y^*$ satisfying $D_\alpha(P_{XY}\|Q_X^* Q_Y^*) = 0$.
If, in addition, $\alpha \in (0,\infty]$, then $P_{XY} = Q_X^* Q_Y^*$ by \propref{prop:renyidivproperties}, and hence $X$ and $Y$ are independent.
\end{proof}

\begin{Lemma}\label{lma:jalphasymmetric}
For all $\alpha \in [0,\infty]$, $J_\alpha(X;Y) = J_\alpha(Y;X)$ (symmetry).
\end{Lemma}

\begin{proof}
The definition of $J_\alpha(X;Y)$ is symmetric in $X$ and $Y$.
\end{proof}

\begin{Lemma}\label{lma:jalphadpi}
If $X \markov Y \markov Z$ form a Markov chain, then $J_\alpha(X;Z) \le J_\alpha(X;Y)$ for all $\alpha \in [0,\infty]$ (data-processing inequality).
\end{Lemma}

\begin{proof}
Let $X \markov Y \markov Z$ form a Markov chain, and let $\alpha \in [0,\infty]$.
Let $\hat{Q}_X$ and $\hat{Q}_Y$ be PMFs that achieve the minimum in the definition of $J_\alpha(X;Y)$, so
\begin{align}
J_\alpha(X;Y) = D_\alpha(P_{XY}\|\hat{Q}_X \hat{Q}_Y).\label{eq:jalphadpiqxqy}
\end{align}
Define the PMF $\hat{Q}_Z$ as
\begin{align}
\hat{Q}_Z(z) \triangleq \sum_y \hat{Q}_Y(y) \mulspace P_{Z|Y}(z|y).\label{eq:jalphadpiqz}
\end{align}
(As noted in the preliminaries, we define $P_{Z|Y}(z|y) \triangleq 1 / \card{\set{Z}}$ when $P_Y(y) = 0$.)
We show below that
\begin{align}
D_\alpha(P_{XZ}\|\hat{Q}_X \hat{Q}_Z) \le D_\alpha(P_{XY}\|\hat{Q}_X \hat{Q}_Y),\label{eq:jalphadpiqxqzleqxqya}
\end{align}
which implies the data-processing inequality because
\begin{align}
J_\alpha(X;Z) &\le D_\alpha(P_{XZ}\|\hat{Q}_X \hat{Q}_Z)\label{eq:jalphadpia}\\
&\le D_\alpha(P_{XY}\|\hat{Q}_X \hat{Q}_Y)\label{eq:jalphadpib}\\
&= J_\alpha(X;Y),\label{eq:jalphadpic}
\end{align}
where \eqref{eq:jalphadpia} holds by the definition of $J_\alpha(X;Z)$;
\eqref{eq:jalphadpib} follows from \eqref{eq:jalphadpiqxqzleqxqya};
and \eqref{eq:jalphadpic} follows from \eqref{eq:jalphadpiqxqy}.

The proof of \eqref{eq:jalphadpiqxqzleqxqya} is based on the data-processing inequality for the R\'enyi divergence.
Define the conditional PMF $A_{X'Z'|XY}$ as
\begin{align}
A_{X'Z'|XY}(x',z'|x,y) \triangleq \mathbbm{1}\{x' = x\} \mulspace P_{Z|Y}(z'|y).\label{eq:jalphadpidefa}
\end{align}
If $(X,Y) \sim P_{XY}$, then the marginal distribution of $X'$ and $Z'$ is
\begin{align}
(P_{XY} A_{X'Z'|XY})(x',z') &= \sum_{x,y} P_{XY}(x,y) \mulspace A_{X'Z'|XY}(x',z'|x,y)\\
&= \sum_y P_{XY}(x',y) \mulspace P_{Z|Y}(z'|y)\label{eq:jalphadpipxyab}\\
&= \sum_y P_{XY}(x',y) \mulspace P_{Z|XY}(z'|x',y)\label{eq:jalphadpipxyac}\\
&= P_{XZ}(x',z'),\label{eq:jalphadpipxyad}
\end{align}
where \eqref{eq:jalphadpipxyab} follows from \eqref{eq:jalphadpidefa}; and \eqref{eq:jalphadpipxyac} holds because $X$, $Y$, and $Z$ form a Markov chain.
If $(X,Y) \sim \hat{Q}_X \hat{Q}_Y$, then the marginal distribution of $X'$ and $Z'$ is
\begin{align}
(\hat{Q}_X \hat{Q}_Y A_{X'Z'|XY})(x',z') &= \sum_{x,y} \hat{Q}_X(x) \mulspace \hat{Q}_Y(y) \mulspace A_{X'Z'|XY}(x',z'|x,y)\\
&= \sum_y \hat{Q}_X(x') \mulspace \hat{Q}_Y(y) \mulspace P_{Z|Y}(z'|y)\label{eq:jalphadpiqxqyab}\\
&= \hat{Q}_X(x') \mulspace \hat{Q}_Z(z'),\label{eq:jalphadpiqxqyac}
\end{align}
where \eqref{eq:jalphadpiqxqyab} follows from \eqref{eq:jalphadpidefa}, and \eqref{eq:jalphadpiqxqyac} follows from \eqref{eq:jalphadpiqz}.
Finally, we are ready to prove \eqref{eq:jalphadpiqxqzleqxqya}:
\begin{align}
D_\alpha(P_{XZ}\|\hat{Q}_X \hat{Q}_Z) &= D_\alpha\bigl((P_{XY} A_{X'Z'|XY})\|(\hat{Q}_X \hat{Q}_Y A_{X'Z'|XY})\bigr)\label{eq:jalphadpiqxqzleqxqyd}\\
&\le D_\alpha(P_{XY}\|\hat{Q}_X \hat{Q}_Y),\label{eq:jalphadpiqxqzleqxqye}
\end{align}
where \eqref{eq:jalphadpiqxqzleqxqyd} follows from \eqref{eq:jalphadpipxyad} and \eqref{eq:jalphadpiqxqyac}, and
where \eqref{eq:jalphadpiqxqzleqxqye} follows from the data-processing inequality for the R\'enyi divergence \mdpicite[Theorem~9]{VanErvenHarremoes}.
\end{proof}

\begin{Lemma}\label{lma:jzeroiszero}
$J_0(X;Y) = 0$.
\end{Lemma}

\begin{proof}
By \lmaref{lma:jalphanonnegative}, $J_0(X;Y) \ge 0$, so it suffices to show that $J_0(X;Y) \le 0$.
Let $(\hat{x},\hat{y}) \in \set{X} \times \set{Y}$ satisfy $P_{XY}(\hat{x},\hat{y}) > 0$.
Define the PMF $\hat{Q}_X$ as $\hat{Q}_X(x) \triangleq \mathbbm{1}\{x = \hat{x}\}$ and the PMF $\hat{Q}_Y$ as $\hat{Q}_Y(y) \triangleq \mathbbm{1}\{y = \hat{y}\}$.
Then, $D_0(P_{XY}\|\hat{Q}_X \hat{Q}_Y) = 0$, so $J_0(X;Y) \le 0$ by the definition of $J_0(X;Y)$.
\end{proof}

\begin{Lemma}\label{lma:jonehalfsingularvalue}
Let $f\colon \{1,\ldots,\card{\set{X}}\} \to \set{X}$ and $g\colon \{1,\ldots,\card{\set{Y}}\} \to \set{Y}$ be bijective functions, and let $\mat{A}$ be the $\card{\set{X}} \times \card{\set{Y}}$ matrix whose Row-$i$ Column-$j$ entry $\mat{A}_{i,j}$ equals $\sqrt{P_{XY}(f(i),g(j))}$.
Then,
\begin{align}
J_\frac{1}{2}(X;Y) = -2 \log \sigma_1(\mat{A}),
\end{align}
where $\sigma_1(\mat{A})$ denotes the largest singular value of $\mat{A}$.
(Because the singular values of a matrix are invariant under row and column permutations, the result does not depend on $f$ or $g$.)
\end{Lemma}

\begin{proof}
By the definitions of $J_\alpha(X;Y)$ and the R\'enyi divergence,
\begin{align}
J_\frac{1}{2}(X;Y) = -2 \log \max_{\kernedqxcommaqy} \sum_{x,y} \sqrt{Q_X(x)} \mulspace \sqrt{P(x,y)} \mulspace \sqrt{Q_Y(y)}.\label{eq:jonehalfa}
\end{align}
The claim follows from \eqref{eq:jonehalfa} because
\begin{align}
\max_{\kernedqxcommaqy} \sum_{x,y} \sqrt{Q_X(x)} \mulspace \sqrt{P(x,y)} \mulspace \sqrt{Q_Y(y)} &= \max_{\norm{\vect{u}}_2 = \norm{\vect{v}}_2 = 1} \trans{\vect{u}} \vthinnegspace \mat{A} \mulspace \vect{v}\label{eq:jonehalfd}\\*
&= \max_{\norm{\vect{v}}_2 = 1} \norm{\mat{A} \mulspace \vect{v}}_2\label{eq:jonehalfe}\\*
&= \sigma_1(\mat{A}),\label{eq:jonehalff}
\end{align}
where $\vect{u}$ and $\vect{v}$ are column vectors with $\card{\set{X}}$ and $\card{\set{Y}}$ elements, respectively;
\eqref{eq:jonehalfd} is shown below;
\eqref{eq:jonehalfe}~follows from the Cauchy--Schwarz inequality $\abs{\trans{\vect{u}} \vthinnegspace \mat{A} \mulspace \vect{v}} \le \norm{\vect{u}}_2 \mulspace \norm{\mat{A} \mulspace \vect{v}}_2$, which holds with equality if $\vect{u}$ and $\mat{A} \mulspace \vect{v}$ are linearly dependent;
and \eqref{eq:jonehalff} holds because the spectral norm of a matrix is equal to its largest singular value \mdpicite[Example~5.6.6]{MatrixAnalysis}.

We now prove \eqref{eq:jonehalfd}.
Let $\vect{u}$ and $\vect{v}$ be vectors that satisfy $\norm{\vect{u}}_2 = \norm{\vect{v}}_2 = 1$, and define the PMFs $\hat{Q}_X$ and $\hat{Q}_Y$ as $\hat{Q}_X(x) \triangleq \vect{u}_{f^{-1}(x)}^2$ and $\hat{Q}_Y(y) \triangleq \vect{v}_{g^{-1}(y)}^2$, where $f^{-1}$ and $g^{-1}$ denote the inverse functions of $f$ and $g$, respectively.
Then,
\begin{align}
\trans{\vect{u}} \vthinnegspace \mat{A} \mulspace \vect{v} &= \sum_{i,j} \vect{u}_i \mulspace \mat{A}_{i,j} \mulspace \vect{v}_j\\
&\le \sum_{i,j} \abs{\vect{u}_i} \, \mat{A}_{i,j} \, \abs{\vect{v}_j}\label{eq:jonehalfj}\\
&= \sum_{x,y} \sqrt{\hat{Q}_X(x)} \mulspace \sqrt{P(x,y)} \mulspace \sqrt{\hat{Q}_Y(y)}\label{eq:jonehalfk}\\
&\le \max_{\kernedqxcommaqy} \sum_{x,y} \sqrt{Q_X(x)} \mulspace \sqrt{P(x,y)} \mulspace \sqrt{Q_Y(y)},\label{eq:jonehalfl}
\end{align}
where \eqref{eq:jonehalfj} holds because all the entries of $\mat{A}$ are nonnegative, and
in \eqref{eq:jonehalfk}, we changed the summation variables to $x \triangleq f(i)$ and $y \triangleq g(j)$.
It remains to show that equality can be achieved in \eqref{eq:jonehalfj} and~\eqref{eq:jonehalfl}.
To that end, let $Q_X^*$ and $Q_Y^*$ be PMFs that achieve the maximum on the RHS of \eqref{eq:jonehalfl}, and define the vectors $\vect{u}$~and $\vect{v}$ as $\vect{u}_i \triangleq Q_X^*(f(i))^{1/2}$ and $\vect{v}_j \triangleq Q_Y^*(g(j))^{1/2}$.
Then, $\norm{\vect{u}}_2 = \norm{\vect{v}}_2 = 1$, and \eqref{eq:jonehalfj} and~\eqref{eq:jonehalfl} hold with equality, which proves \eqref{eq:jonehalfd}.
\end{proof}

\begin{Lemma}\label{lma:jonemutualinformation}
$J_1(X;Y) = I(X;Y)$.
\end{Lemma}

\begin{proof}
This follows from \propref{prop:minqxqydpxyqxqy} because $D_1(P_{XY}\|Q_X Q_Y)$ in the definition of $J_1(X;Y)$ is equal to $D(P_{XY}\|Q_X Q_Y)$.
\end{proof}

\begin{Lemma}\label{lma:oneminusalphajalpharxy}
For all $\alpha > 0$,
\begin{align}
(1 - \alpha) \mulspace J_\alpha(X;Y) = \min_{R_{XY} \in \set{P}(\set{X} \times \set{Y})} \bigl[(1 - \alpha) \mulspace D(R_{XY}\|R_X R_Y) + \alpha \mulspace D(R_{XY}\|P_{XY})\bigr].\label{eq:oneminusalphajalpharxy}
\end{align}
Thus, being the minimum of concave functions in $\alpha$, the mapping $\alpha \mapsto (1 - \alpha) \mulspace J_\alpha(X;Y)$ is concave on $(0,\infty)$.
\end{Lemma}

\begin{proof}
For $\alpha = 1$, \eqref{eq:oneminusalphajalpharxy} holds because $D(R_{XY}\|P_{XY}) \ge 0$ with equality if $R_{XY} = P_{XY}$.
For $\alpha \in (0,1)$,
\begin{align}
(1 - \alpha) \mulspace J_\alpha(X;Y) &= \min_{\kernedqxcommaqy} (1 - \alpha) \mulspace D_\alpha(P_{XY}\|Q_X Q_Y)\label{eq:oneminusalphajalphasmalla}\\
&= \min_{\kernedqxcommaqy} \min_{R_{XY}} \vthinspace \bigl[(1 - \alpha) \mulspace D(R_{XY}\|Q_X Q_Y) + \alpha \mulspace D(R_{XY}\|P_{XY})\bigr]\label{eq:oneminusalphajalphasmallb}\\
&= \min_{R_{XY}} \vthinspace \bigl[(1 - \alpha) \mulspace D(R_{XY}\|R_X R_Y) + \alpha \mulspace D(R_{XY}\|P_{XY})\bigr],\label{eq:oneminusalphajalphasmallc}
\end{align}
where \eqref{eq:oneminusalphajalphasmalla} holds by the definition of $J_\alpha(X;Y)$;
\eqref{eq:oneminusalphajalphasmallb} follows from \mdpicite[Theorem~30]{VanErvenHarremoes}; and
\eqref{eq:oneminusalphajalphasmallc} follows from \propref{prop:minqxqydpxyqxqy} after swapping the minima.

For $\alpha > 1$, define the sets
\begin{align}
\set{Q} &\triangleq \{(Q_X,Q_Y) \in \set{P}(\set{X}) \times \set{P}(\set{Y}) : \supp(Q_X Q_Y) = \set{X} \times \set{Y}\},\\*
\set{R} &\triangleq \{R_{XY} \in \set{P}(\set{X} \times \set{Y}) : \supp(R_{XY}) \subseteq \supp(P_{XY})\}.
\end{align}
Then,
\begin{align}
(1 - \alpha) \mulspace J_\alpha(X;Y) &= \sup_{(\kernedqxcommaqy) \in \set{Q}} \vthinspace (1 - \alpha) \mulspace D_\alpha(P_{XY}\|Q_X Q_Y)\label{eq:oneminusalphajalphabiga}\\
&= \sup_{(\kernedqxcommaqy) \in \set{Q}} \, \minvphantomsup_{R_{XY} \in \set{R}\vphantom{(\kernedqxcommaqy) \in \set{Q}}} \vthinspace \bigl[(1 - \alpha) \mulspace D(R_{XY}\|Q_X Q_Y) + \alpha \mulspace D(R_{XY}\|P_{XY})\bigr]\label{eq:oneminusalphajalphabigb}\\
&= \minvphantomsup_{R_{XY} \in \set{R}\vphantom{(\kernedqxcommaqy) \in \set{Q}}} \, \sup_{(\kernedqxcommaqy) \in \set{Q}} \vthinspace \bigl[(1 - \alpha) \mulspace D(R_{XY}\|Q_X Q_Y) + \alpha \mulspace D(R_{XY}\|P_{XY})\bigr]\label{eq:oneminusalphajalphabigc}\\
&= \min_{R_{XY} \in \set{P}(\set{X} \times \set{Y})} \vthinspace \bigl[(1 - \alpha) \mulspace D(R_{XY}\|R_X R_Y) + \alpha \mulspace D(R_{XY}\|P_{XY})\bigr],\label{eq:oneminusalphajalphabigd}
\end{align}
where \eqref{eq:oneminusalphajalphabiga} follows from the definition of $J_\alpha(X;Y)$ because $1 - \alpha < 0$ and because the mapping $(Q_X,Q_Y) \mapsto D_\alpha(P_{XY}\|Q_X Q_Y)$ is continuous;
\eqref{eq:oneminusalphajalphabigb} follows from \mdpicite[Theorem~30]{VanErvenHarremoes};
\eqref{eq:oneminusalphajalphabigc} follows from a minimax theorem and is justified below; and
\eqref{eq:oneminusalphajalphabigd} follows from \propref{prop:minqxqydpxyqxqy}, a continuity argument, and the observation that $D(R_{XY}\|P_{XY})$ is infinite if $R_{XY} \notin \set{R}$.

We now verify the conditions of Ky Fan's minimax theorem \mdpicite[Theorem~2]{KyFan}, which will establish~\eqref{eq:oneminusalphajalphabigc}.
(We use Ky Fan's minimax theorem because it does not require that the set $\set{Q}$ be compact, and~having a noncompact set $\set{Q}$ helps to guarantee that the function $f$ defined next takes on finite values only.
A brief proof of Ky Fan's minimax theorem appears in \cite{KyFanAltProof}.)
Let the function $f\colon \set{R} \times \set{Q} \to \reals$ be defined by the expression in square brackets in \eqref{eq:oneminusalphajalphabigc}, i.e.,
\begin{align}
f(R_{XY}, Q_X, Q_Y) \triangleq (1 - \alpha) \mulspace D(R_{XY}\|Q_X Q_Y) + \alpha \mulspace D(R_{XY}\|P_{XY}).\label{eq:minimaxdeff}
\end{align}
We check that
\begin{enumerate}[beginpenalty=10000,midpenalty=10000,label=(\roman*)]
\item
the sets $\set{Q}$ and $\set{R}$ are convex;
\item
the set $\set{R}$ is compact;
\item
the function $f$ is real-valued;
\item
for every $(Q_X,Q_Y) \in \set{Q}$, the function $f$ is continuous in $R_{XY}$;
\item
for every $(Q_X,Q_Y) \in \set{Q}$, the function $f$ is convex in $R_{XY}$; and
\item
for every $R_{XY} \in \set{R}$, the function $f$ is concave in the pair $(Q_X,Q_Y)$.
\end{enumerate}
Indeed, Parts~(i) and (ii) are easy to see;
Part~(iii) holds because both relative entropies on the RHS of~\eqref{eq:minimaxdeff} are finite by our definitions of $\set{Q}$ and $\set{R}$;
and to show Parts~(iv)--(vi), we rewrite $f$ as:
\begin{align}
f(R_{XY}, Q_X, Q_Y) &= -H(R_{XY}) - \alpha \sum_{x,y} R_{XY}(x,y) \log P(x,y)\nonumber\\*
&\mathrel{\hphantom{=}} \binaryemptyleft + (\alpha - 1) \sum_x R_X(x) \log Q_X(x) + (\alpha - 1) \sum_y R_Y(y) \log Q_Y(y).\label{eq:oneminusalphadrxyqxqyplusalphadrxypxy}
\end{align}
From \eqref{eq:oneminusalphadrxyqxqyplusalphadrxypxy}, we see that Part~(iv) holds by our definitions of $\set{Q}$ and $\set{R}$;
Part~(v) holds because the entropy is a concave function (so $-H(R_{XY})$ is convex), because linear functionals of $R_{XY}$ are convex, and because the sum of convex functions is convex; and
Part~(vi) holds because the logarithm is a concave function and because a nonnegative weighted sum of concave functions is concave.
(In Ky Fan's theorem, weaker conditions than Parts~(i)--(vi) are required, but it is not difficult to see that Parts~(i)--(vi) are sufficient.)

The last claim, namely, that the mapping $\alpha \mapsto (1 - \alpha) \mulspace J_\alpha(X;Y)$ is concave on $(0,\infty)$,
is true because the expression in square brackets on the RHS of \eqref{eq:oneminusalphajalpharxy} is concave in $\alpha$ for every $R_{XY}$ and because the pointwise minimum preserves the concavity.
\end{proof}

\begin{Lemma}\label{lma:jalphanondecreasing}
The mapping $\alpha \mapsto J_\alpha(X;Y)$ is nondecreasing on $[0,\infty]$.
\end{Lemma}

\begin{proof}
This is true because for every $\alpha,\alpha' \in [0,\infty]$ with $\alpha \le \alpha'$,
\begin{align}
\min_{\kernedqxcommaqy} D_\alpha(P_{XY}\|Q_X Q_Y) \le \min_{\kernedqxcommaqy} D_{\alpha'}(P_{XY}\|Q_X Q_Y),
\end{align}
which holds because the R\'enyi divergence is nondecreasing in $\alpha$ (\propref{prop:renyidivproperties}).
\end{proof}

\begin{Lemma}\label{lma:jalphacontinuous}
The mapping $\alpha \mapsto J_\alpha(X;Y)$ is continuous on $[0,\infty]$.
\end{Lemma}

\begin{proof}
By \lmaref{lma:oneminusalphajalpharxy}, the mapping $\alpha \mapsto (1 - \alpha) \mulspace J_\alpha(X;Y)$ is concave on $(0,\infty)$, thus it is continuous on $(0,\infty)$, which implies that $\alpha \mapsto J_\alpha(X;Y)$ is continuous on $(0,1) \cup (1,\infty)$.

We next prove the continuity at $\alpha = 0$.
Let $Q_X^*$ and $Q_Y^*$ be PMFs that achieve the minimum in the definition of $J_0(X;Y)$.
Then, for all $\alpha \ge 0$,
\begin{align}
D_0(P_{XY}\|Q_X^* Q_Y^*) &= J_0(X;Y)\label{eq:jzerocontinuousa}\\
&\le J_\alpha(X;Y)\label{eq:jzerocontinuousb}\\
&\le D_\alpha(P_{XY}\|Q_X^* Q_Y^*),\label{eq:jzerocontinuousc}
\end{align}
where \eqref{eq:jzerocontinuousb} holds because $\alpha \mapsto J_\alpha(X;Y)$ is nondecreasing (\lmaref{lma:jalphanondecreasing}), and \eqref{eq:jzerocontinuousc} holds by the definition of $J_\alpha(X;Y)$.
The R\'enyi divergence is continuous in $\alpha$ (\propref{prop:renyidivproperties}), so \eqref{eq:jzerocontinuousa}--\eqref{eq:jzerocontinuousc} and the sandwich theorem imply that $J_\alpha(X;Y)$ is continuous at $\alpha = 0$.

We continue with the continuity at $\alpha = \infty$.
Define
\begin{align}
\tau \triangleq \min_{(x,y) \in \supp(P_{XY})} P(x,y).
\end{align}
Then, for all $\alpha > 1$,
\begin{align}
J_\infty(X;Y) &\ge J_\alpha(X;Y)\label{eq:jinftycontinuousa}\\
&= \min_{\kernedqxcommaqy} \frac{1}{\alpha - 1} \log \sum_{x,y} P(x,y) \mulspace \frac{P(x,y)^{\alpha - 1}}{[Q_X(x) \mulspace Q_Y(y)]^{\alpha - 1}}\label{eq:jinftycontinuousb}\\
&\ge \min_{\kernedqxcommaqy} \frac{1}{\alpha - 1} \log \max_{x,y} \frac{\tau \mulspace P(x,y)^{\alpha - 1}}{[Q_X(x) \mulspace Q_Y(y)]^{\alpha - 1}}\\
&= \frac{1}{\alpha - 1} \log \tau + \min_{\kernedqxcommaqy} \log \max_{x,y} \frac{P(x,y)}{Q_X(x) \mulspace Q_Y(y)}\\
&= \frac{1}{\alpha - 1} \log \tau + J_\infty(X;Y),\label{eq:jinftycontinuouse}
\end{align}
where \eqref{eq:jinftycontinuousa} holds because $\alpha \mapsto J_\alpha(X;Y)$ is nondecreasing (\lmaref{lma:jalphanondecreasing}), and \eqref{eq:jinftycontinuousb} and \eqref{eq:jinftycontinuouse} hold by the definitions of $J_\alpha(X;Y)$ and the R\'enyi divergence.
The RHS of \eqref{eq:jinftycontinuouse} tends to $J_\infty(X;Y)$ as $\alpha$ tends to infinity, so $J_\alpha(X;Y)$ is continuous at $\alpha = \infty$ by the sandwich theorem.

It remains to show the continuity at $\alpha = 1$.
Let $\alpha \in (\frac{3}{4},1) \cup (1,\frac{5}{4})$, and let $\delta \triangleq \abs{1 - \alpha} \in (0,\frac{1}{4})$.
Then, for all PMFs $Q_X$ and $Q_Y$,
\begin{align}
2^{-\delta D_\alpha(P_{XY}\|Q_X Q_Y)} &\le 2^{-\delta D_{1 - \delta}(P_{XY}\|Q_X Q_Y)}\label{eq:jonecontinuitya}\\*
&= \sum_{x,y} P(x,y) \mleft[\frac{Q_X(x) \mulspace Q_Y(y)}{P(x,y)}\mright]^\delta\label{eq:jonecontinuityb}\\
&= \sum_{x,y} P(x,y) \mleft[\frac{P_X(x) \mulspace P_Y(y)}{P(x,y)}\mright]^\delta \mulspace \mleft[\frac{Q_X(x) \mulspace Q_Y(y)}{P_X(x) \mulspace P_Y(y)}\mright]^\delta\label{eq:jonecontinuityc}\\
&\le \mleft\{\sum_{x,y} P(x,y) \mleft[\frac{P_X(x) \mulspace P_Y(y)}{P(x,y)}\mright]^{2 \delta}\mright\}^\frac{1}{2} \cdot \mleft\{\sum_{x,y} P(x,y) \mleft[\frac{Q_X(x) \mulspace Q_Y(y)}{P_X(x) \mulspace P_Y(y)}\mright]^{2 \delta}\mright\}^\frac{1}{2}\label{eq:jonecontinuityd}\\
&\le \mleft\{\sum_{x,y} P(x,y) \mleft[\frac{P_X(x) \mulspace P_Y(y)}{P(x,y)}\mright]^{2 \delta}\mright\}^\frac{1}{2}\label{eq:jonecontinuitye}\\
&= 2^{-\delta D_{1 - 2 \delta}(P_{XY}\|P_X P_Y)},\label{eq:jonecontinuityf}
\end{align}
where \eqref{eq:jonecontinuitya} holds because $1 - \delta \le \alpha$ and because the R\'enyi divergence is nondecreasing in $\alpha$ (\propref{prop:renyidivproperties});
\eqref{eq:jonecontinuityd} follows from the Cauchy--Schwarz inequality; and
\eqref{eq:jonecontinuitye} holds because
\begin{align}
\mleft\{\sum_{x,y} P(x,y) \mleft[\frac{Q_X(x) \mulspace Q_Y(y)}{P_X(x) \mulspace P_Y(y)}\mright]^{2 \delta}\mright\}^\frac{1}{2} &\le \mleft\{\sum_x P_X(x) \mleft[\frac{Q_X(x)}{P_X(x)}\mright]^{4 \delta}\mright\}^\frac{1}{4} \cdot \mleft\{\sum_y P_Y(y) \mleft[\frac{Q_Y(y)}{P_Y(y)}\mright]^{4 \delta}\mright\}^\frac{1}{4}\label{eq:jonecontinuityi}\\
&= 2^{-\delta D_{1 - 4 \delta}(P_X\|Q_X)} \cdot 2^{-\delta D_{1 - 4 \delta}(P_Y\|Q_Y)}\label{eq:jonecontinuityj}\\
&\le 1,\label{eq:jonecontinuityk}
\end{align}
where \eqref{eq:jonecontinuityi} follows from the Cauchy--Schwarz inequality, and
\eqref{eq:jonecontinuityk} holds because $1 - 4 \delta > 0$ and because the R\'enyi divergence is nonnegative for positive orders (\propref{prop:renyidivproperties}).
Thus, for all $\alpha \in (\frac{3}{4},\frac{5}{4})$,
\begin{align}
D_{1 - 2 \abs{1 - \alpha}}(P_{XY}\|P_X P_Y) &\le \min_{\kernedqxcommaqy} D_\alpha(P_{XY}\|Q_X Q_Y)\label{eq:jonecontinuitym}\\
&= J_\alpha(X;Y)\label{eq:jonecontinuityn}\\
&\le D_\alpha(P_{XY}\|P_X P_Y),\label{eq:jonecontinuityo}
\end{align}
where \eqref{eq:jonecontinuitym} follows from \eqref{eq:jonecontinuityf} if $\alpha \ne 1$ and from \propref{prop:minqxqydpxyqxqy} if $\alpha = 1$; and \eqref{eq:jonecontinuityo} holds by the definition of $J_\alpha(X;Y)$.
The R\'enyi divergence is continuous in $\alpha$ (\propref{prop:renyidivproperties}), thus \eqref{eq:jonecontinuitym}--\eqref{eq:jonecontinuityo} and the sandwich theorem imply that $J_\alpha(X;Y)$ is continuous at $\alpha = 1$.
\end{proof}

\begin{Lemma}\label{lma:jalphaselfinformation}
If $X = Y$ with probability one, then
\begin{align}
J_\alpha(X;Y) = \begin{cases}\frac{\alpha}{1 - \alpha} \mulspace H_\infty(X) & \text{if $\alpha \in [0,\frac{1}{2}]$,}\\
H_\frac{\alpha}{2 \alpha - 1}(X) & \text{if $\alpha > \frac{1}{2}$,}\\
H_\frac{1}{2}(X) & \text{if $\alpha = \infty$.}\end{cases}\label{eq:jalphaselfinformation}
\end{align}
\end{Lemma}

\begin{proof}
We show below that \eqref{eq:jalphaselfinformation} holds for $\alpha \in (0,1) \cup (1,\infty)$.
Thus, \eqref{eq:jalphaselfinformation} holds also for $\alpha \in \{0,1,\infty\}$ because both its sides are continuous in $\alpha$:
its LHS by \lmaref{lma:jalphacontinuous}, and its RHS by the continuity of the R\'enyi entropy (\propref{prop:renyientropyproperties}).

Fix $\alpha \in (0,1) \cup (1,\infty)$.
Then,
\begin{align}
J_\alpha(X;Y) &= \min_{Q_X} \min_{Q_Y} D_\alpha(P_{XY}\|Q_X Q_Y)\\*
&= \min_{Q_X} \frac{\alpha}{\alpha - 1} \log \sum_y \mleft[\sum_x P(x,y)^\alpha \mulspace Q_X(x)^{1 - \alpha}\mright]^\frac{1}{\alpha}\label{eq:jalphaselfinformationb}\\
&= \min_{Q_X} \frac{\alpha}{\alpha - 1} \log \sum_x P_X(x) \mulspace Q_X(x)^\frac{1 - \alpha}{\alpha},\label{eq:jalphaselfinformationc}
\end{align}
where \eqref{eq:jalphaselfinformationb} follows from \propref{prop:infqydalphapxyqxqy}, and \eqref{eq:jalphaselfinformationc} holds because
\begin{align}
P_{XY}(x,y) = \begin{cases}P_X(x) & \text{if $x = y$,}\\
0 & \text{otherwise.}\end{cases}
\end{align}

First consider the case $\alpha > \frac{1}{2}$.
Define $\gamma \triangleq \sum_x P_X(x)^\frac{\alpha}{2 \alpha - 1}$.
Then, for all $Q_X \in \set{P}(\set{X})$,
\begin{align}
\frac{\alpha}{\alpha - 1} \log \sum_x P_X(x) \mulspace Q_X(x)^\frac{1 - \alpha}{\alpha} &= \frac{\alpha}{\alpha - 1} \log \sum_x \bigl[\gamma \mulspace \gamma^{-1} \mulspace P_X(x)^\frac{\alpha}{2 \alpha - 1}\bigr]^\frac{2 \alpha - 1}{\alpha} Q_X(x)^\frac{1 - \alpha}{\alpha}\\
&= \frac{2 \alpha - 1}{\alpha - 1} \log \gamma + D_\frac{2 \alpha - 1}{\alpha}(\gamma^{-1} \mulspace P_X\sepsubandsup{}^\frac{\alpha}{2 \alpha - 1}\|Q_X)\label{eq:jalphaselfinformationg}\\
&= H_\frac{\alpha}{2 \alpha - 1}(X) + D_\frac{2 \alpha - 1}{\alpha}(\gamma^{-1} \mulspace P_X\sepsubandsup{}^\frac{\alpha}{2 \alpha - 1}\|Q_X),\label{eq:jalphaselfinformationh}
\end{align}
where \eqref{eq:jalphaselfinformationg} holds because $x \mapsto \gamma^{-1} \mulspace P_X(x)^\frac{\alpha}{2 \alpha - 1}$ is a PMF.
Because $\frac{2 \alpha - 1}{\alpha} > 0$, \propref{prop:renyidivproperties} implies that $D_{(2 \alpha - 1) / \alpha}(P\|Q) \ge 0$ with equality if $Q = P$.
This together with \eqref{eq:jalphaselfinformationc} and \eqref{eq:jalphaselfinformationh} establishes \eqref{eq:jalphaselfinformation}.

Now consider the case $\alpha \in (0,\frac{1}{2}]$.
For all $Q_X \in \set{P}(\set{X})$,
\begin{align}
\sum_x P_X(x) \mulspace Q_X(x)^\frac{1 - \alpha}{\alpha} &\le \sum_x P_X(x) \mulspace Q_X(x)\label{eq:jalphaselfinformationk}\\
&\le \sum_x \mleft[\max_{x'} P_X(x')\mright] Q_X(x)\label{eq:jalphaselfinformationl}\\
&= \max_x P_X(x),\label{eq:jalphaselfinformationm}
\end{align}
where \eqref{eq:jalphaselfinformationk} holds because $Q_X(x) \in [0,1]$ for all $x \in \set{X}$ and because $\frac{1 - \alpha}{\alpha} \ge 1$.
The inequalities~\eqref{eq:jalphaselfinformationk} and \eqref{eq:jalphaselfinformationl} both hold with equality when $Q_X(x) = \mathbbm{1}\{x = x^*\}$, where $x^* \in \set{X}$ is such that $P_X(x^*) = \max_x P_X(x)$.
Thus,
\begin{align}
\max_{Q_X} \sum_x P_X(x) \mulspace Q_X(x)^\frac{1 - \alpha}{\alpha} = \max_x P_X(x).\label{eq:jalphaselfinformationn}
\end{align}
Now \eqref{eq:jalphaselfinformation} follows:
\begin{align}
J_\alpha(X;Y) &= \min_{Q_X} \frac{\alpha}{\alpha - 1} \log \sum_x P_X(x) \mulspace Q_X(x)^\frac{1 - \alpha}{\alpha}\label{eq:jalphaselfinformationp}\\*
&= \frac{\alpha}{\alpha - 1} \log \max_{Q_X} \sum_x P_X(x) \mulspace Q_X(x)^\frac{1 - \alpha}{\alpha}\label{eq:jalphaselfinformationq}\\
&= \frac{\alpha}{\alpha - 1} \log \max_x P_X(x)\label{eq:jalphaselfinformationr}\\
&= \frac{\alpha}{1 - \alpha} \mulspace H_\infty(X),\label{eq:jalphaselfinformations}
\end{align}
where \eqref{eq:jalphaselfinformationp} follows from \eqref{eq:jalphaselfinformationc};
\eqref{eq:jalphaselfinformationq} holds because $\frac{\alpha}{\alpha - 1} < 0$;
\eqref{eq:jalphaselfinformationr} follows from \eqref{eq:jalphaselfinformationn}; and
\eqref{eq:jalphaselfinformations} follows from the definition of $H_\infty(X)$.
\end{proof}

\begin{Lemma}\label{lma:jalphaadditive}
If the pairs $(X_1,Y_1)$ and $(X_2,Y_2)$ are independent, then $J_\alpha(X_1,X_2;Y_1,Y_2) = J_\alpha(X_1;Y_1) + J_\alpha(X_2;Y_2)$ for all $\alpha \in [0,\infty]$ (additivity).
\end{Lemma}

\begin{proof}
Let the pairs $(X_1,Y_1)$ and $(X_2,Y_2)$ be independent.
For $\alpha \in (0,1) \cup (1,\infty)$, we establish the lemma by showing the following two inequalities:
\begin{align}
J_\alpha(X_1,X_2;Y_1,Y_2) &\le J_\alpha(X_1;Y_1) + J_\alpha(X_2;Y_2),\label{eq:jalphaadditivem}\\
J_\alpha(X_1,X_2;Y_1,Y_2) &\ge J_\alpha(X_1;Y_1) + J_\alpha(X_2;Y_2).\label{eq:jalphaadditivesecondpart}
\end{align}
Because $J_\alpha(X;Y)$ is continuous in $\alpha$ (\lmaref{lma:jalphacontinuous}), this will also establish the lemma for $\alpha \in \{0,1,\infty\}$.

To show \eqref{eq:jalphaadditivem}, let $Q_{X_1}^*$ and $Q_{Y_1}^*$ be PMFs that achieve the minimum in the definition of $J_\alpha(X_1;Y_1)$, and let $Q_{X_2}^*$ and $Q_{Y_2}^*$ be PMFs that achieve the minimum in the definition of $J_\alpha(X_2;Y_2)$, so
\begin{align}
J_\alpha(X_1;Y_1) &= D_\alpha(P_{X_1 Y_1}\|Q_{X_1}^* Q_{Y_1}^*),\label{eq:jalphaadditiveqxonestarqyonestar}\\
J_\alpha(X_2;Y_2) &= D_\alpha(P_{X_2 Y_2}\|Q_{X_2}^* Q_{Y_2}^*).\label{eq:jalphaadditiveqxtwostarqytwostar}
\end{align}
Then, \eqref{eq:jalphaadditivem} holds because
\begin{align}
J_\alpha(X_1,X_2;Y_1,Y_2) &\le D_\alpha(P_{X_1 X_2 Y_1 Y_2}\|Q_{X_1}^* Q_{X_2}^* Q_{Y_1}^* Q_{Y_2}^*)\label{eq:jalphaadditiveo}\\*
&= D_\alpha(P_{X_1 Y_1}\|Q_{X_1}^* Q_{Y_1}^*) + D_\alpha(P_{X_2 Y_2}\|Q_{X_2}^* Q_{Y_2}^*)\label{eq:jalphaadditivep}\\
&= J_\alpha(X_1;Y_1) + J_\alpha(X_2;Y_2),\label{eq:jalphaadditiveq}
\end{align}
where \eqref{eq:jalphaadditiveo} holds by the definition of $J_\alpha(X_1,X_2;Y_1,Y_2)$ as a minimum;
\eqref{eq:jalphaadditivep} follows from a simple computation using the independence hypothesis $P_{X_1 X_2 Y_1 Y_2} = P_{X_1 Y_1} P_{X_2 Y_2}$; and
\eqref{eq:jalphaadditiveq} follows from \eqref{eq:jalphaadditiveqxonestarqyonestar} and \eqref{eq:jalphaadditiveqxtwostarqytwostar}.

To establish \eqref{eq:jalphaadditivesecondpart}, we consider the cases $\alpha > 1$ and $\alpha < 1$ separately, starting with $\alpha > 1$.
Let $\hat{Q}_{X_1 X_2}$ and $\hat{Q}_{Y_1 Y_2}$ be PMFs that achieve the minimum in the definition of $J_\alpha(X_1,X_2;Y_1,Y_2)$, so
\begin{align}
J_\alpha(X_1,X_2;Y_1,Y_2) = D_\alpha(P_{X_1 X_2 Y_1 Y_2}\|\hat{Q}_{X_1 X_2} \hat{Q}_{Y_1 Y_2}).\label{eq:jalphaadditivedalpha}
\end{align}
Define the function $f\colon \set{X}_1 \times \set{Y}_1 \to \reals \cup \{\infty\}$ as
\begin{align}
f(x_1,y_1) \triangleq \sum_{x_2,y_2} P_{X_2 Y_2}(x_2,y_2)^\alpha \bigl[\hat{Q}_{X_2|X_1}(x_2|x_1) \mulspace \hat{Q}_{Y_2|Y_1}(y_2|y_1)\bigr]^{1 - \alpha},\label{eq:jalphaadditivedeff}
\end{align}
and let $(x_1',y_1') \in \set{X}_1 \times \set{Y}_1$ be such that
\begin{align}
f(x_1',y_1') = \min_{x_1,y_1} f(x_1,y_1).\label{eq:jalphaadditivexoneprimeyoneprime}
\end{align}
Define the PMFs $Q_{X_2}'$ and $Q_{Y_2}'$ as
\begin{align}
Q_{X_2}'(x_2) &\triangleq \hat{Q}_{X_2|X_1}(x_2|x_1'),\label{eq:jalphaadditiveqxtwoprime}\\
Q_{Y_2}'(y_2) &\triangleq \hat{Q}_{Y_2|Y_1}(y_2|y_1').\label{eq:jalphaadditiveqytwoprime}
\end{align}
Then,
\begin{align}
2^{(\alpha - 1) J_\alpha(X_1,X_2;Y_1,Y_2)} &= 2^{(\alpha - 1) D_\alpha(P_{X_1 X_2 Y_1 Y_2}\|\hat{Q}_{X_1 X_2} \hat{Q}_{Y_1 Y_2})}\label{eq:jalphaadditiveaa}\\*
&= \sum_{x_1,x_2,y_1,y_2} \bigl[P_{X_1 Y_1}(x_1,y_1) \mulspace P_{X_2 Y_2}(x_2,y_2)\bigr]^\alpha \bigl[\hat{Q}_{X_1 X_2}(x_1,x_2) \mulspace \hat{Q}_{Y_1 Y_2}(y_1,y_2)\bigr]^{1 - \alpha}\label{eq:jalphaadditivea}\\
&= \sum_{x_1,y_1} P_{X_1 Y_1}(x_1,y_1)^\alpha \bigl[\hat{Q}_{X_1}(x_1) \mulspace \hat{Q}_{Y_1}(y_1)\bigr]^{1 - \alpha} \mulspace f(x_1,y_1)\label{eq:jalphaadditiveb}\\
&\ge \sum_{x_1,y_1} P_{X_1 Y_1}(x_1,y_1)^\alpha \bigl[\hat{Q}_{X_1}(x_1) \mulspace \hat{Q}_{Y_1}(y_1)\bigr]^{1 - \alpha} \mulspace f(x_1',y_1')\label{eq:jalphaadditivec}\\
&= 2^{(\alpha - 1) D_\alpha(P_{X_1 Y_1}\|\hat{Q}_{X_1} \hat{Q}_{Y_1}) + (\alpha - 1) D_\alpha(P_{X_2 Y_2}\|Q_{X_2}' Q_{Y_2}')},\label{eq:jalphaadditived}
\end{align}
where \eqref{eq:jalphaadditiveaa} follows from \eqref{eq:jalphaadditivedalpha};
\eqref{eq:jalphaadditivea} holds by the independence hypothesis $P_{X_1 X_2 Y_1 Y_2} = P_{X_1 Y_1} P_{X_2 Y_2}$;
\eqref{eq:jalphaadditiveb} follows from \eqref{eq:jalphaadditivedeff};
\eqref{eq:jalphaadditivec} follows from \eqref{eq:jalphaadditivexoneprimeyoneprime};
and \eqref{eq:jalphaadditived} follows from \eqref{eq:jalphaadditiveqxtwoprime} and \eqref{eq:jalphaadditiveqytwoprime}.
Taking the logarithm and multiplying by $\frac{1}{\alpha - 1} > 0$ establishes \eqref{eq:jalphaadditivesecondpart}:
\begin{align}
J_\alpha(X_1,X_2;Y_1,Y_2) &\ge D_\alpha(P_{X_1 Y_1}\|\hat{Q}_{X_1} \hat{Q}_{Y_1}) + D_\alpha(P_{X_2 Y_2}\|Q_{X_2}' Q_{Y_2}')\label{eq:jalphaadditiveg}\\
&\ge J_\alpha(X_1;Y_1) + J_\alpha(X_2;Y_2),\label{eq:jalphaadditiveh}
\end{align}
where \eqref{eq:jalphaadditiveh} holds by the definition of $J_\alpha(X_1;Y_1)$ and $J_\alpha(X_2;Y_2)$.

The proof of \eqref{eq:jalphaadditivesecondpart} for $\alpha \in (0,1)$ is essentially the same as for $\alpha > 1$:
Replace the minimum in \eqref{eq:jalphaadditivexoneprimeyoneprime}~by a maximum.
Inequality \eqref{eq:jalphaadditivec} is then reversed, but \eqref{eq:jalphaadditiveg} continues to hold because $\frac{1}{\alpha - 1} < 0$.
Inequality \eqref{eq:jalphaadditiveh} also continues to hold, and \eqref{eq:jalphaadditiveg} and \eqref{eq:jalphaadditiveh} together imply \eqref{eq:jalphaadditivesecondpart}.
\end{proof}

\begin{Lemma}\label{lma:jalphacardinalityupperbound}
For all $\alpha \in [0,\infty]$, $J_\alpha(X;Y) \le \logcard{\set{X}}$ with equality if and only if $\bigl(\alpha \in [\frac{1}{2},\infty]$, $X$ is distributed uniformly over $\set{X}$, and $H(X|Y) = 0\bigr)$.
\end{Lemma}

\begin{proof}
Throughout the proof, define $X' \triangleq X$.
We first show that $J_\alpha(X;Y) \le \logcard{\set{X}}$ for all $\alpha \in [0,\infty]$:
\begin{align}
J_\alpha(X;Y) &\le J_\alpha(X;X')\label{eq:jalphacardinalitybounda}\\*
&\le J_\infty(X;X')\label{eq:jalphacardinalityboundb}\\
&= H_\frac{1}{2}(X)\label{eq:jalphacardinalityboundc}\\
&\le \logcard{\set{X}},\label{eq:jalphacardinalityboundd}
\end{align}
where \eqref{eq:jalphacardinalitybounda} follows from the data-processing inequality (\lmaref{lma:jalphadpi}) because $X \markov X' \markov Y$ form a Markov chain;
\eqref{eq:jalphacardinalityboundb} holds because $J_\alpha(X;X')$ is nondecreasing in $\alpha$ (\lmaref{lma:jalphanondecreasing});
\eqref{eq:jalphacardinalityboundc} follows from \lmaref{lma:jalphaselfinformation}; and
\eqref{eq:jalphacardinalityboundd} follows from \propref{prop:renyientropyproperties}.

We now show that \eqref{eq:jalphacardinalitybounda}--\eqref{eq:jalphacardinalityboundd} can hold with equality only if the following conditions all hold:
\begin{enumerate}[label=(\roman*)]
\item
$\alpha \in [\frac{1}{2},\infty]$;
\item
$X$ is distributed uniformly over $\set{X}$; and
\item
$H(X|Y) = 0$, i.e., for every $y \in \supp(P_Y)$, there exists an $x \in \set{X}$ for which $P(x|y) = 1$.
\end{enumerate}
Indeed, if $\alpha < \frac{1}{2}$, then \lmaref{lma:jalphaselfinformation} implies that
\begin{align}
J_\alpha(X;X') = \frac{\alpha}{1 - \alpha} \mulspace H_\infty(X).\label{eq:jalphacardboundnec}
\end{align}
Because $\frac{\alpha}{1 - \alpha} < 1$ for such $\alpha$'s and because $H_\infty(X) \le \logcard{\set{X}}$ (\propref{prop:renyientropyproperties}), the RHS of \eqref{eq:jalphacardboundnec} is strictly smaller than $\logcard{\set{X}}$.
This, together with \eqref{eq:jalphacardinalitybounda}, shows that Part~(i) is a necessary condition.
The~necessity of Part~(ii) follows from \eqref{eq:jalphacardinalityboundd}: if $X$ is not distributed uniformly over $\set{X}$, then \eqref{eq:jalphacardinalityboundd} holds with strict inequality (\propref{prop:renyientropyproperties}).
As to the necessity of Part~(iii),
\begin{align}
J_\alpha(X;Y) &\le J_\infty(X;Y)\label{eq:jalphacardinalityboundh}\\
&= \min_{Q_X} \min_{Q_Y} D_\infty(P_{XY}\|Q_X Q_Y)\label{eq:jalphacardinalityboundi}\\
&= \min_{Q_X} \log \sum_y \max_x \frac{P(x,y)}{Q_X(x)}\label{eq:jalphacardinalityboundj}\\
&\le \log \sum_y \max_x \frac{P(y) \mulspace P(x|y)}{1 / \card{\set{X}}}\label{eq:jalphacardinalityboundk}\\
&= \logcard{\set{X}} + \log \sum_y P(y) \max_x P(x|y)\label{eq:jalphacardinalityboundl}\\
&\le \logcard{\set{X}},\label{eq:jalphacardinalityboundm}
\end{align}
where \eqref{eq:jalphacardinalityboundh} holds because $J_\alpha(X;Y)$ is nondecreasing in $\alpha$ (\lmaref{lma:jalphanondecreasing});
\eqref{eq:jalphacardinalityboundj} follows from \propref{prop:infqydalphapxyqxqy}; and
\eqref{eq:jalphacardinalityboundk} follows from choosing $Q_X$ to be the uniform distribution.
The inequality \eqref{eq:jalphacardinalityboundm} is strict when Part~(iii) does not hold, so Part~(iii) is a necessary condition.

It remains to show that when Parts~(i)--(iii) all hold, $J_\alpha(X;Y) = \logcard{\set{X}}$.
By \eqref{eq:jalphacardinalityboundd}, $J_\alpha(X;Y) \le \logcard{\set{X}}$ always holds, so it suffices to show that Parts~(i)--(iii) together imply $J_\alpha(X;Y) \ge \logcard{\set{X}}$.
Indeed,
\begin{align}
J_\alpha(X;Y) &\ge J_\frac{1}{2}(X;Y)\label{eq:jalphacardinalityboundp}\\
&\ge J_\frac{1}{2}(X;X')\label{eq:jalphacardinalityboundq}\\
&= H_\infty(X)\label{eq:jalphacardinalityboundr}\\
&= \logcard{\set{X}},\label{eq:jalphacardinalitybounds}
\end{align}
where \eqref{eq:jalphacardinalityboundp} holds because Part~(i) implies that $\alpha \ge \frac{1}{2}$ and because $J_\alpha(X;Y)$ is nondecreasing in $\alpha$ (\lmaref{lma:jalphanondecreasing});
\eqref{eq:jalphacardinalityboundq} follows from the data-processing inequality (\lmaref{lma:jalphadpi}) because Part~(iii) implies that $X \markov Y \markov X'$ form a Markov chain;
\eqref{eq:jalphacardinalityboundr} follows from \lmaref{lma:jalphaselfinformation}; and
\eqref{eq:jalphacardinalitybounds} follows from Part~(ii).
\end{proof}

\begin{Lemma}\label{lma:jalphaconcavepx}
For every $\alpha \in [1,\infty]$, $J_\alpha(X;Y)$ is concave in $P_X$ for fixed $P_{Y|X}$.
\end{Lemma}

\begin{proof}
We prove the claim for $\alpha \in (1,\infty)$; for $\alpha \in \{1,\infty\}$ the claim will then hold because $J_\alpha(X;Y)$ is continuous in $\alpha$ (\lmaref{lma:jalphacontinuous}).

Fix $\alpha \in (1,\infty)$.
Let $\lambda,\lambda' \in [0,1]$ with $\lambda + \lambda' = 1$, let $P_X$ and $P_X'$ be PMFs, let $P_{Y|X}$ be a conditional PMF, and define $f\colon \set{X} \times \set{P}(\set{Y}) \to \reals \cup \{\infty\}$ as
\begin{align}
f(x,Q_Y) \triangleq \mleft[\sum_y P_{Y|X}(y|x)^\alpha \mulspace Q_Y(y)^{1 - \alpha}\mright]^\frac{1}{\alpha}.
\end{align}
Denoting $J_\alpha(X;Y)$ by $J_\alpha(P_X P_{Y|X})$,
\begin{align}
&J_\alpha\bigl((\lambda \mulspace P_X + \lambda' \mulspace P_X') P_{Y|X}\bigr)\nonumber\\*
&\qquad = \min_{Q_Y} \min_{Q_X} D_\alpha\bigl((\lambda \mulspace P_X + \lambda' \mulspace P_X') P_{Y|X}\|Q_X Q_Y\bigr)\label{eq:jalphaconcavepxa}\\[-1ex]
&\qquad = \min_{Q_Y} \frac{\alpha}{\alpha - 1} \log \sum_x \mleft[\sum_y \vthinspace \mleft[\lambda \mulspace P_X(x) + \lambda' \mulspace P_X'(x)\mright]^\alpha \mulspace P_{Y|X}(y|x)^\alpha \mulspace Q_Y(y)^{1 - \alpha}\mright]^\frac{1}{\alpha}\label{eq:jalphaconcavepxb}\\
&\qquad = \min_{Q_Y} \frac{\alpha}{\alpha - 1} \log \sum_x \vthinspace \mleft[\lambda \mulspace P_X(x) + \lambda' \mulspace P_X'(x)\mright] \mleft[\sum_y P_{Y|X}(y|x)^\alpha \mulspace Q_Y(y)^{1 - \alpha}\mright]^\frac{1}{\alpha}\label{eq:jalphaconcavepxbis}\\
&\qquad = \min_{Q_Y} \frac{\alpha}{\alpha - 1} \log \mleft[\lambda \sum_x P_X(x) \mulspace f(x,Q_Y) + \lambda' \sum_x P_X'(x) \mulspace f(x,Q_Y)\mright]\label{eq:jalphaconcavepxc}\\
&\qquad \ge \min_{Q_Y} \frac{\alpha}{\alpha - 1} \mleft[\lambda \log \sum_x P_X(x) \mulspace f(x,Q_Y) + \lambda' \log \sum_x P_X'(x) \mulspace f(x,Q_Y)\mright]\label{eq:jalphaconcavepxd}\\
&\qquad \ge \lambda \min_{Q_Y} \frac{\alpha}{\alpha - 1} \log \sum_x P_X(x) \mulspace f(x,Q_Y) + \lambda' \min_{Q_Y} \frac{\alpha}{\alpha - 1} \log \sum_x P_X'(x) \mulspace f(x,Q_Y)\label{eq:jalphaconcavepxe}\\
&\qquad = \lambda \mulspace J_\alpha(P_X P_{Y|X}) + \lambda' \mulspace J_\alpha(P_X' P_{Y|X}),\label{eq:jalphaconcavepxf}
\end{align}
where \eqref{eq:jalphaconcavepxb} follows from \propref{prop:infqydalphapxyqxqy} with the roles of $Q_X$ and $Q_Y$ swapped;
\eqref{eq:jalphaconcavepxd} holds because $\log(\cdot)$ is concave;
\eqref{eq:jalphaconcavepxe} holds because optimizing $Q_Y$ separately cannot be worse than optimizing a common $Q_Y$; and
\eqref{eq:jalphaconcavepxf} can be established using steps similar to \eqref{eq:jalphaconcavepxa}--\eqref{eq:jalphaconcavepxbis}.
\end{proof}

\begin{Lemma}\label{lma:dalphapxyqxqyconvexgeonehalf}
For every $\alpha \in [\frac{1}{2},\infty]$, the mapping $(Q_X,Q_Y) \mapsto D_\alpha(P_{XY}\|Q_X Q_Y)$ is convex, i.e., for all $\lambda,\lambda' \in [0,1]$ with $\lambda + \lambda' = 1$, all $Q_X,Q_X' \in \set{P}(\set{X})$, and all $Q_Y,Q_Y' \in \set{P}(\set{Y})$,
\begin{align}
D_\alpha\bigl(P_{XY}\|(\lambda Q_X + \lambda' Q_X') (\lambda Q_Y + \lambda' Q_Y')\bigr) \le \lambda \mulspace D_\alpha(P_{XY}\|Q_X Q_Y) + \lambda' \mulspace D_\alpha(P_{XY}\|Q_X' Q_Y').\label{eq:dalphapxyqxqyconvex}
\end{align}
For $\alpha \in [0,\frac{1}{2})$, the mapping need not be convex.
\end{Lemma}

\begin{proof}
We establish \eqref{eq:dalphapxyqxqyconvex} for $\alpha \in [\frac{1}{2},1)$ and for $\alpha \in (1,\infty)$,
which also establishes \eqref{eq:dalphapxyqxqyconvex} for $\alpha \in \{1,\infty\}$ because the R\'enyi divergence is continuous in $\alpha$ (\propref{prop:renyidivproperties}).
Afterwards, we provide an example where \eqref{eq:dalphapxyqxqyconvex} is violated for all $\alpha \in [0,\frac{1}{2})$.

We begin with the case where $\alpha \in [\frac{1}{2},1)$:
\begin{align}
&2^{(\alpha - 1) \lambda D_\alpha(P_{XY}\|Q_X Q_Y) + (\alpha - 1) \lambda' D_\alpha(P_{XY}\|Q_X' Q_Y')}\nonumber\\*[-0.5ex]
&\qquad = \mleft[\sum_{x,y} P(x,y)^\alpha \mulspace [Q_X(x) \mulspace Q_Y(y)]^{1 - \alpha}\mright]^\lambda \cdot \mleft[\sum_{x,y} P(x,y)^\alpha \mulspace \mleft[Q_X'(x) \mulspace Q_Y'(y)\mright]^{1 - \alpha}\mright]^{\lambda'}\\
&\qquad \le \lambda \sum_{x,y} P(x,y)^\alpha \mulspace [Q_X(x) \mulspace Q_Y(y)]^{1 - \alpha} + \lambda' \sum_{x,y} P(x,y)^\alpha \mulspace \mleft[Q_X'(x) \mulspace Q_Y'(y)\mright]^{1 - \alpha}\label{eq:dalphapxyqxqyconvexsmalla}\\
&\qquad = \sum_{x,y} P(x,y)^\alpha \mleft[\sqrt{\lambda} \mulspace Q_X(x)^{1 - \alpha} \mulspace \sqrt{\lambda} \mulspace Q_Y(y)^{1 - \alpha} + \sqrt{\lambda'} \mulspace Q_X'(x)^{1 - \alpha} \mulspace \sqrt{\lambda'} \mulspace Q_Y'(y)^{1 - \alpha}\mright]\label{eq:dalphapxyqxqyconvexsmallb}\\
&\qquad \le \sum_{x,y} P(x,y)^\alpha \sqrt{\lambda \mulspace Q_X(x)^{2(1 - \alpha)} + \lambda' \mulspace Q_X'(x)^{2(1 - \alpha)}} \mulspace \sqrt{\lambda \mulspace Q_Y(y)^{2(1 - \alpha)} + \lambda' \mulspace Q_Y'(y)^{2(1 - \alpha)}}\label{eq:dalphapxyqxqyconvexsmallc}\\
&\qquad \le \sum_{x,y} P(x,y)^\alpha \mleft[\lambda \mulspace Q_X(x) + \lambda' \mulspace Q_X'(x)\mright]^{1 - \alpha} \mulspace \sqrt{\lambda \mulspace Q_Y(y)^{2(1 - \alpha)} + \lambda' \mulspace Q_Y'(y)^{2(1 - \alpha)}}\label{eq:dalphapxyqxqyconvexsmalld}\\
&\qquad \le \sum_{x,y} P(x,y)^\alpha \mleft[\lambda \mulspace Q_X(x) + \lambda' \mulspace Q_X'(x)\mright]^{1 - \alpha} \mulspace \mleft[\lambda \mulspace Q_Y(y) + \lambda' \mulspace Q_Y'(y)\mright]^{1 - \alpha}\label{eq:dalphapxyqxqyconvexsmalle}\\
&\qquad = 2^{(\alpha - 1) D_\alpha(P_{XY}\|(\lambda Q_X + \lambda' Q_X') (\lambda Q_Y + \lambda' Q_Y'))},
\end{align}
where \eqref{eq:dalphapxyqxqyconvexsmalla} follows from the arithmetic mean-geometric mean inequality;
\eqref{eq:dalphapxyqxqyconvexsmallc} follows from the Cauchy--Schwarz inequality; and
\eqref{eq:dalphapxyqxqyconvexsmalld} and \eqref{eq:dalphapxyqxqyconvexsmalle} hold because the mapping $z \mapsto z^{2 (1 - \alpha)}$ is concave on $\reals_{\ge 0}$ for $\alpha \in [\frac{1}{2},1)$.
Taking the logarithm and multiplying by $\frac{1}{\alpha - 1} < 0$ establishes \eqref{eq:dalphapxyqxqyconvex}.

Now, consider $\alpha \in (1,\infty)$.
Then,
\begin{align}
&2^{(\alpha - 1) D_\alpha(P_{XY}\|(\lambda Q_X + \lambda' Q_X') (\lambda Q_Y + \lambda' Q_Y'))}\nonumber\\*[-0.5ex]
&\qquad = \sum_{x,y} P(x,y)^\alpha \mleft[\lambda \mulspace Q_X(x) + \lambda' \mulspace Q_X'(x)\mright]^{1 - \alpha} \mleft[\lambda \mulspace Q_Y(y) + \lambda' \mulspace Q_Y'(y)\mright]^{1 - \alpha}\\
&\qquad \le \sum_{x,y} P(x,y)^\alpha \mleft[Q_X(x)^\lambda \mulspace Q_X'(x)^{\lambda'}\mright]^{1 - \alpha} \mleft[Q_Y(y)^\lambda \mulspace Q_Y'(y)^{\lambda'}\mright]^{1 - \alpha}\label{eq:dalphapxyqxqyconvexbiga}\\
&\qquad = \sum_{x,y} P(x,y)^\alpha \bigl[Q_X(x) \mulspace Q_Y(y)\bigr]^{(1 - \alpha) \lambda} \bigl[Q_X'(x) \mulspace Q_Y'(y)\bigr]^{(1 - \alpha) \lambda'}\\[-0.5ex]
&\qquad \le \mleft[\sum_{x,y} P(x,y)^\alpha \bigl[Q_X(x) \mulspace Q_Y(y)\bigr]^{1 - \alpha}\mright]^\lambda \cdot \mleft[\sum_{x,y} P(x,y)^\alpha \bigl[Q_X'(x) \mulspace Q_Y'(y)\bigr]^{1 - \alpha}\mright]^{\lambda'}\label{eq:dalphapxyqxqyconvexbigc}\\
&\qquad = 2^{(\alpha - 1) \lambda D_\alpha(P_{XY}\|Q_X Q_Y) + (\alpha - 1) \lambda' D_\alpha(P_{XY}\|Q_X' Q_Y')},
\end{align}
where \eqref{eq:dalphapxyqxqyconvexbiga} follows from the arithmetic mean-geometric mean inequality and the fact that the mapping $z \mapsto z^{1 - \alpha}$ is decreasing on $\reals_{>0}$ for $\alpha > 1$, and
\eqref{eq:dalphapxyqxqyconvexbigc} follows from H\"older's inequality.
Taking the logarithm and multiplying by $\frac{1}{\alpha - 1} > 0$ establishes \eqref{eq:dalphapxyqxqyconvex}.

Finally, we show that the mapping $(Q_X,Q_Y) \mapsto D_\alpha(P_{XY}\|Q_X Q_Y)$ does not need to be convex for $\alpha \in [0,\frac{1}{2})$.
Let $X$ be uniformly distributed over $\{0,1\}$, and let $Y = X$.
Then, for all $\alpha \in [0,\frac{1}{2})$,
\begin{align}
D_\alpha\bigl(P_{XY}\|(0.5,0.5) (0.5,0.5)\bigr) > 0.5 \mulspace D_\alpha\bigl(P_{XY}\|(1,0) (1,0)\bigr) + 0.5 \mulspace D_\alpha\bigl(P_{XY}\|(0,1) (0,1)\bigr),\label{eq:dalphapxyqxqynotconvex}
\end{align}
because the LHS of \eqref{eq:dalphapxyqxqynotconvex} is equal to $\log 2$, and the RHS of \eqref{eq:dalphapxyqxqynotconvex} is equal to $\frac{\alpha}{1 - \alpha} \log 2$.
\end{proof}

\begin{Lemma}\label{lma:dalphapxyqxqyqxqyconsistency}
Let $\alpha \in (0,1) \cup (1,\infty)$.
If $(Q_X^*,Q_Y^*)$ achieves the minimum in the definition of $J_\alpha(X;Y)$, then~there exist positive normalization constants $c$ and $d$ such that
\begin{align}
Q_X^*(x) &= c \mleft[\sum_y P(x,y)^\alpha \mulspace Q_Y^*(y)^{1 - \alpha}\mright]^\frac{1}{\alpha} \quad \forall \vthinspace x \in \set{X},\label{eq:dalphapxyqxstarforqystar}\\*
Q_Y^*(y) &= d \mleft[\sum_x P(x,y)^\alpha \mulspace Q_X^*(x)^{1 - \alpha}\mright]^\frac{1}{\alpha} \quad \forall \vthinspace y \in \set{Y},\label{eq:dalphapxyqystarforqxstar}
\end{align}
with the conventions of \eqref{eq:zeroinftyconventions}.
The case $\alpha = \infty$ is similar: if $(Q_X^*,Q_Y^*)$ achieves the minimum in the definition of $J_\infty(X;Y)$, then there exist positive normalization constants $c$ and $d$ such that
\begin{align}
Q_X^*(x) &= c \max_y \frac{P(x,y)}{Q_Y^*(y)} \quad \forall \vthinspace x \in \set{X},\label{eq:dinftypxyqxstarforqystar}\\*
Q_Y^*(y) &= d \max_x \frac{P(x,y)}{Q_X^*(x)} \quad \forall \vthinspace y \in \set{Y},\label{eq:dinftypxyqystarforqxstar}
\end{align}
with the conventions of \eqref{eq:zeroinftyconventions}.
(If $\alpha = 1$, then $Q_X^* = P_X$ and $Q_Y^* = P_Y$ by \propref{prop:minqxqydpxyqxqy}.)
Thus, for all $\alpha \in (0,\infty]$, both inclusions $\supp(Q_X^*) \subseteq \supp(P_X)$ and $\supp(Q_Y^*) \subseteq \supp(P_Y)$ hold.
\end{Lemma}

\begin{proof}
If $(Q_X^*,Q_Y^*)$ achieves the minimum in the definition of $J_\alpha(X;Y)$, then
\begin{align}
\min_{Q_Y} D_\alpha(P_{XY}\|Q_X^* Q_Y) = D_\alpha(P_{XY}\|Q_X^* Q_Y^*).
\end{align}
Hence, \eqref{eq:dalphapxyqystarforqxstar} and \eqref{eq:dinftypxyqystarforqxstar} follow from \eqref{eq:infqydalphapxyqxqyminimizer} and \eqref{eq:infqydinftypxyqxqyminimizer} of \propref{prop:infqydalphapxyqxqy} because $D_\alpha(P_{XY}\|Q_X^* Q_Y^*) = J_\alpha(X;Y)$ is finite.
Swapping the roles of $Q_X$ and $Q_Y$ establishes \eqref{eq:dalphapxyqxstarforqystar} and \eqref{eq:dinftypxyqxstarforqystar}.
For $\alpha \in (0,1) \cup (1,\infty)$ the claimed inclusions follow from \eqref{eq:dalphapxyqxstarforqystar} and \eqref{eq:dalphapxyqystarforqxstar}; for $\alpha = \infty$ from \eqref{eq:dinftypxyqxstarforqystar} and \eqref{eq:dinftypxyqystarforqxstar}; and for $\alpha = 1$ from \propref{prop:minqxqydpxyqxqy}.
\end{proof}

\begin{Lemma}\label{lma:jalphaminimizationqx}
For all $\alpha \in (0,\infty]$,
\begin{align}
J_\alpha(X;Y) = \min_{Q_X} \phi_\alpha(Q_X),\label{eq:jalphaminimizationqx}
\end{align}
where $\phi_\alpha(Q_X)$ is defined as
\begin{align}
\phi_\alpha(Q_X) \triangleq \min_{Q_Y} D_\alpha(P_{XY}\|Q_X Q_Y)\label{eq:jalphaminimizationqxdefphi}
\end{align}
and is given explicitly as follows: for $\alpha \in (0,1) \cup (1,\infty)$,
\begin{align}
\phi_\alpha(Q_X) = \frac{\alpha}{\alpha - 1} \log \sum_y \mleft[\sum_x P(x,y)^\alpha \mulspace Q_X(x)^{1 - \alpha}\mright]^\frac{1}{\alpha},\label{eq:jalphaminimizationqxphi}
\end{align}
with the conventions of \eqref{eq:zeroinftyconventions}; and for $\alpha \in \{1,\infty\}$,
\begin{align}
\phi_1(Q_X) &= D(P_{XY}\|Q_X P_Y),\label{eq:joneminimizationqxphi}\\
\phi_\infty(Q_X) &= \log \sum_y \max_x \frac{P(x,y)}{Q_X(x)},\label{eq:jinftyminimizationqxphi}
\end{align}
with the conventions of \eqref{eq:zeroinftyconventions}.
For every $\alpha \in [\frac{1}{2},\infty]$, the mapping $Q_X \mapsto \phi_\alpha(Q_X)$ is convex.
For $\alpha \in (0,\frac{1}{2})$, the~mapping need not be convex.
\end{Lemma}

\begin{proof}
We first establish \eqref{eq:jalphaminimizationqx} and \eqref{eq:jalphaminimizationqxphi}--\eqref{eq:jinftyminimizationqxphi}:
\eqref{eq:jalphaminimizationqx} follows from the definition of $J_\alpha(X;Y)$;
\eqref{eq:jalphaminimizationqxphi} and~\eqref{eq:jinftyminimizationqxphi} follow from \propref{prop:infqydalphapxyqxqy}; and
\eqref{eq:joneminimizationqxphi} holds because
\begin{align}
\min_{Q_Y} D(P_{XY}\|Q_X Q_Y) &= \min_{Q_Y} \vthinspace \mleft[D(P_{XY}\|Q_X P_Y) + D(P_Y\|Q_Y)\mright]\label{eq:joneminia}\\*
&= D(P_{XY}\|Q_X P_Y),\label{eq:joneminib}
\end{align}
where \eqref{eq:joneminia} follows from a simple computation, and \eqref{eq:joneminib} holds because $D(P_Y\|Q_Y) \ge 0$ with equality if $Q_Y = P_Y$.

We now show that the mapping $Q_X \mapsto \phi_\alpha(Q_X)$ is convex for every $\alpha \in [\frac{1}{2},\infty]$.
To that end, let~$\alpha \in [\frac{1}{2},\infty]$, let $\lambda,\lambda' \in [0,1]$ with $\lambda + \lambda' = 1$, and let $Q_X,Q_X' \in \set{P}(\set{X})$.
Let $\hat{Q}_Y$ and $\hat{Q}_Y'$ be PMFs that achieve the minimum in the definitions of $\phi_\alpha(Q_X)$ and $\phi_\alpha(Q_X')$, respectively.
Then,
\begin{align}
\phi_\alpha(\lambda \mulspace Q_X + \lambda' \mulspace Q_X') &\le D_\alpha\bigl(P_{XY}\|(\lambda Q_X + \lambda' Q_X') (\lambda \hat{Q}_Y + \lambda' \hat{Q}_Y')\bigr)\label{eq:phialphaconvexd}\\
&\le \lambda \mulspace D_\alpha(P_{XY}\|Q_X \hat{Q}_Y) + \lambda' \mulspace D_\alpha(P_{XY}\|Q_X' \hat{Q}_Y')\label{eq:phialphaconvexe}\\
&= \lambda \mulspace \phi_\alpha(Q_X) + \lambda' \mulspace \phi_\alpha(Q_X'),\label{eq:phialphaconvexf}
\end{align}
where \eqref{eq:phialphaconvexd} holds by the definition of $\phi_\alpha(\cdot)$;
\eqref{eq:phialphaconvexe} holds because $D_\alpha(P_{XY}\|Q_X Q_Y)$ is convex in the pair $(Q_X,Q_Y)$ for $\alpha \in [\frac{1}{2},\infty]$ (\lmaref{lma:dalphapxyqxqyconvexgeonehalf}); and
\eqref{eq:phialphaconvexf} follows from our choice of $\hat{Q}_Y$ and $\hat{Q}_Y'$.

Finally, we show that the mapping $Q_X \mapsto \phi_\alpha(Q_X)$ need not be convex for $\alpha \in (0,\frac{1}{2})$.
Let $X$ be uniformly distributed over $\{0,1\}$, and let $Y = X$.
Then, for all $\alpha \in (0,\frac{1}{2})$,
\begin{align}
\phi_\alpha\bigl((0.5,0.5)\bigr) > 0.5 \mulspace \phi_\alpha\bigl((1,0)\bigr) + 0.5 \mulspace \phi_\alpha\bigl((0,1)\bigr),\label{eq:phialphanotconvex}
\end{align}
because the LHS of \eqref{eq:phialphanotconvex} is equal to $\log 2$, and the RHS of \eqref{eq:phialphanotconvex} is equal to $\frac{\alpha}{1 - \alpha} \log 2$.
\end{proof}

\begin{Lemma}\label{lma:jalpharxypsi}
For all $\alpha \in (0,1) \cup (1,\infty]$,
\begin{align}
J_\alpha(X;Y) = \begin{cases}\displaystyle \min_{R_{XY} \in \set{P}(\set{X} \times \set{Y})} \psi_\alpha(R_{XY}) & \text{if $\alpha \in (0,1)$,}\\
\displaystyle \max_{R_{XY} \in \set{P}(\set{X} \times \set{Y})} \psi_\alpha(R_{XY}) & \text{if $\alpha \in (1,\infty]$,}\end{cases}\label{eq:jalpharxypsi}
\end{align}
where
\begin{align}
\psi_\alpha(R_{XY}) \triangleq \begin{cases}\displaystyle D(R_{XY}\|R_X R_Y) + \frac{\alpha}{1 - \alpha} D(R_{XY}\|P_{XY}) & \text{if $\alpha \in (0,1) \cup (1,\infty)$,}\\[1ex]
\displaystyle D(R_{XY}\|R_X R_Y) - D(R_{XY}\|P_{XY}) & \text{if $\alpha = \infty$.}\end{cases}\label{eq:jalpharxypsidefpsi}
\end{align}
For every $\alpha \in (1,\infty]$, the mapping $R_{XY} \mapsto \psi_\alpha(R_{XY})$ is concave.
For all $\alpha \in (1,\infty]$ and all $R_{XY} \in \set{P}(\set{X} \times \set{Y})$, the statement $J_\alpha(X;Y) = \psi_\alpha(R_{XY})$ is equivalent to $\psi_\alpha(R_{XY}) = D_\alpha(P_{XY}\|R_X R_Y)$.
\end{Lemma}

\begin{proof}
For $\alpha \in (0,1) \cup (1,\infty)$, \eqref{eq:jalpharxypsi} follows from \lmaref{lma:oneminusalphajalpharxy} by dividing by $1 - \alpha$, which is positive or negative depending on whether $\alpha$ is smaller than or greater than one.
For $\alpha = \infty$, we establish \eqref{eq:jalpharxypsi} as follows:
By \lmaref{lma:jalphacontinuous}, its LHS is continuous at $\alpha = \infty$.
We argue below that its RHS is continuous at $\alpha = \infty$, i.e., that
\begin{align}
\lim_{\alpha \to \infty} \max_{R_{XY}} \psi_\alpha(R_{XY}) = \max_{R_{XY}} \psi_\infty(R_{XY}).\label{eq:limalphainftymaxpsi}
\end{align}
Because \eqref{eq:jalpharxypsi} holds for $\alpha \in (1,\infty)$ and because both its sides are continuous at $\alpha = \infty$, it must also hold for $\alpha = \infty$.

We now establish \eqref{eq:limalphainftymaxpsi}.
Let $R_{XY}^*$ be a PMF that achieves the maximum on the RHS of \eqref{eq:limalphainftymaxpsi}.
Then, for all $\alpha > 1$,
\begin{align}
\psi_\infty(R_{XY}^*) &= \max_{R_{XY}} \psi_\infty(R_{XY})\\
&\ge \max_{R_{XY}} \psi_\alpha(R_{XY})\label{eq:psiinftylimb}\\
&\ge \psi_\alpha(R_{XY}^*),\label{eq:psiinftylimc}
\end{align}
where \eqref{eq:psiinftylimb} holds because, by \eqref{eq:jalpharxypsidefpsi}, $\psi_\infty(R_{XY}) = \psi_\alpha(R_{XY}) + \frac{1}{\alpha - 1} \mulspace D(R_{XY}\|P_{XY}) \ge \psi_\alpha(R_{XY})$ for all $R_{XY} \in \set{P}(\set{X} \times \set{Y})$.
By \eqref{eq:jalpharxypsidefpsi}, $\alpha \mapsto \psi_\alpha(R_{XY}^*)$ is continuous at $\alpha = \infty$, so the RHS of \eqref{eq:psiinftylimc} approaches $\psi_\infty(R_{XY}^*)$ as $\alpha$ tends to infinity, and \eqref{eq:limalphainftymaxpsi} follows from the sandwich theorem.

We now show that $R_{XY} \mapsto \psi_\alpha(R_{XY})$ is concave for $\alpha \in (1,\infty]$.
A simple computation reveals that for all $\alpha \in (1,\infty)$,
\begin{align}
\psi_\alpha(R_{XY}) = H(R_X) + H(R_Y) + \frac{1}{\alpha - 1} \mulspace H(R_{XY}) + \frac{\alpha}{\alpha - 1} \sum_{x,y} R_{XY}(x,y) \log P(x,y).\label{eq:psialpharxyconcavity}
\end{align}
Because the entropy is a concave function and because a nonnegative weighted sum of concave functions is concave, this implies that $\psi_\alpha(R_{XY})$ is concave in $R_{XY}$ for $\alpha \in (1,\infty)$.
By \eqref{eq:jalpharxypsidefpsi}, $\alpha \mapsto \psi_\alpha(R_{XY})$ is continuous at $\alpha = \infty$, so $\psi_\alpha(R_{XY})$ is concave in $R_{XY}$ also for $\alpha = \infty$.

We next show that if $\alpha \in (1,\infty]$ and $\psi_\alpha(R_{XY}) = D_\alpha(P_{XY}\|R_X R_Y)$, then $J_\alpha(X;Y) = \psi_\alpha(R_{XY})$.
Let $\alpha \in (1,\infty]$, and let $R_{XY}$ be a PMF that satisfies $\psi_\alpha(R_{XY}) = D_\alpha(P_{XY}\|R_X R_Y)$.
Then,
\begin{align}
\psi_\alpha(R_{XY}) &\le J_\alpha(X;Y)\label{eq:psialphaeqja}\\
&\le D_\alpha(P_{XY}\|R_X R_Y),\label{eq:psialphaeqjb}
\end{align}
where \eqref{eq:psialphaeqja} follows from \eqref{eq:jalpharxypsi}, and \eqref{eq:psialphaeqjb} holds by the definition of $J_\alpha(X;Y)$.
Because $\psi_\alpha(R_{XY})$ is equal to $D_\alpha(P_{XY}\|R_X R_Y)$, both inequalities hold with equality, which implies the claim.

Finally, we show that if $\alpha \in (1,\infty]$ and $J_\alpha(X;Y) = \psi_\alpha(R_{XY})$, then $\psi_\alpha(R_{XY}) = D_\alpha(P_{XY}\|R_X R_Y)$.
We first consider $\alpha \in (1,\infty)$.
Let $R_{XY}$ be a PMF that satisfies $J_\alpha(X;Y) = \psi_\alpha(R_{XY})$, and let $Q_X^*$ and $Q_Y^*$ be PMFs that achieve the minimum in the definition of $J_\alpha(X;Y)$.
Then,
\begin{align}
J_\alpha(X;Y) &= \psi_\alpha(R_{XY})\label{eq:psialpharxyeqdalphaa}\\
&= D(R_{XY}\|R_X R_Y) + \frac{\alpha}{1 - \alpha} \mulspace D(R_{XY}\|P_{XY})\\
&\le D(R_{XY}\|Q_X^* Q_Y^*) + \frac{\alpha}{1 - \alpha} \mulspace D(R_{XY}\|P_{XY})\label{eq:psialpharxyeqdalphab}\\
&\le D_\alpha(P_{XY}\|Q_X^* Q_Y^*)\label{eq:psialpharxyeqdalphad}\\
&= J_\alpha(X;Y),\label{eq:psialpharxyeqdalphaz}
\end{align}
where \eqref{eq:psialpharxyeqdalphab} follows from \propref{prop:minqxqydpxyqxqy}, and \eqref{eq:psialpharxyeqdalphad} follows from \mdpicite[Theorem~30]{VanErvenHarremoes}.
Thus, all inequalities hold with equality.
Because \eqref{eq:psialpharxyeqdalphab} holds with equality, $Q_X^* = R_X$ and $Q_Y^* = R_Y$ by \propref{prop:minqxqydpxyqxqy}.
Hence, $\psi_\alpha(R_{XY}) = D_\alpha(P_{XY}\|Q_X^* Q_Y^*) = D_\alpha(P_{XY}\|R_X R_Y)$ as desired.
We now consider $\alpha = \infty$.
Here, \eqref{eq:psialpharxyeqdalphaa}--\eqref{eq:psialpharxyeqdalphaz} remain valid after replacing $\frac{\alpha}{1 - \alpha}$ by $-1$. (Now, \eqref{eq:psialpharxyeqdalphad} follows from a short computation.)
Consequently, $\psi_\alpha(R_{XY}) = D_\alpha(P_{XY}\|R_X R_Y)$ holds also for $\alpha = \infty$.
\end{proof}

\begin{Lemma}\label{lma:jalpharxezero}
For all $\alpha \in (0,1) \cup (1,\infty)$,
\begin{align}
J_\alpha(X;Y) = \min_{R_X \ll P_X} \frac{1}{\alpha - 1} \Bigl[D_\frac{\alpha}{\alpha - 1}(P_X\|R_X) - \alpha \mulspace E_0\bigl(\tfrac{1 - \alpha}{\alpha},R_X\bigr)\Bigr],\label{eq:jalpharxezero}
\end{align}
where the minimization is over all PMFs $R_X$ satisfying $R_X \ll P_X$ $\bigl(\text{i.e., $\supp(R_X) \subseteq \supp(P_X)$}\bigr)$;
$D_\alpha(P\|Q)$ for negative $\alpha$ is given by \eqref{eq:defrenyidivnegative};
and Gallager's $E_0$ function \cite{Gallager} is defined as
\begin{align}
E_0(\rho,R_X) \triangleq -\log \sum_y \mleft[\sum_x R_X(x) \mulspace P(y|x)^\frac{1}{1 + \rho}\mright]^{1 + \rho}.
\end{align}
\end{Lemma}

\begin{proof}
Let $\alpha \in (0,1) \cup (1,\infty)$, and define the set $\set{R} \triangleq \{R_X \in \set{P}(\set{X}) : \supp(R_X) \subseteq \supp(P_X)\}$.
We establish \eqref{eq:jalpharxezero} by showing that for all $R_X \in \set{R}$,
\begin{align}
\frac{1}{\alpha - 1} \Bigl[D_\frac{\alpha}{\alpha - 1}(P_X\|R_X) - \alpha \mulspace E_0\bigl(\tfrac{1 - \alpha}{\alpha},R_X\bigr)\Bigr] \ge J_\alpha(X;Y),\label{eq:jalpharxezeroa}
\end{align}
with equality for some $R_X \in \set{R}$.

Fix $R_X \in \set{R}$.
If the LHS of \eqref{eq:jalpharxezeroa} is infinite, then \eqref{eq:jalpharxezeroa} holds trivially.
Otherwise, define the PMF $\hat{Q}_X$ as
\begin{align}
\hat{Q}_X(x) \triangleq \frac{P_X(x)^\frac{\alpha}{\alpha - 1} R_X(x)^\frac{1}{1 - \alpha}}{\sum_{x' \in \set{X}} P_X(x')^\frac{\alpha}{\alpha - 1} R_X(x')^\frac{1}{1 - \alpha}},\label{eq:jalpharxezerod}
\end{align}
where we use the convention that $0^\frac{\alpha}{\alpha - 1} \cdot 0^\frac{1}{1 - \alpha} = 0$.
(The RHS of \eqref{eq:jalpharxezerod} is finite whenever the LHS of~\eqref{eq:jalpharxezeroa} is finite.)
Then, \eqref{eq:jalpharxezeroa} holds because
\begin{align}
J_\alpha(X;Y) &= \min_{Q_X} \frac{\alpha}{\alpha - 1} \log \sum_y \mleft[\sum_x P(x,y)^\alpha \mulspace Q_X(x)^{1 - \alpha}\mright]^\frac{1}{\alpha}\label{eq:jalpharxezerog}\\
&\le \frac{\alpha}{\alpha - 1} \log \sum_y \mleft[\sum_x P(x,y)^\alpha \mulspace \hat{Q}_X(x)^{1 - \alpha}\mright]^\frac{1}{\alpha}\label{eq:jalpharxezeroh}\\
&= \log \sum_x P_X(x)^\frac{\alpha}{\alpha - 1} R_X(x)^\frac{1}{1 - \alpha} + \frac{\alpha}{\alpha - 1} \log \sum_y \mleft[\sum_x R_X(x) \mulspace P(y|x)^\alpha\mright]^\frac{1}{\alpha}\label{eq:jalpharxezeroi}\\
&= \frac{1}{\alpha - 1} \Bigl[D_\frac{\alpha}{\alpha - 1}(P_X\|R_X) - \alpha \mulspace E_0\bigl(\tfrac{1 - \alpha}{\alpha},R_X\bigr)\Bigr],
\end{align}
where \eqref{eq:jalpharxezerog} follows from \lmaref{lma:jalphaminimizationqx}, and
\eqref{eq:jalpharxezeroi} follows from \eqref{eq:jalpharxezerod} using some algebra.
It remains to show that there exists an $R_X \in \set{R}$ for which \eqref{eq:jalpharxezeroh} holds with equality.
To that end, let $Q_X^*$ be a PMF that achieves the minimum on the RHS of \eqref{eq:jalpharxezerog}, and define the PMF $R_X$ as
\begin{align}
R_X(x) \triangleq \frac{P_X(x)^\alpha \mulspace Q_X^*(x)^{1 - \alpha}}{\sum_{x' \in \set{X}} P_X(x')^\alpha \mulspace Q_X^*(x')^{1 - \alpha}},\label{eq:jalpharxezerom}
\end{align}
where we use the convention that $0^\alpha \cdot 0^{1 - \alpha} = 0$.
Because $\supp(Q_X^*) \subseteq \supp(P_X)$ (\lmaref{lma:dalphapxyqxqyqxqyconsistency}), the definitions \eqref{eq:jalpharxezerom} and \eqref{eq:jalpharxezerod} imply that $\hat{Q}_X = Q_X^*$.
Hence, \eqref{eq:jalpharxezeroh} holds with equality for this $R_X \in \set{R}$.
\end{proof}

\begin{Lemma}\label{lma:jalphauniqueminimizer}
For every $\alpha \in (\frac{1}{2},\infty]$, the mapping $(Q_X,Q_Y) \mapsto D_\alpha(P_{XY}\|Q_X Q_Y)$ has a unique minimizer.
This need not be the case when $\alpha \in [0,\frac{1}{2}]$.
\end{Lemma}

\begin{proof}
First consider $\alpha \in (\frac{1}{2},1)$.
Let $(Q_X^*,Q_Y^*)$ and $(\hat{Q}_X,\hat{Q}_Y)$ be pairs of PMFs that both minimize $(Q_X,Q_Y) \mapsto D_\alpha(P_{XY}\|Q_X Q_Y)$.
We establish uniqueness by arguing that $(Q_X^*,Q_Y^*)$ and $(\hat{Q}_X,\hat{Q}_Y)$ must be identical.
Observe that
\begin{align}
J_\alpha(X;Y) &\le D_\alpha\bigl(P_{XY}\|(0.5 \mulspace Q_X^* + 0.5 \mulspace \hat{Q}_X) (0.5 \mulspace Q_Y^* + 0.5 \mulspace \hat{Q}_Y)\bigr)\label{eq:jalphauniqueminsmalla}\\
&\le 0.5 \mulspace D_\alpha(P_{XY}\|Q_X^* Q_Y^*) + 0.5 \mulspace D_\alpha(P_{XY}\|\hat{Q}_X \hat{Q}_Y)\label{eq:jalphauniqueminsmallb}\\
&= J_\alpha(X;Y),\label{eq:jalphauniqueminsmallc}
\end{align}
where \eqref{eq:jalphauniqueminsmalla} holds by the definition of $J_\alpha(X;Y)$, and \eqref{eq:jalphauniqueminsmallb} follows from \lmaref{lma:dalphapxyqxqyconvexgeonehalf}.
Hence, \eqref{eq:jalphauniqueminsmallb} holds with equality, which implies that \eqref{eq:dalphapxyqxqyconvexsmalld} in the proof of \lmaref{lma:dalphapxyqxqyconvexgeonehalf} holds with equality, i.e.,
\begin{align}
&\sum_{x,y} P(x,y)^\alpha \sqrt{0.5 \mulspace Q_X^*(x)^{2(1 - \alpha)} + 0.5 \mulspace \hat{Q}_X(x)^{2(1 - \alpha)}} \mulspace \sqrt{0.5 \mulspace Q_Y^*(y)^{2(1 - \alpha)} + 0.5 \mulspace \hat{Q}_Y(y)^{2(1 - \alpha)}}\nonumber\\*
&\qquad = \sum_{x,y} P(x,y)^\alpha \mleft[0.5 \mulspace Q_X^*(x) + 0.5 \mulspace \hat{Q}_X(x)\mright]^{1 - \alpha} \mulspace \sqrt{0.5 \mulspace Q_Y^*(y)^{2(1 - \alpha)} + 0.5 \mulspace \hat{Q}_Y(y)^{2(1 - \alpha)}}.\label{eq:jalphauniqueminsmallf}
\end{align}
We first argue that $Q_X^* = \hat{Q}_X$.
Since $Q_X^*$ and $\hat{Q}_X$ are PMFs, it suffices to show that $Q_X^*(x) = \hat{Q}_X(x)$ for every $x \in \supp(\hat{Q}_X)$.
Let $\hat{x} \in \supp(\hat{Q}_X)$.
Because $\supp(\hat{Q}_X) \subseteq \supp(P_X)$ (\lmaref{lma:dalphapxyqxqyqxqyconsistency}), there exists a $\hat{y} \in \set{Y}$ such that $P(\hat{x},\hat{y}) > 0$.
Again by \lmaref{lma:dalphapxyqxqyqxqyconsistency}, this implies that $\hat{Q}_Y(\hat{y}) > 0$.
Because the mapping $z \mapsto z^{2 (1 - \alpha)}$ is strictly concave on $\reals_{\ge 0}$ for $\alpha \in (\frac{1}{2},1)$, it follows from \eqref{eq:jalphauniqueminsmallf} that $Q_X^*(\hat{x}) = \hat{Q}_X(\hat{x})$.
Swapping the roles of $Q_X$ and $Q_Y$, we obtain that $Q_Y^* = \hat{Q}_Y$.

For $\alpha = 1$, the minimizer is unique by \propref{prop:minqxqydpxyqxqy} because $D_1(P_{XY}\|Q_X Q_Y) = D(P_{XY}\|Q_X Q_Y)$.

Now consider $\alpha \in (1,\infty]$.
Here, we establish uniqueness via the characterization of $J_\alpha(X;Y)$ provided by \lmaref{lma:jalpharxypsi}.
Let $\psi_\alpha(R_{XY})$ be defined as in \lmaref{lma:jalpharxypsi}.
Let $R_{XY}$ be a PMF that satisfies $J_\alpha(X;Y) = \psi_\alpha(R_{XY})$, and let $(Q_X^*,Q_Y^*)$ be a pair of PMFs that minimizes $(Q_X,Q_Y) \mapsto D_\alpha(P_{XY}\|Q_X Q_Y)$.
If $\alpha \in (1,\infty)$, then \eqref{eq:psialpharxyeqdalphab} in the proof of \lmaref{lma:jalpharxypsi} holds with equality, i.e.,
\begin{align}
D(R_{XY}\|R_X R_Y) + \frac{\alpha}{1 - \alpha} \mulspace D(R_{XY}\|P_{XY}) = D(R_{XY}\|Q_X^* Q_Y^*) + \frac{\alpha}{1 - \alpha} \mulspace D(R_{XY}\|P_{XY}).\label{eq:jalphauniqueminbiga}
\end{align}
Because the LHS of \eqref{eq:jalphauniqueminbiga} is finite, \propref{prop:minqxqydpxyqxqy} implies that $Q_X^* = R_X$ and $Q_Y^* = R_Y$, thus the minimizer is unique.
As shown in the proof of \lmaref{lma:jalpharxypsi}, \eqref{eq:jalphauniqueminbiga} remains valid for $\alpha = \infty$ after replacing $\frac{\alpha}{1 - \alpha}$ by $-1$, thus the same argument establishes the uniqueness for $\alpha = \infty$.

Finally, we show that, for $\alpha \in [0,\frac{1}{2}]$, the mapping $(Q_X,Q_Y) \mapsto D_\alpha(P_{XY}\|Q_X Q_Y)$ can have more than one minimizer.
Let $X$ be uniformly distributed over $\{0,1\}$, and let $Y = X$.
Then, for all $\alpha \in [0,\frac{1}{2}]$,
\begin{align}
J_\alpha(X;Y) &= \frac{\alpha}{1 - \alpha} \log 2\label{eq:jalphauniqueminimizeri}\\
&= D_\alpha\bigl(P_{XY}\|(1,0) (1,0)\bigr)\\
&= D_\alpha\bigl(P_{XY}\|(0,1) (0,1)\bigr),
\end{align}
where \eqref{eq:jalphauniqueminimizeri} follows from \lmaref{lma:jalphaselfinformation}.
\end{proof}

\begin{Lemma}\label{lma:kalphawelldefinedfinite}
For every $\alpha \in [0,\infty]$, the minimum in the definition of $K_\alpha(X;Y)$ in \eqref{eq:defkalpha} exists and is finite.
\end{Lemma}

\begin{proof}
Let $\alpha \in [0,\infty]$, and denote by $U_X$ and $U_Y$ the uniform distribution over $\set{X}$ and $\set{Y}$, respectively.
Then $\inf_{\kernedqxcommaqy} \Delta_\alpha(P_{XY}\|Q_X Q_Y)$ is finite because $\Delta_\alpha(P_{XY}\|U_X U_Y)$ is finite and because the relative $\alpha$-entropy is nonnegative (\propref{prop:relalphaentropyproperties}).
For $\alpha \in (0,\infty)$, the minimum exists because the set $\set{P}(\set{X}) \times \set{P}(\set{Y})$ is compact and the mapping $(Q_X,Q_Y) \mapsto \Delta_\alpha(P_{XY}\|Q_X Q_Y)$ is continuous.
For $\alpha \in \{0,\infty\}$, the~minimum exists because $(Q_X,Q_Y) \mapsto \Delta_\alpha(P_{XY}\|Q_X Q_Y)$ takes on only a finite number of values:
if~$\alpha = 0$, then $\Delta_\alpha(P_{XY}\|Q_X Q_Y)$ depends on $Q_X Q_Y$ only via $\supp(Q_X Q_Y) \subseteq \set{X} \times \set{Y}$; and
if $\alpha = \infty$, then~$\Delta_\alpha(P_{XY}\|Q_X Q_Y)$ depends on $Q_X Q_Y$ only via $\argmax(Q_X Q_Y) \subseteq \set{X} \times \set{Y}$.
\end{proof}

\begin{Lemma}\label{lma:kalphanonnegative}
For all $\alpha \in [0,\infty]$, $K_\alpha(X;Y) \ge 0$.
If $\alpha \in (0,\infty)$, then $K_\alpha(X;Y) = 0$ if and only if $X$ and $Y$ are independent (nonnegativity).
\end{Lemma}

\begin{proof}
The nonnegativity follows from the definition of $K_\alpha(X;Y)$ because the relative $\alpha$-entropy is nonnegative for $\alpha \in [0,\infty]$ (\propref{prop:relalphaentropyproperties}).
If $X$ and $Y$ are independent, then $P_{XY} = P_X P_Y$, and the choice $Q_X = P_X$ and $Q_Y = P_Y$ in the definition of $K_\alpha(X;Y)$ achieves $K_\alpha(X;Y) = 0$.
Conversely, if $K_\alpha(X;Y) = 0$, then there exist PMFs $Q_X^*$ and $Q_Y^*$ satisfying $\Delta_\alpha(P_{XY}\|Q_X^* Q_Y^*) = 0$.
If, in addition, $\alpha \in (0,\infty)$, then $P_{XY} = Q_X^* Q_Y^*$ by \propref{prop:relalphaentropyproperties}, and hence $X$ and $Y$ are independent.
\end{proof}

\begin{Lemma}\label{lma:kalphasymmetric}
For all $\alpha \in [0,\infty]$, $K_\alpha(X;Y) = K_\alpha(Y;X)$ (symmetry).
\end{Lemma}

\begin{proof}
The definition of $K_\alpha(X;Y)$ is symmetric in $X$ and $Y$.
\end{proof}

\begin{Lemma}\label{lma:kalphapowermean}
For all $\alpha \in (0,\infty)$,
\begin{align}
K_\alpha(X;Y) + H_\alpha(X,Y) = \min_{\kernedqxcommaqy} -\log M_\frac{\alpha - 1}{\alpha}(Q_X,Q_Y),\label{eq:kalphaminpowermean}
\end{align}
where $M_\beta(Q_X,Q_Y)$ is the following weighted power mean \mdpicite[Chapter~III]{HandbookMeansInequalities}:
For $\beta \in \reals \setminus \{0\}$,
\begin{align}
M_\beta(Q_X,Q_Y) \triangleq \mleft[\sum_{x,y} P(x,y) [Q_X(x) \mulspace Q_Y(y)]^\beta\mright]^\frac{1}{\beta},\label{eq:defmbeta}
\end{align}
where for $\beta < 0$, we read $P(x,y) [Q_X(x) \mulspace Q_Y(y)]^\beta$ as $P(x,y) / [Q_X(x) \mulspace Q_Y(y)]^{-\beta}$ and use the conventions \eqref{eq:zeroinftyconventions};
and for $\beta = 0$, using the convention $0^0 = 1$,
\begin{align}
M_0(Q_X,Q_Y) \triangleq \prod_{x,y} [Q_X(x) \mulspace Q_Y(y)]^{P(x,y)}.\label{eq:defmzero}
\end{align}
\end{Lemma}

\begin{proof}
Let $\alpha \in (0,\infty)$, and define the PMF $\widetilde{P}_{XY}$ as
\begin{align}
\widetilde{P}_{XY}(x,y) \triangleq \frac{P_{XY}(x,y)^\alpha}{\sum_{(x',y') \in \set{X} \times \set{Y}} P_{XY}(x',y')^\alpha}.
\end{align}
Then,
\begin{align}
K_\alpha(X;Y) &= J_\frac{1}{\alpha}(\widetilde{X};\widetilde{Y})\label{eq:kalphapowermeanzza}\\
&= \min_{\kernedqxcommaqy} D_\frac{1}{\alpha}(\widetilde{P}_{XY}\|Q_X Q_Y),\label{eq:kalphapowermeanzzb}
\end{align}
where \eqref{eq:kalphapowermeanzza} follows from \propref{prop:jalphavskalpha}, and \eqref{eq:kalphapowermeanzzb} follows from the definition of $J_{1 / \alpha}(\widetilde{X};\widetilde{Y})$.
A simple computation reveals that for all PMFs $Q_X$ and $Q_Y$,
\begin{align}
D_\frac{1}{\alpha}(\widetilde{P}_{XY}\|Q_X Q_Y) = -\log M_\frac{\alpha - 1}{\alpha}(Q_X,Q_Y) - H_\alpha(X,Y).\label{eq:kalphapowermeanzze}
\end{align}
Hence, \eqref{eq:kalphaminpowermean} follows from \eqref{eq:kalphapowermeanzzb} and \eqref{eq:kalphapowermeanzze}.
\end{proof}

\begin{Lemma}\label{lma:kzerovalueandlimit}
For $\alpha = 0$,
\begin{align}
K_0(X;Y) &= \log \frac{\card{\supp(P_X P_Y)}}{\card{\supp(P_{XY})}}\label{eq:kzerovalueandlimita}\\*[-0.5ex]
&\ge \min_{\kernedqxcommaqy} \log \max_{(x,y) \in \supp(P_{XY})} \frac{1}{Q_X(x) \mulspace Q_Y(y)} - \logcard{\supp(P_{XY})}\label{eq:kzerovalueandlimitb}\\
&= \lim_{\alpha \downarrow 0} K_\alpha(X;Y),\label{eq:kzerovalueandlimitc}
\end{align}
where in the RHS of \eqref{eq:kzerovalueandlimitb}, we use the conventions \eqref{eq:zeroinftyconventions}.
The inequality can be strict, so $\alpha \mapsto K_\alpha(X;Y)$ need not be continuous at $\alpha = 0$.
\end{Lemma}

\begin{proof}
We first prove \eqref{eq:kzerovalueandlimita}.
Recall that
\begin{align}
\Delta_0(P_{XY}\|Q_X Q_Y) = \begin{cases}\log \frac{\card{\supp(Q_X Q_Y)}}{\card{\supp(P_{XY})}} & \text{if $\supp(P_{XY}) \subseteq \supp(Q_X Q_Y)$,}\\
\infty & \text{otherwise.}\end{cases}
\end{align}
Observe that $\Delta_0(P_{XY}\|Q_X Q_Y)$ is finite only if $\supp(P_X) \subseteq \supp(Q_X)$ and $\supp(P_Y) \subseteq \supp(Q_Y)$.
For such PMFs $Q_X$ and $Q_Y$, we have $\card{\supp(Q_X Q_Y)} \ge \card{\supp(P_X P_Y)}$.
Thus, for all PMFs $Q_X$ and $Q_Y$,
\begin{align}
\Delta_0(P_{XY}\|Q_X Q_Y) \ge \log \frac{\card{\supp(P_X P_Y)}}{\card{\supp(P_{XY})}}.\label{eq:kzerovalueandlimitdeltazero}
\end{align}
Choosing $Q_X = P_X$ and $Q_Y = P_Y$ achieves equality in \eqref{eq:kzerovalueandlimitdeltazero}, which establishes \eqref{eq:kzerovalueandlimita}.

We now show \eqref{eq:kzerovalueandlimitb}.
Let $Q_X$ and $Q_Y$ be the uniform distributions over $\supp(P_X)$ and $\supp(P_Y)$, respectively.
Then,
\begin{align}
\log \max_{(x,y) \in \supp(P_{XY})} \frac{1}{Q_X(x) \mulspace Q_Y(y)} - \logcard{\supp(P_{XY})} = \log \frac{\card{\supp(P_X P_Y)}}{\card{\supp(P_{XY})}},
\end{align}
and hence \eqref{eq:kzerovalueandlimitb} holds.

We next establish \eqref{eq:kzerovalueandlimitc}.
To that end, define
\begin{align}
\tau \triangleq \min_{(x,y) \in \supp(P_{XY})} P(x,y).
\end{align}
We bound $K_\alpha(X;Y) + H_\alpha(X,Y)$ as follows:
For all $\alpha \in (0,1)$,
\begin{align}
K_\alpha(X;Y) + H_\alpha(X,Y) &= \min_{\kernedqxcommaqy} \frac{\alpha}{1 - \alpha} \log \sum_{x,y} P(x,y) [Q_X(x) \mulspace Q_Y(y)]^\frac{\alpha - 1}{\alpha}\label{eq:kzerolimita}\\*
&\ge \min_{\kernedqxcommaqy} \frac{\alpha}{1 - \alpha} \log \sum_{(x,y) \in \supp(P_{XY})} \tau \mulspace [Q_X(x) \mulspace Q_Y(y)]^\frac{\alpha - 1}{\alpha}\\
&\ge \min_{\kernedqxcommaqy} \frac{\alpha}{1 - \alpha} \log \max_{(x,y) \in \supp(P_{XY})} \tau \mulspace [Q_X(x) \mulspace Q_Y(y)]^\frac{\alpha - 1}{\alpha}\\
&= \min_{\kernedqxcommaqy} \log \max_{(x,y) \in \supp(P_{XY})} \frac{1}{Q_X(x) \mulspace Q_Y(y)} - \frac{\alpha}{1 - \alpha} \log \frac{1}{\tau},\label{eq:kzerolimitd}
\end{align}
where \eqref{eq:kzerolimita} follows from \lmaref{lma:kalphapowermean}.
Similarly, for all $\alpha \in (0,1)$,
\begin{align}
K_\alpha(X;Y) + H_\alpha(X,Y) &= \min_{\kernedqxcommaqy} \frac{\alpha}{1 - \alpha} \log \sum_{x,y} P(x,y) [Q_X(x) \mulspace Q_Y(y)]^\frac{\alpha - 1}{\alpha}\label{eq:kzerolimitg}\\
&\le \min_{\kernedqxcommaqy} \frac{\alpha}{1 - \alpha} \log \max_{(x,y) \in \supp(P_{XY})} [Q_X(x) \mulspace Q_Y(y)]^\frac{\alpha - 1}{\alpha}\\
&= \min_{\kernedqxcommaqy} \log \max_{(x,y) \in \supp(P_{XY})} \frac{1}{Q_X(x) \mulspace Q_Y(y)},\label{eq:kzerolimiti}
\end{align}
where \eqref{eq:kzerolimitg} is the same as \eqref{eq:kzerolimita}.
Now \eqref{eq:kzerovalueandlimitc} follows from \eqref{eq:kzerolimitd}, \eqref{eq:kzerolimiti}, and the sandwich theorem because $\lim_{\alpha \downarrow 0} \frac{\alpha}{1 - \alpha} \log \frac{1}{\tau} = 0$ and because $\lim_{\alpha \downarrow 0} H_\alpha(X,Y) = \logcard{\supp(P_{XY})}$ (\propref{prop:renyientropyproperties}).

Finally, we provide an example for which \eqref{eq:kzerovalueandlimitb} holds with strict inequality.
Let $\set{X} = \{1,2,3\}$, let $\set{Y} = \{1,2\}$, and let $(X,Y)$ be uniformly distributed over $\{(1,1),(2,2),(3,1)\}$.
The LHS of \eqref{eq:kzerovalueandlimitb} then equals $\log 2$.
Using
\begin{align}
Q_X(x) &\triangleq \begin{cases}0.28 & \text{if $x \in \{1,3\}$,}\\
0.44 & \text{if $x = 2$,}\end{cases}\\
Q_Y(y) &\triangleq \begin{cases}0.60 & \text{if $y = 1$,}\\
0.40 & \text{if $y = 2$,}\end{cases}
\end{align}
we see that the RHS of \eqref{eq:kzerovalueandlimitb} is upper bounded by $\log \frac{5.952\ldots}{3}$, which is smaller than $\log 2$.
\end{proof}

\begin{Lemma}\label{lma:konemutualinformation}
$K_1(X;Y) = I(X;Y)$.
\end{Lemma}

\begin{proof}
The claim follows from \propref{prop:minqxqydpxyqxqy} because $\Delta_1(P_{XY}\|Q_X Q_Y)$ in the definition of $K_1(X;Y)$ is equal to $D(P_{XY}\|Q_X Q_Y)$.
\end{proof}

\begin{Lemma}\label{lma:ktwosingularvalue}
Let $f\colon \{1,\ldots,\card{\set{X}}\} \to \set{X}$ and $g\colon \{1,\ldots,\card{\set{Y}}\} \to \set{Y}$ be bijective functions, and let $\mat{B}$ be the $\card{\set{X}} \times \card{\set{Y}}$ matrix whose Row-$i$ Column-$j$ entry $\mat{B}_{i,j}$ equals $P_{XY}(f(i),g(j))$.
Then,
\begin{align}
K_2(X;Y) = -2 \log \sigma_1(\mat{B}) - H_2(X,Y),
\end{align}
where $\sigma_1(\mat{B})$ denotes the largest singular value of $\mat{B}$.
(Because the singular values of a matrix are invariant under row and column permutations, the result does not depend on $f$ or $g$.)
\end{Lemma}

\begin{proof}
Let $(\widetilde{X},\widetilde{Y})$ be distributed according to the joint PMF
\begin{align}
\widetilde{P}_{XY}(x,y) \triangleq \bigl[\beta \mulspace P_{XY}(x,y)\bigr]^2,\label{eq:ktwosingularvaluedefpxytilde}
\end{align}
where
\begin{align}
\beta \triangleq \mleft[\sum_{x,y} P_{XY}(x,y)^2\mright]^{-\frac{1}{2}}.
\end{align}
Then,
\begin{align}
K_2(X;Y) &= J_\frac{1}{2}(\widetilde{X};\widetilde{Y})\label{eq:ktwosingularvaluea}\\
&= -2 \log \sigma_1(\beta \mulspace \mat{B})\label{eq:ktwosingularvalueb}\\
&= -2 \log \bigl[\beta \mulspace \sigma_1(\mat{B})\bigr]\label{eq:ktwosingularvaluec}\\
&= -2 \log \sigma_1(\mat{B}) - H_2(X,Y),\label{eq:ktwosingularvalued}
\end{align}
where \eqref{eq:ktwosingularvaluea} follows from \propref{prop:jalphavskalpha};
\eqref{eq:ktwosingularvalueb} follows from \lmaref{lma:jonehalfsingularvalue} and \eqref{eq:ktwosingularvaluedefpxytilde};
\eqref{eq:ktwosingularvaluec} holds because $\beta > 0$;
and \eqref{eq:ktwosingularvalued} follows from the definition of $H_2(X,Y)$.
\end{proof}

\begin{Lemma}\label{lma:kinftyiszero}
$K_\infty(X;Y) = 0$.
\end{Lemma}

\begin{proof}
Let the pair $(\hat{x},\hat{y})$ be such that $P(\hat{x},\hat{y}) = \max_{x,y} P(x,y)$, and define the PMFs $\hat{Q}_X$ and $\hat{Q}_Y$ as $\hat{Q}_X(x) = \mathbbm{1}\{x = \hat{x}\}$ and $\hat{Q}_Y(y) = \mathbbm{1}\{y = \hat{y}\}$.
Then, $\Delta_\infty(P_{XY}\|\hat{Q}_X \hat{Q}_Y) = 0$, so $K_\infty(X;Y) \le 0$.
Because $K_\infty(X;Y) \ge 0$ (\lmaref{lma:kalphanonnegative}), this implies $K_\infty(X;Y) = 0$.
\end{proof}

\begin{Lemma}\label{lma:kalphanotmonotonic}
The mapping $\alpha \mapsto K_\alpha(X;Y)$ need not be monotonic on $[0,\infty]$.
\end{Lemma}

\begin{proof}
Let $P_{XY}$ be such that $\supp(P_{XY}) = \set{X} \times \set{Y}$ and $I(X;Y) > 0$.
Then,
\begin{align}
K_0(X;Y) &= 0,\\
K_1(X;Y) &> 0,\\
K_\infty(X;Y) &= 0,
\end{align}
which follow from Lemmas \ref{lma:kzerovalueandlimit}, \ref{lma:konemutualinformation}, and \ref{lma:kinftyiszero}, respectively.
Thus, $\alpha \mapsto K_\alpha(X;Y)$ is not monotonic on $[0,\infty]$.
\end{proof}

\begin{Lemma}\label{lma:kalphaplushalphanonincreasing}
The mapping $\alpha \mapsto K_\alpha(X;Y) + H_\alpha(X,Y)$ is nonincreasing on $[0,\infty]$.
\end{Lemma}

\begin{proof}
We first show the monotonicity for $\alpha \in (0,\infty)$.
To that end, let $\alpha,\alpha' \in (0,\infty)$ with $\alpha \le \alpha'$, and let $M_\beta(Q_X,Q_Y)$ be defined as in \eqref{eq:defmbeta} and \eqref{eq:defmzero}.
Then, for all PMFs $Q_X$ and $Q_Y$,
\begin{align}
M_\frac{\alpha - 1}{\alpha}(Q_X,Q_Y) \le M_\frac{\alpha' - 1}{\alpha'}(Q_X,Q_Y),\label{eq:kalphaplushalphanoninca}
\end{align}
which follows from the power mean inequality \mdpicite[III 3.1.1 Theorem~1]{HandbookMeansInequalities} because $\frac{\alpha - 1}{\alpha} \le \frac{\alpha' - 1}{\alpha'}$.
Hence,
\begin{align}
K_\alpha(X;Y) + H_\alpha(X,Y) &= \min_{\kernedqxcommaqy} -\log M_\frac{\alpha - 1}{\alpha}(Q_X,Q_Y)\label{eq:kalphaplushalphanonincd}\\
&\ge \min_{\kernedqxcommaqy} -\log M_\frac{\alpha' - 1}{\alpha'}(Q_X,Q_Y)\label{eq:kalphaplushalphanonince}\\
&= K_{\alpha'}(X;Y) + H_{\alpha'}(X,Y),\label{eq:kalphaplushalphanonincf}
\end{align}
where \eqref{eq:kalphaplushalphanonincd} and \eqref{eq:kalphaplushalphanonincf} follow from \lmaref{lma:kalphapowermean}, and
\eqref{eq:kalphaplushalphanonince} follows from \eqref{eq:kalphaplushalphanoninca}.

The monotonicity extends to $\alpha = 0$ because
\begin{align}
K_0(X;Y) + H_0(X,Y) &\ge \lim_{\alpha \downarrow 0} K_\alpha(X;Y) + H_0(X,Y)\label{eq:kzeroplushzerononincreasinga}\\*
&= \lim_{\alpha \downarrow 0} [K_\alpha(X;Y) + H_\alpha(X,Y)],\label{eq:kzeroplushzerononincreasingb}
\end{align}
where \eqref{eq:kzeroplushzerononincreasinga} follows from \lmaref{lma:kzerovalueandlimit}, and
\eqref{eq:kzeroplushzerononincreasingb} holds because $\alpha \mapsto H_\alpha(X,Y)$ is continuous at $\alpha = 0$ (\propref{prop:renyientropyproperties}).

The monotonicity extends to $\alpha = \infty$ because for all $\alpha \in (0,\infty)$,
\begin{align}
K_\alpha(X;Y) + H_\alpha(X,Y) &\ge H_\alpha(X,Y)\label{eq:kinftyplushinftynonincreasinga}\\
&\ge H_\infty(X,Y)\label{eq:kinftyplushinftynonincreasingb}\\
&= K_\infty(X;Y) + H_\infty(X,Y),\label{eq:kinftyplushinftynonincreasingc}
\end{align}
where \eqref{eq:kinftyplushinftynonincreasinga} holds because $K_\alpha(X;Y) \ge 0$ (\lmaref{lma:kalphanonnegative});
\eqref{eq:kinftyplushinftynonincreasingb} holds because $H_\alpha(X,Y)$ is nonincreasing in $\alpha$ (\propref{prop:renyientropyproperties}); and
\eqref{eq:kinftyplushinftynonincreasingc} holds because $K_\infty(X;Y) = 0$ (\lmaref{lma:kinftyiszero}).
\end{proof}

\begin{Lemma}\label{lma:kalphacontinuousalpha}
The mapping $\alpha \mapsto K_\alpha(X;Y)$ is continuous on $(0,\infty]$.
(See \lmaref{lma:kzerovalueandlimit} for the behavior at $\alpha = 0$.)
\end{Lemma}

\begin{proof}
Because $\alpha \mapsto H_\alpha(X,Y)$ is continuous on $[0,\infty]$ (\propref{prop:renyientropyproperties}), it suffices to show that the mapping $\alpha \mapsto K_\alpha(X;Y) + H_\alpha(X,Y)$ is continuous on $(0,\infty]$.
We first show that it is continuous on $(0,1) \cup (1,\infty)$ by showing that $\alpha \mapsto \bigl(1 - \tfrac{1}{\alpha}\bigr) \bigl[K_\alpha(X;Y) + H_\alpha(X,Y)\bigr]$ is concave and hence continuous on $(0,\infty)$.
For a fixed $\alpha \in (0,\infty)$, let $(\widetilde{X},\widetilde{Y})$ be distributed according to the joint PMF
\begin{align}
\widetilde{P}_{XY}(x,y) \triangleq \frac{P_{XY}(x,y)^\alpha}{\sum_{(x',y') \in \set{X} \times \set{Y}} P_{XY}(x',y')^\alpha}.
\end{align}
Then, for all $\alpha \in (0,\infty)$,
\begin{align}
&\bigl(1 - \tfrac{1}{\alpha}\bigr) \bigl[K_\alpha(X;Y) + H_\alpha(X,Y)\bigr]\nonumber\\*[0.5ex]
&\qquad = \bigl(1 - \tfrac{1}{\alpha}\bigr) J_\frac{1}{\alpha}(\widetilde{X};\widetilde{Y}) + \bigl(1 - \tfrac{1}{\alpha}\bigr) H_\alpha(X,Y)\label{eq:omodalphakplushconcavea}\\
&\qquad = \min_{R_{XY}} \Bigl[\bigl(1 - \tfrac{1}{\alpha}\bigr) D(R_{XY}\|R_X R_Y) + \tfrac{1}{\alpha} \mulspace D(R_{XY}\|\widetilde{P}_{XY}) + \bigl(1 - \tfrac{1}{\alpha}\bigr) H_\alpha(X,Y)\Bigr]\label{eq:omodalphakplushconcaveb}\\
&\qquad = \min_{R_{XY}} \Bigl[\bigl(1 - \tfrac{1}{\alpha}\bigr) D(R_{XY}\|R_X R_Y) + \bigl(1 - \tfrac{1}{\alpha}\bigr) H(R_{XY}) + D(R_{XY}\|P_{XY})\Bigr],\label{eq:omodalphakplushconcavec}
\end{align}
where \eqref{eq:omodalphakplushconcavea} follows from \propref{prop:jalphavskalpha};
\eqref{eq:omodalphakplushconcaveb} follows from \lmaref{lma:oneminusalphajalpharxy}; and
\eqref{eq:omodalphakplushconcavec} follows from a short computation.
For every $R_{XY} \in \set{P}(\set{X} \times \set{Y})$, the expression in square brackets on the RHS of \eqref{eq:omodalphakplushconcavec} is concave in $\alpha$ because the mapping $\alpha \mapsto 1 - \tfrac{1}{\alpha}$ is concave on $(0,\infty)$ and because $D(R_{XY}\|R_X R_Y)$ and $H(R_{XY})$ are nonnegative.
The pointwise minimum preserves the concavity, thus the LHS of \eqref{eq:omodalphakplushconcavea} is concave in $\alpha$ and hence continuous in $\alpha \in (0,\infty)$.
This implies that $\alpha \mapsto K_\alpha(X;Y) + H_\alpha(X,Y)$ and hence $\alpha \mapsto K_\alpha(X;Y)$ is continuous on $(0,1) \cup (1,\infty)$.

We now establish continuity at $\alpha = \infty$.
Let $(\hat{x},\hat{y})$ be such that $P(\hat{x},\hat{y}) = \max_{x,y} P(x,y)$; define the PMFs $\hat{Q}_X$ and $\hat{Q}_Y$ as $\hat{Q}_X(x) \triangleq \mathbbm{1}\{x = \hat{x}\}$ and $\hat{Q}_Y(y) \triangleq \mathbbm{1}\{y = \hat{y}\}$; and let $M_\beta(Q_X,Q_Y)$ be defined as in~\eqref{eq:defmbeta}.
Then, for all $\alpha \in (1,\infty)$,
\begin{align}
K_\infty(X;Y) + H_\infty(X,Y) &\le K_\alpha(X;Y) + H_\alpha(X,Y)\label{eq:kinftyplushinftycontinuousa}\\
&\le -\log M_\frac{\alpha - 1}{\alpha}(\hat{Q}_X,\hat{Q}_Y)\label{eq:kinftyplushinftycontinuousb}\\
&= \frac{\alpha}{\alpha - 1} \mulspace H_\infty(X,Y)\label{eq:kinftyplushinftycontinuousc}\\
&= K_\infty(X;Y) + \frac{\alpha}{\alpha - 1} \mulspace H_\infty(X,Y),\label{eq:kinftyplushinftycontinuousd}
\end{align}
where \eqref{eq:kinftyplushinftycontinuousa} holds because $K_\alpha(X;Y) + H_\alpha(X,Y)$ is nonincreasing in $\alpha$ (\lmaref{lma:kalphaplushalphanonincreasing});
\eqref{eq:kinftyplushinftycontinuousb} follows from \lmaref{lma:kalphapowermean};
\eqref{eq:kinftyplushinftycontinuousc} follows from the definitions of $M_\beta(Q_X,Q_Y)$ in \eqref{eq:defmbeta} and $H_\infty(X,Y)$ in \eqref{eq:defhinfinity}; and~\eqref{eq:kinftyplushinftycontinuousd}~holds because $K_\infty(X;Y) = 0$ (\lmaref{lma:kinftyiszero}).
Because $\lim_{\alpha \to \infty} \frac{\alpha}{\alpha - 1} = 1$, \eqref{eq:kinftyplushinftycontinuousa}--\eqref{eq:kinftyplushinftycontinuousd} and the sandwich theorem imply that $\alpha \mapsto K_\alpha(X;Y) + H_\alpha(X,Y)$ is continuous at $\alpha = \infty$.
This and the continuity of $\alpha \mapsto H_\alpha(X,Y)$ at $\alpha = \infty$ (\propref{prop:renyientropyproperties}) establish the continuity of $\alpha \mapsto K_\alpha(X;Y)$ at $\alpha = \infty$.

It remains to show the continuity at $\alpha = 1$.
Let $\alpha \in (\frac{4}{5},1) \cup (1,\frac{4}{3})$, and define $\delta \triangleq \frac{\abs{\alpha - 1}}{\alpha} \in (0,\frac{1}{4})$.
(These definitions ensure that on the RHS of \eqref{eq:koneplushonecontinuousi} ahead, $1 - 4 \delta$ will be positive.)
Let $M_\beta(Q_X,Q_Y)$ be defined as in \eqref{eq:defmbeta} and \eqref{eq:defmzero}.
Then, for all PMFs $Q_X$ and $Q_Y$,
\begin{align}
M_\frac{\alpha - 1}{\alpha}(Q_X,Q_Y) &\le M_\delta(Q_X,Q_Y)\label{eq:koneplushonecontinuousa}\\[-0.5ex]
&= \mleft[\sum_{x,y} P(x,y) \mulspace [P_X(x) \mulspace P_Y(y)]^\delta \mleft[\frac{Q_X(x) \mulspace Q_Y(y)}{P_X(x) \mulspace P_Y(y)}\mright]^\delta\vthinspace\mright]^\frac{1}{\delta}\label{eq:koneplushonecontinuousb}\\
&\le \mleft[\sum_{x,y} P(x,y) \mulspace [P_X(x) \mulspace P_Y(y)]^{2 \delta}\mright]^\frac{1}{2 \delta} \cdot \mleft[\sum_{x,y} P(x,y) \mleft[\frac{Q_X(x) \mulspace Q_Y(y)}{P_X(x) \mulspace P_Y(y)}\mright]^{2 \delta}\vthinspace\mright]^\frac{1}{2 \delta}\label{eq:koneplushonecontinuousc}\\
&\le \mleft[\sum_{x,y} P(x,y) \mulspace [P_X(x) \mulspace P_Y(y)]^{2 \delta}\mright]^\frac{1}{2 \delta}\label{eq:koneplushonecontinuousd}\\
&= M_{2 \delta}(P_X,P_Y),\label{eq:koneplushonecontinuouse}
\end{align}
where \eqref{eq:koneplushonecontinuousa} follows from the power mean inequality \mdpicite[III 3.1.1 Theorem~1]{HandbookMeansInequalities} because $\frac{\alpha - 1}{\alpha} \le \delta$;
\eqref{eq:koneplushonecontinuousc}~follows from the Cauchy--Schwarz inequality; and
\eqref{eq:koneplushonecontinuousd} holds because
\begin{align}
&\mleft[\sum_{x,y} P(x,y) \mleft[\frac{Q_X(x)}{P_X(x)}\mright]^{2 \delta} \mleft[\frac{Q_Y(y)}{P_Y(y)}\mright]^{2 \delta}\vthinspace\mright]^\frac{1}{2 \delta}\nonumber\\*
&\qquad \le \mleft[\sum_x P_X(x) \mleft[\frac{Q_X(x)}{P_X(x)}\mright]^{4 \delta}\vthinspace\mright]^\frac{1}{4 \delta} \cdot \mleft[\sum_y P_Y(y) \mleft[\frac{Q_Y(y)}{P_Y(y)}\mright]^{4 \delta}\vthinspace\mright]^\frac{1}{4 \delta}\label{eq:koneplushonecontinuoush}\\
&\qquad = 2^{-D_{1 - 4\delta}(P_X\|Q_X)} \cdot 2^{-D_{1 - 4\delta}(P_Y\|Q_Y)}\label{eq:koneplushonecontinuousi}\\
&\qquad \le 1,\label{eq:koneplushonecontinuousj}
\end{align}
where \eqref{eq:koneplushonecontinuoush} follows from the Cauchy--Schwarz inequality, and
\eqref{eq:koneplushonecontinuousj} holds because $1 - 4\delta > 0$ and because the R\'enyi divergence is nonnegative for positive orders (\propref{prop:renyidivproperties}).
Thus, for all $\alpha \in (\frac{4}{5},\frac{4}{3})$,
\begin{align}
-\log M_\frac{2 \abs{\alpha - 1}}{\alpha}(P_X,P_Y) &\le \min_{\kernedqxcommaqy} -\log M_\frac{\alpha - 1}{\alpha}(Q_X,Q_Y)\label{eq:koneplushonecontinuousm}\\
&\le -\log M_\frac{\alpha - 1}{\alpha}(P_X,P_Y),
\end{align}
where \eqref{eq:koneplushonecontinuousm} follows from \eqref{eq:koneplushonecontinuouse} if $\alpha \ne 1$ and from \propref{prop:minqxqydpxyqxqy} and a simple computation if $\alpha = 1$.
By \lmaref{lma:kalphapowermean}, this implies that for all $\alpha \in (\frac{4}{5},\frac{4}{3})$,
\begin{align}
-\log M_\frac{2 \abs{\alpha - 1}}{\alpha}(P_X,P_Y) &\le K_\alpha(X;Y) + H_\alpha(X,Y)\label{eq:kalphaplushalphasanda}\\[-0.5ex]
&\le -\log M_\frac{\alpha - 1}{\alpha}(P_X,P_Y).\label{eq:kalphaplushalphasandb}
\end{align}
Because $\beta \mapsto M_\beta(P_X,P_Y)$ is continuous at $\beta = 0$ \mdpicite[III 1 Theorem~2(b)]{HandbookMeansInequalities}, \eqref{eq:kalphaplushalphasanda}--\eqref{eq:kalphaplushalphasandb} and the sandwich theorem imply that $\alpha \mapsto K_\alpha(X;Y) + H_\alpha(X,Y)$ is continuous at $\alpha = 1$.
This and the continuity of $\alpha \mapsto H_\alpha(X,Y)$ at $\alpha = 1$ (\propref{prop:renyientropyproperties}) establish the continuity of $\alpha \mapsto K_\alpha(X;Y)$ at $\alpha = 1$.
\end{proof}

\begin{Lemma}\label{lma:kalphaselfinformation}
If $X = Y$ with probability one, then
\begin{align}
K_\alpha(X;Y) = \begin{cases}2 H_\frac{\alpha}{2 - \alpha}(X) - H_\alpha(X) & \text{if $\alpha \in [0,2)$,}\\
\frac{\alpha}{\alpha - 1} \mulspace H_\infty(X) - H_\alpha(X) & \text{if $\alpha \ge 2$,}\\
0 & \text{if $\alpha = \infty$.}\end{cases}\label{eq:kalphaselfinformation}
\end{align}
\end{Lemma}

\begin{proof}
We first treat the cases $\alpha = 0$, $\alpha = 1$, and $\alpha = \infty$.
For $\alpha = 0$, \eqref{eq:kalphaselfinformation} holds because
\begin{align}
K_0(X;Y) &= \log \frac{\card{\supp(P_X P_Y)}}{\card{\supp(P_{XY})}}\label{eq:kzeroselfinformationa}\\
&= \logcard{\supp(P_X)}\label{eq:kzeroselfinformationb}\\
&= H_0(X),\label{eq:kzeroselfinformationc}
\end{align}
where \eqref{eq:kzeroselfinformationa} follows from \lmaref{lma:kzerovalueandlimit}, and
\eqref{eq:kzeroselfinformationb} holds because the hypothesis $\Prv{X = Y} = 1$ implies that $\card{\supp(P_X P_Y)} = \card{\supp(P_X)}^2$ and $\card{\supp(P_{XY})} = \card{\supp(P_X)}$.
For $\alpha = 1$, \eqref{eq:kalphaselfinformation} holds because $K_1(X;Y) = I(X;Y)$ (\lmaref{lma:konemutualinformation}) and because $\Prv{X = Y} = 1$ implies that $I(X;Y) = H(X) = H_1(X)$.
For $\alpha = \infty$, \eqref{eq:kalphaselfinformation} holds because $K_\infty(X;Y) = 0$ (\lmaref{lma:kinftyiszero}).

Now let $\alpha \in (0,1) \cup (1,\infty)$, and let $(\widetilde{X},\widetilde{Y})$ be distributed according to the joint PMF
\begin{align}
\widetilde{P}_{XY}(x,y) &\triangleq \frac{P_{XY}(x,y)^\alpha}{\sum_{(x',y') \in \set{X} \times \set{Y}} P_{XY}(x',y')^\alpha}\\*
&= \frac{P_X(x)^\alpha}{\sum_{x' \in \set{X}} P_X(x')^\alpha} \mulspace \mathbbm{1}\{x = y\},\label{eq:kalphaselfinformationtildeb}
\end{align}
where \eqref{eq:kalphaselfinformationtildeb} holds because $P_{XY}(x,y) = P_X(x) \mulspace \mathbbm{1}\{x = y\}$ for all $x \in \set{X}$ and all $y \in \set{Y}$.
If $\alpha < 2$, then~\eqref{eq:kalphaselfinformation}~holds because
\begin{align}
K_\alpha(X;Y) &= J_\frac{1}{\alpha}(\widetilde{X};\widetilde{Y})\label{eq:kalphaselfinformationa}\\
&= H_\frac{1}{2 - \alpha}(\widetilde{X})\label{eq:kalphaselfinformationb}\\[-0.5ex]
&= \frac{2 - \alpha}{1 - \alpha} \log \sum_x \mleft[\frac{P_X(x)^\alpha}{\sum_{x' \in \set{X}} P_X(x')^\alpha}\mright]^\frac{1}{2 - \alpha}\label{eq:kalphaselfinformationc}\\
&= 2 H_\frac{\alpha}{2 - \alpha}(X) - H_\alpha(X),\label{eq:kalphaselfinformationd}
\end{align}
where \eqref{eq:kalphaselfinformationa} follows from \propref{prop:jalphavskalpha};
\eqref{eq:kalphaselfinformationb} follows from \lmaref{lma:jalphaselfinformation} because $\Prv{\widetilde{X} = \widetilde{Y}} = 1$ and because $\frac{1}{\alpha} > \frac{1}{2}$; and
\eqref{eq:kalphaselfinformationd} follows from a simple computation.
If $\alpha \ge 2$, then \eqref{eq:kalphaselfinformation} holds because
\begin{align}
K_\alpha(X;Y) &= J_\frac{1}{\alpha}(\widetilde{X};\widetilde{Y})\label{eq:kalphaselfinformationg}\\
&= \frac{1}{\alpha - 1} \mulspace H_\infty(\widetilde{X})\label{eq:kalphaselfinformationh}\\
&= \frac{-1}{\alpha - 1} \log \max_x \frac{P_X(x)^\alpha}{\sum_{x' \in \set{X}} P_X(x')^\alpha}\label{eq:kalphaselfinformationi}\\
&= \frac{\alpha}{\alpha - 1} \mulspace H_\infty(X) - H_\alpha(X),\label{eq:kalphaselfinformationj}
\end{align}
where \eqref{eq:kalphaselfinformationg} follows from \propref{prop:jalphavskalpha};
\eqref{eq:kalphaselfinformationh} follows from \lmaref{lma:jalphaselfinformation} because $\Prv{\widetilde{X} = \widetilde{Y}} = 1$ and because $\frac{1}{\alpha} \le \frac{1}{2}$; and
\eqref{eq:kalphaselfinformationj} follows from a simple computation.
\end{proof}

\begin{Lemma}\label{lma:kalphauniqueminimizer}
For every $\alpha \in (0,2)$, the mapping $(Q_X,Q_Y) \mapsto \Delta_\alpha(P_{XY}\|Q_X Q_Y)$ in the definition of $K_\alpha(X;Y)$ in \eqref{eq:defkalpha} has a unique minimizer.
This need not be the case when $\alpha \in \{0\} \cup [2,\infty]$.
\end{Lemma}

\begin{proof}
Let $\alpha \in (0,2)$.
By \propref{prop:jalphavskalpha}, $K_\alpha(X;Y) = J_{1 / \alpha}(\widetilde{X};\widetilde{Y})$, where the pair $(\widetilde{X},\widetilde{Y})$ is distributed according to the joint PMF $\widetilde{P}_{XY}$ defined in \propref{prop:jalphavskalpha}.
The mapping $(Q_X,Q_Y) \mapsto D_{1 / \alpha}(\widetilde{P}_{XY}\|Q_X Q_Y)$ in the definition of $J_{1 / \alpha}(\widetilde{X};\widetilde{Y})$ has a unique minimizer by \lmaref{lma:jalphauniqueminimizer} because $\frac{1}{\alpha} > \frac{1}{2}$.
By \propref{prop:deltaalphadonedivalpha}, there is a bijection between the minimizers of $D_{1 / \alpha}(\widetilde{P}_{XY}\|Q_X Q_Y)$ and $\Delta_\alpha(P_{XY}\|Q_X Q_Y)$, so the mapping $(Q_X,Q_Y) \mapsto \Delta_\alpha(P_{XY}\|Q_X Q_Y)$ also has a unique minimizer.

We next show that for $\alpha \in \{0\} \cup [2,\infty]$, the mapping $(Q_X,Q_Y) \mapsto \Delta_\alpha(P_{XY}\|Q_X Q_Y)$ can have more than one minimizer.
Let $X$ be uniformly distributed over $\{0,1\}$, and let $Y = X$.
Then, by \lmaref{lma:kalphaselfinformation},
\begin{align}
K_\alpha(X;Y) = \begin{cases} \log 2 & \text{if $\alpha = 0$,}\\
\frac{1}{\alpha - 1} \log 2 & \text{if $\alpha \ge 2$,}\\
0 & \text{if $\alpha = \infty$.}\end{cases}
\end{align}
If $\alpha = 0$, then it follows from the definition of $\Delta_0(P\|Q)$ in \eqref{eq:relalphaentropyzero} that $\Delta_0(P_{XY}\|Q_X Q_Y) = \log 2$ whenever $\supp(Q_X) = \supp(Q_Y) = \{0,1\}$, so the minimizer is not unique.
Otherwise, if $\alpha \in [2,\infty]$, it can be verified that
\begin{align}
\Delta_\alpha\bigl(P_{XY}\|(1,0) (1,0)\bigr) &= \Delta_\alpha\bigl(P_{XY}\|(0,1) (0,1)\bigr)\\
&= \begin{cases} \frac{1}{\alpha - 1} \log 2 & \text{if $\alpha \ge 2$,}\\
0 & \text{if $\alpha = \infty$,}\end{cases}
\end{align}
so the minimizer is not unique in this case either.
\end{proof}

\begin{Lemma}\label{lma:kalphaadditive}
If the pairs $(X_1,Y_1)$ and $(X_2,Y_2)$ are independent, then $K_\alpha(X_1,X_2;Y_1,Y_2) = K_\alpha(X_1;Y_1) + K_\alpha(X_2;Y_2)$ for all $\alpha \in [0,\infty]$ (additivity).
\end{Lemma}

\begin{proof}
We first treat the cases $\alpha = 0$ and $\alpha = \infty$.
For $\alpha = 0$, the claim is true because
\begin{align}
K_0(X_1,X_2;Y_1,Y_2) &= \log \frac{\card{\supp(P_{X_1 X_2} P_{Y_1 Y_2})}}{\card{\supp(P_{X_1 X_2 Y_1 Y_2})}}\label{eq:kzeroadditivea}\\
&= \log \frac{\card{\supp(P_{X_1} P_{Y_1})} \cdot \card{\supp(P_{X_2} P_{Y_2})}}{\card{\supp(P_{X_1 Y_1})} \cdot \card{\supp(P_{X_2 Y_2})}}\label{eq:kzeroadditiveb}\\
&= K_0(X_1;Y_1) + K_0(X_2;Y_2),\label{eq:kzeroadditivec}
\end{align}
where \eqref{eq:kzeroadditivea} and \eqref{eq:kzeroadditivec} follow from \lmaref{lma:kzerovalueandlimit}, and \eqref{eq:kzeroadditiveb} follows from the independence hypothesis $P_{X_1 X_2 Y_1 Y_2} = P_{X_1 Y_1} P_{X_2 Y_2}$.
For $\alpha = \infty$, the claim is true because $K_\infty(X;Y) = 0$ (\lmaref{lma:kinftyiszero}).

Now let $\alpha \in (0,\infty)$, and let $(\widetilde{X}_1,\widetilde{X}_2,\widetilde{Y}_1,\widetilde{Y}_2)$ be distributed according to the joint PMF
\begin{align}
\widetilde{P}_{X_1 X_2 Y_1 Y_2}(x_1,x_2,y_1,y_2) &\triangleq \frac{P_{X_1 X_2 Y_1 Y_2}(x_1,x_2,y_1,y_2)^\alpha}{\sum_{x_1',x_2',y_1',y_2'} P_{X_1 X_2 Y_1 Y_2}(x_1',x_2',y_1',y_2')^\alpha}\\*
&= \frac{P_{X_1 Y_1}(x_1,y_1)^\alpha}{\sum_{x_1',y_1'} P_{X_1 Y_1}(x_1',y_1')^\alpha} \cdot \frac{P_{X_2 Y_2}(x_2,y_2)^\alpha}{\sum_{x_2',y_2'} P_{X_2 Y_2}(x_2',y_2')^\alpha},\label{eq:kalphaadditivetilde}
\end{align}
where \eqref{eq:kalphaadditivetilde} follows from the independence hypothesis $P_{X_1 X_2 Y_1 Y_2} = P_{X_1 Y_1} P_{X_2 Y_2}$.
Then,
\begin{align}
K_\alpha(X_1,X_2;Y_1,Y_2) &= J_\frac{1}{\alpha}(\widetilde{X}_1,\widetilde{X}_2;\widetilde{Y}_1,\widetilde{Y}_2)\label{eq:kalphaadditivea}\\
&= J_\frac{1}{\alpha}(\widetilde{X}_1;\widetilde{Y}_1) + J_\frac{1}{\alpha}(\widetilde{X}_2;\widetilde{Y}_2)\label{eq:kalphaadditiveb}\\
&= K_\alpha(X_1;Y_1) + K_\alpha(X_2;Y_2),\label{eq:kalphaadditivec}
\end{align}
where \eqref{eq:kalphaadditivea} and \eqref{eq:kalphaadditivec} follow from \propref{prop:jalphavskalpha}, and
\eqref{eq:kalphaadditiveb} follows from \lmaref{lma:jalphaadditive} because the pairs $(\widetilde{X}_1,\widetilde{Y}_1)$ and $(\widetilde{X}_2,\widetilde{Y}_2)$ are independent by \eqref{eq:kalphaadditivetilde}.
\end{proof}

\begin{Lemma}\label{lma:kalphacardinalityupperbound}
For all $\alpha \in [0,\infty]$, $K_\alpha(X;Y) \le \logcard{\set{X}}$.
\end{Lemma}

\begin{proof}
For $\alpha = 0$, this is true because
\begin{align}
K_0(X;Y) &= \log \frac{\card{\supp(P_X P_Y)}}{\card{\supp(P_{XY})}}\label{eq:kalphacardinalityube}\\
&\le \log \frac{\card{\set{X}} \cdot \card{\supp(P_Y)}}{\card{\supp(P_{XY})}}\\
&\le \logcard{\set{X}},
\end{align}
where \eqref{eq:kalphacardinalityube} follows from \lmaref{lma:kzerovalueandlimit}.
For $\alpha \in (0,\infty)$, the claim is true because
\begin{align}
K_\alpha(X;Y) &= J_\frac{1}{\alpha}(\widetilde{X};\widetilde{Y})\label{eq:kalphacardinalityuba}\\
&\le \logcard{\set{X}},\label{eq:kalphacardinalityubb}
\end{align}
where \eqref{eq:kalphacardinalityuba} follows from \propref{prop:jalphavskalpha}, and
\eqref{eq:kalphacardinalityubb} follows from \lmaref{lma:jalphacardinalityupperbound}.
For $\alpha = \infty$, the claim is true because $K_\infty(X;Y) = 0$ (\lmaref{lma:kinftyiszero}).
\end{proof}

\begin{Lemma}\label{lma:kalphanodpi}
There exists a Markov chain $X \markov Y \markov Z$ for which $K_2(X;Z) > K_2(X;Y)$.
\end{Lemma}

\begin{proof}
Let the Markov chain $X \markov Y \markov Z$ be given by
\begin{center}
\begin{tabular}{c|cc}
\vphantom{\raisebox{-0.1ex}{$P_{Z|Y}$}}\smash{$P_{XY}(x,y)$} & $y=0$ & $y=1$\\
\hline
$x=0$ & $0.6$ & $0$\\
$x=1$ & $0$ & $0.4$
\end{tabular}
\qquad
\begin{tabular}{c|cc}
\vphantom{\raisebox{-0.1ex}{$P_{Z|Y}$}}\smash{$P_{Z|Y}(z|y)$} & $z=0$ & $z=1$\\
\hline
$y=0$ & $0.9$ & $0.1$\\
$y=1$ & $0$ & $1$
\end{tabular}
\end{center}
Using \lmaref{lma:ktwosingularvalue}, we see that $K_2(X;Z) \approx 0.605$ bits, which is larger than $K_2(X;Y) \approx 0.531$ bits.
\end{proof}

\vspace{6pt}

\authorcontributions{Writing--original draft preparation, A.L. and C.P.; writing--review and editing, A.L.~and~C.P.}

\funding{This research received no external funding.}

\conflictsofinterest{The authors declare no conflict of interest.}

\reftitle{References}


\begin{thebibliography}{999}
\bibitem{Shannon1948} Shannon, C.E. A mathematical theory of communication. \emph{Bell Syst. Tech. J.} \textbf{1948}, \emph{27}, 379--423. [\href{http://dx.doi.org/10.1002/j.1538-7305.1948.tb01338.x}{CrossRef}]
\bibitem{TomamichelHayashi} Tomamichel, M.; Hayashi, M. Operational interpretation of R\'enyi information measures via composite hypothesis testing against product and Markov distributions. \emph{IEEE Trans. Inf. Theory} \textbf{2018}, \emph{64}, 1064--1082. [\href{http://dx.doi.org/10.1109/TIT.2017.2776900}{CrossRef}]
\bibitem{SibsonInformation} Sibson, R. Information radius. \emph{Z. Wahrscheinlichkeitstheorie verw. Geb.} \textbf{1969}, \emph{14}, 149--160. [\href{http://dx.doi.org/10.1007/BF00537520}{CrossRef}]
\bibitem{ArimotoEntropy} Arimoto, S. Information measures and capacity of order $\alpha$ for discrete memoryless channels. In \emph{Topics in Information Theory}; Csisz\'ar, I., Elias, P., Eds.; North-Holland Publishing Company: Amsterdam, The~Netherlands, 1977; pp.~41--52, ISBN 0-7204-0699-4.
\bibitem{CsiszarInformation} Csisz\'ar, I. Generalized cutoff rates and R\'enyi's information measures. \emph{IEEE Trans. Inf. Theory} \textbf{1995}, \emph{41}, 26--34. [\href{http://dx.doi.org/10.1109/18.370121}{CrossRef}]
\bibitem{FehrBerens} Fehr, S.; Berens, S. On the conditional R\'enyi entropy. \emph{IEEE Trans. Inf. Theory} \textbf{2014}, \emph{60}, 6801--6810. [\href{http://dx.doi.org/10.1109/TIT.2014.2357799}{CrossRef}]
\bibitem{SasonVerduArimotoBayesian} Sason, I.; Verd\'u, S. Arimoto--R\'enyi conditional entropy and Bayesian $M$-ary hypothesis testing. \emph{IEEE Trans. Inf. Theory} \textbf{2018}, \emph{64}, 4--25.
[\href{http://dx.doi.org/10.1109/TIT.2017.2757496}{CrossRef}]

\newpage
\bibitem{VerduAlphaMutual} Verd\'u, S. $\alpha$-mutual information. In Proceedings of the 2015 Information Theory and Applications Workshop (ITA), San Diego, CA, USA, 1--6 February 2015; pp.~1--6. [\href{http://dx.doi.org/10.1109/ITA.2015.7308959}{CrossRef}]
\bibitem{TridenskiZamirIngber} Tridenski, S.; Zamir, R.; Ingber, A. The Ziv--Zakai--R\'enyi bound for joint source-channel coding. \emph{IEEE Trans. Inf. Theory} \textbf{2015}, \emph{61}, 4293--4315. [\href{http://dx.doi.org/10.1109/TIT.2015.2445874}{CrossRef}]
\bibitem{AishwaryaMadiman} Aishwarya, G.; Madiman, M. Remarks on R\'enyi versions of conditional entropy and mutual information. In~Proceedings of the 2019 IEEE International Symposium on Information Theory (ISIT), Paris, France, 7--12 July 2019; pp.~1117--1121.
\bibitem{CsiszarShields} Csisz\'ar, I.; Shields, P.C. \emph{Information Theory and Statistics: A Tutorial}; now Publishers: Hanover, MA, USA, 2004; ISBN 978-1-933019-05-5.
\bibitem{LieseVajda} Liese, F.; Vajda, I. On divergences and informations in statistics and information theory. \emph{IEEE Trans. Inf.~Theory} \textbf{2006}, \emph{52}, 4394--4412. [\href{http://dx.doi.org/10.1109/TIT.2006.881731}{CrossRef}]
\bibitem{fDivergenceInequalities} Sason, I.; Verd\'u, S. $f$-divergence inequalities. \emph{IEEE Trans. Inf. Theory} \textbf{2016}, \emph{62}, 5973--6006. [\href{http://dx.doi.org/10.1109/TIT.2016.2603151}{CrossRef}]
\bibitem{JiaoHanWeissman} Jiao, J.; Han, Y.; Weissman, T. Dependence measures bounding the exploration bias for general measurements. In Proceedings of the 2017 IEEE International Symposium on Information Theory (ISIT), Aachen, Germany, 25--30 June 2017; pp.~1475--1479. [\href{http://dx.doi.org/10.1109/ISIT.2017.8006774}{CrossRef}]
\bibitem{ZivZakai} Ziv, J.; Zakai, M. On functionals satisfying a data-processing theorem. \emph{IEEE Trans. Inf. Theory} \textbf{1973}, \emph{19},~275--283. [\href{http://dx.doi.org/10.1109/TIT.1973.1055015}{CrossRef}]
\bibitem{TestingAgainstIndependence} Lapidoth, A.; Pfister, C. Testing against independence and a R\'enyi information measure. In Proceedings of the 2018 IEEE Information Theory Workshop (ITW), Guangzhou, China, 25--29 November 2018; pp.~1--5. [\href{http://dx.doi.org/10.1109/ITW.2018.8613520}{CrossRef}]
\bibitem{HanKobayashi} Han, T.S.; Kobayashi, K. The strong converse theorem for hypothesis testing. \emph{IEEE Trans. Inf. Theory} \textbf{1989}, \emph{35},~178--180. [\href{http://dx.doi.org/10.1109/18.42188}{CrossRef}]
\bibitem{Nakagawa} Nakagawa, K.; Kanaya, F. On the converse theorem in statistical hypothesis testing. \emph{IEEE Trans. Inf. Theory} \textbf{1993}, \emph{39}, 623--628. [\href{http://dx.doi.org/10.1109/18.212293}{CrossRef}]
\bibitem{TaskEncoding} Bunte, C.; Lapidoth, A. Encoding tasks and R\'enyi entropy. \emph{IEEE Trans. Inf. Theory} \textbf{2014}, \emph{60}, 5065--5076. [\href{http://dx.doi.org/10.1109/TIT.2014.2329490}{CrossRef}]
\bibitem{DistributedTaskEncoding} Bracher, A.; Lapidoth, A.; Pfister, C. Distributed task encoding. In Proceedings of the 2017 IEEE International Symposium on Information Theory (ISIT), Aachen, Germany, 25--30 June 2017; pp.~1993--1997. [\href{http://dx.doi.org/10.1109/ISIT.2017.8006878}{CrossRef}]
\bibitem{RenyiEntropyDivergence} R\'enyi, A. On measures of entropy and information. In Proceedings of the Fourth Berkeley Symposium on Mathematical Statistics and Probability, Berkeley, CA, USA, 20 June--30 July 1960; Volume~1, pp.~547--561.
\bibitem{VanErvenHarremoes} van~Erven, T.; Harremo\"es, P. R\'enyi divergence and Kullback--Leibler divergence. \emph{IEEE Trans. Inf. Theory} \textbf{2014}, \emph{60}, 3797--3820. [\href{http://dx.doi.org/10.1109/TIT.2014.2320500}{CrossRef}]
\bibitem{GuessingImprovedBounds} Sason, I.; Verd\'u, S. Improved bounds on lossless source coding and guessing moments via R\'enyi measures. \emph{IEEE Trans. Inf. Theory} \textbf{2018}, \emph{64}, 4323--4346. [\href{http://dx.doi.org/10.1109/TIT.2018.2803162}{CrossRef}]
\bibitem{KumarSundaresanA} Ashok~Kumar, M.; Sundaresan, R. Minimization problems based on relative $\alpha$-entropy I: Forward projection. \emph{IEEE Trans. Inf. Theory} \textbf{2015}, \emph{61}, 5063--5080. [\href{http://dx.doi.org/10.1109/TIT.2015.2449311}{CrossRef}]
\bibitem{KumarSundaresanB} Ashok~Kumar, M.; Sundaresan, R. Minimization problems based on relative $\alpha$-entropy II: Reverse projection. \emph{IEEE Trans. Inf. Theory} \textbf{2015}, \emph{61}, 5081--5095. [\href{http://dx.doi.org/10.1109/TIT.2015.2449312}{CrossRef}]
\bibitem{SundaresanGuessing} Sundaresan, R. Guessing under source uncertainty. \emph{IEEE Trans. Inf. Theory} \textbf{2007}, \emph{53}, 269--287. [\href{http://dx.doi.org/10.1109/TIT.2006.887466}{CrossRef}]
\bibitem{PolyanskiyWu} Polyanskiy, Y.; Wu, Y. {Lecture Notes on Information Theory}. 2017. Available online: \url{http://people.lids.mit.edu/yp/homepage/data/itlectures_v5.pdf } (accessed on 18 August 2017).
\bibitem{CoverThomas} Cover, T.M.; Thomas, J.A. \emph{Elements of Information Theory}, 2nd ed.; John Wiley \& Sons: Hoboken, NJ, USA, 2006; ISBN 978-0-471-24195-9.
\bibitem{Gallager} Gallager, R.G. \emph{Information Theory and Reliable Communication}; John Wiley \& Sons: Hoboken, NJ, USA, 1968; ISBN 978-0-471-29048-3.
\bibitem{HandbookMeansInequalities} Bullen, P.S. \emph{Handbook of Means and Their Inequalities}; Kluwer Academic Publishers: Dordrecht, The~Netherlands, 2003; ISBN 978-1-4020-1522-9.

\bibitem{MatrixAnalysis} Horn, R.A.; Johnson, C.R. \emph{Matrix Analysis}, 2nd ed.; Cambridge University Press: Cambridge, UK, 2013; ISBN~978-0-521-83940-2.
\newpage
\bibitem{KyFan} Fan, K. Minimax theorems. \emph{Proc. Natl. Acad. Sci. USA} \textbf{1953}, \emph{39}, 42--47. [\href{http://dx.doi.org/10.1073/pnas.39.1.42}{CrossRef}] [\href{http://www.ncbi.nlm.nih.gov/pubmed/16589233}{PubMed}]
\bibitem{KyFanAltProof} Borwein, J.M.; Zhuang, D. On Fan's minimax theorem. \emph{Math. Program.} \textbf{1986}, \emph{34}, 232--234. [\href{http://dx.doi.org/10.1007/BF01580587}{CrossRef}]
\end{thebibliography}
\end{document}